\documentstyle[11pt,twoside,epsfig,rotate]{article}
\def\r#1{(\ref{#1})}

\def\R1{\varepsilon_1}
\def\E8{\varepsilon_8}

\newcommand{\nn}{\nonumber}
\newcommand{\mt}{m_{\rm t}}
\newcommand{\mtb}{\overline{m}_{\rm t}}

\newcommand{\bea}{\begin{eqnarray}}
\newcommand{\eea}{\end{eqnarray}}
\newcommand{\be}{\begin{equation}}
\newcommand{\ee}{\end{equation}}
\newcommand{\bi}{\begin{itemize}}
\newcommand{\ei}{\end{itemize}}
\newcommand{\ord}{{\cal O}}

\newcommand{\sss}{\scriptscriptstyle}

\def\npb#1#2#3{    {\it Nucl. Phys. }{\bf B #1} (19#2) #3}
\def\npps#1#2#3{   {\it Nucl. Phys. Proc. Suppl. }{\bf #1} (19#2) #3}
\def\plb#1#2#3{    {\it Phys. Lett. }{\bf B #1} (19#2) #3}
\def\prd#1#2#3{    {\it Phys. Rev. }{\bf D #1} (19#2) #3}
\def\prep#1#2#3{   {\it Phys. Rep. }{\bf #1} (19#2) #3}
\def\prl#1#2#3{    {\it Phys. Rev. Lett. }{\bf #1} (19#2) #3}

\def\rmp#1#2#3{    {\it Rev. Mod. Phys. }{\bf #1} (19#2) #3}
\def\zpc#1#2#3{    {\it Zeit. f{\"u}r Physik }{\bf C #1} (19#2) #3}

\def\sjnp#1#2#3{   {\it Sov. J. Nucl. Phys. }{\bf #1} (19#2) #3}

\begin{document}
\thispagestyle{empty}
\begin{flushright}
 TUM-HEP-339/98 \\
 December 1998
\end{flushright}
\vskip1truecm 
\centerline{\Large\bf Non-Leptonic Two-Body $B$ Decays}
\centerline{\Large\bf Beyond Factorization\footnote[1]{\noindent
    Supported by the German Bundesministerium f\"ur Bildung und
    Forschung under contract 06 TM 874 and by the DFG project Li
    519/2-2.}}  
\vskip1truecm 
\centerline{\large\bf Andrzej J. Buras and Luca Silvestrini} 
\bigskip 
\centerline{\sl Technische
  Universit\"at M\"unchen, Physik Department} 
\centerline{\sl D-85748
  Garching, Germany} 
\vskip1truecm 
\centerline{\bf Abstract} 
We present a general scheme and scale independent parameterization of
two-body non-leptonic $B$ decay amplitudes which includes perturbative
QCD corrections as well as final state interactions in a consistent
way. This parameterization is based on the Next-to-Leading effective
Hamiltonian for non-leptonic $B$ decays and on Wick contractions in
the matrix elements of the local operators. Using this
parameterization, and making no dynamical assumption, we present a
classification of two-body $B$ decay channels in terms of the
parameters entering in the decay amplitudes. This classification can
be considered as the starting point for a model-independent analysis
of non-leptonic $B$ decays and of CP violating asymmetries. We also
propose, on the basis of the large $N$ expansion, a possible hierarchy
among the different effective parameters. We discuss the strategy to
extract the most important effective parameters from the experimental
data. Finally, we establish a connection between our parameterization
and the diagrammatic approach which is widely used in the literature.
  
\vfill 
\newpage 

\section{Introduction}
\label{sec:intro}

A quantitative description of two-body non-leptonic $B$ decays in the
framework of the Standard Model remains as an important challenge for
theorists. Simultaneously these decays play a decisive role in the
study of CP violation and the determination of the
Cabibbo-Kobayashi-Maskawa (CKM) parameters at
$B$-factories and dedicated $B$-physics experiments at hadron
colliders. The studies of these decays should also provide some
insight into the long distance non-perturbative structure of QCD.

The basic theoretical framework for non-leptonic $B$ decays is based
on the Operator Product Expansion (OPE) and renormalization group
methods which allow to write the amplitude for a decay of a given meson
$B$=$B_d$, $B_s$, $B^+$ into a final state $F$=$\pi \pi$, $K \pi$, $D
K$, $KK$, \dots, generally as follows:
\begin{equation}
  \label{eq:ope}
  {\cal A}(B \to F) = \langle F \vert {\cal H}_{\rm eff} \vert B
  \rangle = \frac {G_F}{\sqrt{2}} \sum_i V^{\sss{\rm CKM}}_i C_i(\mu)
  \langle F \vert Q_i(\mu) \vert B \rangle. 
\end{equation}
Here ${\cal H}_{\rm eff}$ is the effective weak Hamiltonian, with
$Q_i$ denoting the relevant local operators which govern the decays in
question. The CKM factors $V^{\sss{\rm CKM}}_i$ and the Wilson
coefficients $C_i(\mu)$ describe the strength with which a given
operator enters the Hamiltonian. In a more intuitive language, the
operators $Q_i(\mu)$ can be regarded as effective vertices and the
coefficients $C_i(\mu)$ as the corresponding effective couplings. The
latter can be calculated in renormalization-group improved
perturbation theory and are known including Next-to-Leading order
(NLO) QCD corrections \cite{NLOQCD1}--\cite{NLOQCD3}. The scale $\mu$
separates the contributions to ${\cal A}(B\to F)$ into short-distance
contributions with energy scales higher than $\mu$ contained in
$C_i(\mu)$ and long-distance contributions with energy scales lower
than $\mu$ contained in the hadronic matrix elements $\langle Q_i(\mu)
\rangle$.  The scale $\mu$ is usually chosen to be $O(m_b)$ but is
otherwise arbitrary. The $\mu$-dependence of $C_i(\mu)$ has to cancel
the $\mu$-dependence of $\langle Q_i(\mu) \rangle$ so that the
physical amplitude ${\cal A}(B \to F)$ is $\mu$-independent.
Similarly, the renormalization scheme dependence of $C_i(\mu)$ cancels
the one of $\langle Q_i(\mu) \rangle$. It should be stressed that
these cancellations involve generally several terms in the expansion
(\ref{eq:ope}).

The great challenge for theorists is a reliable calculation of the
matrix elements $\langle Q_i(\mu) \rangle$ in QCD. Unfortunately, due
to the non-perturbative nature of the problem, the progress towards
this goal has been very slow and it is fair to say that no reliable
calculations of the hadronic matrix elements $\langle Q_i(\mu) \rangle$
in QCD exist at present. In the case of two-body non-leptonic $B$
decays, the $\langle Q_i(\mu) \rangle$ cannot in general be calculated
from first principles in lattice QCD, due to the Maiani-Testa no-go
theorem \cite{MTNGT}, and even the feasibility of a model-dependent
estimate of these matrix elements by numerical simulations still has
to be verified \cite{modlat}. 

In view of this situation a number of strategies for the calculation
of ${\cal A}(B \to F)$ have been used in the literature. The most
extensive analyses have been done in the factorization approach
\cite{fact}--\cite{LNF} in which the hadronic matrix elements
$\langle Q_i(\mu) \rangle$ are replaced by the products of the matrix
elements of weak currents. The latter can be expressed in terms of
various meson decay constants and generally model-dependent form
factors. Thus the decay amplitude in this framework is given simply by
\begin{equation}
  \label{eq:afact}
  {\cal A}_{\rm I,II} = \frac{G_F}{\sqrt{2}} V^{\sss{\rm CKM}}
  a_{1,2}(\mu)  \langle Q_{1,2}(\mu) \rangle_F,
\end{equation}
with
\begin{equation}
  \label{eq:a12f}
  a_{1,2}(\mu) = C_{1,2}(\mu) + \frac{1}{N} C_{2,1}(\mu)
\end{equation}
and $N$ being the number of colours.

Here $\langle Q_i(\mu) \rangle_F$ denote the factorized matrix
elements of the current-current operators $Q_{1,2}$ given explicitly
in Section \ref{sec:amplitudes}. The indices I and II distinguish
between the so-called class I and class II decays. Since in this
approach the matrix elements $\langle Q_i(\mu) \rangle_F$ are scheme-
and $\mu$-independent, the resulting amplitudes have these
dependencies, which are clearly unphysical.

Two ways of remedying these difficulties have been suggested, which
led to the concept of generalized factorization
\cite{NS97}--\cite{soares}. In the formulation due to Neubert and Stech
\cite{NS97}, the $\mu$-dependent parameters $a_{1,2}(\mu)$ are
replaced by the $\mu$- and scheme-independent effective parameters
$a_{1,2}^{\rm eff}$. The latter depend formally on $C_i(\mu)$ and
non-factorizable contributions to $\langle Q_i(\mu) \rangle$ which are
supposed to cancel the $\mu$- and renormalization scheme dependences
of $a_i(\mu)$. In this framework there is no explicit calculation of
non-factorizable contributions and $a_{1,2}^{\rm eff}$ are treated as
free parameters to be extracted from the data. With the assumption of
universality and the neglect of a class of non-factorizable
contributions represented by penguin diagrams and Final State
Interactions (FSI), two-body decays in this approach are parameterized
by the two real free parameters $a_{1,2}^{\rm eff}$.

The generalized factorization presented in \cite{GNF}--\cite{Cheng} is
similar in spirit but includes more dynamics than the formulation in
\cite{NS97}. Here the non-factorizable contributions to the matrix
elements are calculated in a perturbative framework at the one-loop
level. Subsequently these non-factorizable contributions are combined
with the coefficients $C_i(\mu)$ to obtain effective $\mu$ and
renormalization scheme independent coefficients $C_i^{\rm eff}$. The
effective parameters $a^{\rm eff}_i$ are given in this formulation as
follows:
\begin{equation}\label{BS23F}
a_1^{\rm eff}=C^{\rm eff}_1+\frac{1}{N^{\rm eff}} C^{\rm eff}_2 \qquad
a_2^{\rm eff}=C^{\rm eff}_2+\frac{1}{N^{\rm eff}} C^{\rm eff}_1
\end{equation}
with analogous expressions for $a_{i}^{\rm eff}$ ($i=3-10$)
parameterizing penguin contributions.  Here $N^{\rm eff}$ is treated as
a phenomenological parameter which models the non-factorizable
contributions to the hadronic matrix elements.  In particular it has
been suggested in \cite{GNF}--\cite{Cheng} that the values for $N^{\rm
  eff}$ extracted from the data on two-body non-leptonic decays should
teach us about the pattern of non-factorizable contributions.

A critical analysis of these two approaches has been presented by us
in ref.~\cite{BSI}. In particular we have pointed out that the
effective coefficients $C_i^{\rm eff}$ advocated in \cite{GNF,Cheng}
are gauge and infrared regulator dependent. This implies that the
effective number of colours extracted in \cite{GNF,Cheng} also carries
these dependencies, and therefore it cannot have any physical meaning.
Concerning the approach of Neubert and Stech \cite{NS97}, we do not
think that this approach is most suitable for the study of
non-factorizable contributions to non-leptonic decays. In particular,
in the present formulation the contributions from penguin operators,
penguin diagram insertions and final state interactions are not
included. This implies, for instance, that a large number of
penguin-dominated decays cannot be properly described in the present
formulation of this approach. Another important issue in the analyses
performed in generalized factorization is the dependence on the form
factors used. See ref.~\cite{rom} for a detailed discussion of this
point.

Another, more general, approach to non-leptonic decays is the
diagrammatic approach, in which the decay amplitudes are decomposed
into various contributions corresponding to certain flavour-flow
topologies which in the literature appear under the names of
``trees'', ``colour-suppressed trees'', ``penguins'',
``annihilations'' etc. \cite{diagr1,diagr2}. Supplemented by isospin
symmetry, the approximate $SU(3)$ flavour symmetry and various
``plausible'' dynamical assumptions the diagrammatic approach has been
used extensively for non-leptonic $B$ decays in the nineties.

Recently the usefulness of the diagrammatic approach has been
questioned with respect to the effects of final state interactions. In
particular various ``plausible'' diagrammatic arguments to neglect
certain flavour-flow topologies may not hold in the presence of FSI,
which mix up different classes of diagrams
\cite{FSI}--\cite{CHARMING1}.  Another criticism which one may add is
the lack of an explicit relation of this approach to the basic
framework for non-leptonic decays represented by the effective weak
Hamiltonian and OPE in eq.~(\ref{eq:ope}). In particular, the
diagrammatic approach is governed by Feynman drawings with $W$-, $Z$-
and top-quark exchanges.  Yet such Feynman diagrams with full
propagators of heavy fields represent really the situation at very
short distance scales $O(M_{W,Z},m_t)$, whereas the true picture of a
decaying meson with a mass $O(m_b)$ is more properly described by
effective point-like vertices represented by the local operators
$Q_i$. The effect of $W,Z$ and top quark exchanges is then described
by the values of the Wilson coefficients of these operators. The only
explicit fundamental degrees of freedom in the effective theory are
the quarks $u$, $d$, $s$, $c$, $b$, the gluons and the photon.

In view of this situation it is desirable to develop another
phenomenological approach based directly on the OPE which does not
have the limitations of generalized factorization and allows
a systematic description of non-factorizable contributions such as
penguin contributions and final state interactions. Simultaneously one
would like to have an approach that does not lose the intuition of the
diagrammatic approach while avoiding the limitations of the latter.

First steps in this direction have been made in
refs.~\cite{bfm,CHARMING1,flei}. In refs.~\cite{bfm,flei} some of the
parameters of the diagrammatic approach have been, in the case of $B
\to K \pi$, expressed in terms of matrix elements of local operators.
On the other hand in ref.~\cite{CHARMING1} the amplitudes for $B \to
\pi \pi$, $B \to K \pi$ and $B \to K K$ have been given in terms of
diagrams representing Wick contractions of the operators of the
effective Hamiltonian between the relevant hadronic states.

Now the formulation of non-leptonic decays given in
ref.~\cite{CHARMING1} involves the explicit expansion of the decay
amplitudes in terms of Wick contractions; therefore, with ten
operators entering the basic formula (\ref{eq:ope}) and several
possible contractions, one ends up with over hundred different
contributions, each of them being scale and renormalization scheme
dependent.

In the present paper we make the approach of ref.~\cite{CHARMING1}
manifestly scale and scheme independent. In this context we introduce
a set of effective scale and scheme independent parameters which are
given as linear combinations of particular Wick contractions of the
operators times the corresponding Wilson coefficients. This
reformulation of the approach in \cite{CHARMING1} in terms of
effective parameters results in more transparent formulae for the
decay amplitudes than given in \cite{CHARMING1}, establishes some
connection with the usual diagrammatic approach and is more suitable
for approximations. Moreover we include additional Zweig-suppressed
Wick contractions, not considered in the literature, and we study
in addition to charmless final states also those including the charm
flavour.

The formulae for the effective parameters given here allow in
principle their calculation in QCD by means of lattice techniques or
other non-perturbative methods. Since this is not possible at present
they have to be considered as free parameters to be determined from
the data.

It turns out that the full description of two-body $B$-decays requires
the introduction of fourteen flavour-dependent effective
parameters. With the help of large $N$ ideas and plausible dynamical
assumptions one can argue that several of these parameters play only a
minor role in two-body non-leptonic decays.

The approach presented here allows for a general phenomenological
description of non-leptonic decays which with more data could teach us
about the role of non-factorizable contributions and about the flavour
structure of non-leptonic decays. In order to be more predictive some
symmetry relations, as $SU(3)$ relations \cite{GL}, and dynamical
input are needed. In this context we would like to mention an
interesting work on non-factorizable contributions to non-leptonic
decays within the QCD sum rules approach \cite{QCDSR}. We will return
to several of these issues in a subsequent publication.

Our paper is organized as follows. In Section \ref{sec:amplitudes} we
recall the complete effective weak Hamiltonian. In Section
\ref{sec:class} we classify various topologies of Wick
contractions. We identify fourteen flavour topologies of which only
nine have been considered in the literature. The new topologies
correspond to Zweig-suppressed transitions. They may play a role only
in certain decays but strictly speaking they have to be included for
consistency.

In Section \ref{sec:general} we introduce the effective parameters in
general terms. In Section \ref{sec:details} we derive by means of a
diagrammatic technique the explicit expressions for these parameters
including flavour dependence. In Section \ref{sec:approx} we propose,
using the $1/N$ expansion, a hierarchical structure for the effective
parameters. In Section \ref{sec:classific} we classify the $B_d$ and
$B^+$ decays into suitable classes and we give explicit expressions
for a large number of decays in terms of the most important effective
parameters, following the hierarchy proposed in Section
\ref{sec:approx}. In Section \ref{sec:classificbs} we give analogous
expressions for $B_s$ decays.  The contributions neglected in the
analysis of Sections \ref{sec:classific} and \ref{sec:classificbs} are
collected in Appendix \ref{sec:appew} for completeness. In Section
\ref{sec:pheno} we discuss briefly strategies for the determination of
some of the effective parameters from the data and in Section
\ref{sec:comparison} we compare the present approach with the
diagrammatic approach of refs.~\cite{diagr1,diagr2}. We end our paper
with a brief summary and conclusions. A detailed application of this
formalism to two-body $B$ decays, in particular to CP asymmetries, will
be presented elsewhere.

\section{Effective  Hamiltonian}
\label{sec:amplitudes}

The complete effective weak Hamiltonian for non-leptonic $B$ decays is
given by: 
\begin{eqnarray} 
{\cal H}_{\rm eff} &=& \frac {G_F} {\sqrt{2}} 
\Biggl\{ V_{ub} V^*_{ud} \biggl[C_1(\mu)\Bigl( Q^{duu}_1(\mu) 
  - Q_1^{dcc}(\mu) \Bigr) + C_2(\mu)\Bigl( Q^{duu}_2(\mu) 
  - Q_2^{dcc}(\mu) \Bigr)  \biggr] \nn  \\
&& -V_{tb} V^*_{td} \, \biggl[C_1(\mu) Q_1^{dcc}(\mu) + 
C_2(\mu) Q_2^{dcc}(\mu) + \sum_{i=3,10} C_i(\mu) Q^{d}_i(\mu)
\biggr] \nn \\
&& + V_{ub} V^*_{us} \biggl[C_1(\mu)\Bigl( Q^{suu}_1(\mu) 
  - Q_1^{scc}(\mu) \Bigr) + C_2(\mu)\Bigl( Q^{suu}_2(\mu) 
  - Q_2^{scc}(\mu) \Bigr)  \biggr] \nn  \\
&& -V_{tb} V^*_{ts} \, \biggl[C_1(\mu) Q_1^{scc}(\mu) + 
C_2(\mu) Q_2^{scc}(\mu) + \sum_{i=3,10} C_i(\mu) Q^{s}_i(\mu)
\biggr] \nn \\ 
&& + V_{ub} V^*_{cs} \biggl[C_1(\mu) Q^{scu}_1(\mu) 
+ C_2(\mu) Q^{scu}_2(\mu) \biggr]
 + V_{cb} V^*_{us} \biggl[C_1(\mu) Q^{suc}_1(\mu) 
+ C_2(\mu) Q^{suc}_2(\mu) \biggr] \nn  \\
&& + V_{ub} V^*_{cd} \biggl[C_1(\mu) Q^{dcu}_1(\mu) 
+ C_2(\mu) Q^{dcu}_2(\mu) \biggr] 
 + V_{cb} V^*_{ud} \biggl[C_1(\mu) Q^{duc}_1(\mu) 
+ C_2(\mu) Q^{duc}_2(\mu) \biggr] \Biggr\}\,.
\protect\label{eh} 
\end{eqnarray}
A basis of operators convenient for our considerations is the one in
which all operators are written in the colour singlet form:
\begin{equation}
  \begin{array}{ll}
    Q^{d_i u_j u_k}_{ 1}=({\bar b}u_k)_{ (V-A)}
    ({\bar u_j }d_i)_{ (V-A)}\,,
    &
    Q^{d_i u_j u_k}_{ 2}=({\bar b} d_i)_{ (V-A)}
    ({\bar u_j }u_k)_{ (V-A)}\,,
    \\
    Q^{d_i}_{J=3,...,10} = \sum_{q} Q^{d_i
    q}_{J} \,, & \\
    Q^{d_i q}_{ 3,5} = ({\bar b}d_i)_{ (V-A)}
    ({\bar q}q)_{ (V\mp A)}\,,
    &
    Q^{d_i q}_{ 4} = ({\bar b}q)_{ (V-A)}
    ({\bar q}d_i)_{ (V - A)}\,,
    \\
    Q^{d_i q}_{6} = -2 ({\bar b}q)_{ (S+P)}
    ({\bar q}d_i)_{ (S-P)}\,,
    &
    Q^{d_i q}_{ 7,9} = \frac{3}{2}({\bar b}d_i)_
    { (V-A)} e_{ q}({\bar q}q)_
    { (V\pm A)}\,,  
    \\
    Q^{d_i q}_{8} = - 3 e_q({\bar b}q)_
    { (S+P)} ({\bar q}d_i)_{ (S-P)}\,, 
    &
    Q^{d_i q}_{ 10} = \frac{3}{2} e_{ q}({\bar b}q)_
    { (V-A)}({\bar q}d_i)_{ (V -  A)}\,,
  \end{array}  
  \label{eq:basis}
\end{equation}
where the subscripts $(V \pm A)$ and $(S \pm P)$ 
indicate the chiral structures, $d_i=\{d,s\}$, $u_i=\{u,c\}$ and $e_q$
denotes the quark electric charge ($e_u=2/3$, $e_d=-1/3$, etc.).
The sum over the quarks $q$  runs over the active flavours at the
scale $\mu$. Note that we use here the labeling of the operators as
given in 
\cite{BAUER,NEUBERT} which differs from 
\cite{NLOQCD1}--\cite{NLOQCD3} by the interchange
$1\leftrightarrow 2$.

$Q_1$ and $Q_2$ are the so-called current-current operators, $Q_{3-6}$
the QCD-penguin operators and $Q_{7-10}$ the electroweak penguin
operators. $C_i(\mu)$ are the Wilson coefficients evaluated at $\mu =
O(m_b)$. They depend generally on the renormalization scheme for the
operators. For example in the HV scheme one has, including NLO
corrections and setting $\mtb(\mt)=170$ GeV, $\mu=4.4$ GeV and
$\alpha_s^{\overline{MS}} (M_Z)=0.118$ \cite{BBL}:
\begin{equation}
  \label{eq:coeffi}
  \begin{array}{ccccc}
    C_1=1.105, & C_2=-0.228, &
    C_3=0.013, & C_4=-0.029, &
    C_5=0.009, \\
    C_6=-0.033, &
    C_7/\alpha= 0.005, & C_8/\alpha = 0.060, &
    C_9/\alpha=-1.283, & C_{10}/\alpha = 0.266,
  \end{array}
\end{equation}
where $\alpha$ is the electromagnetic coupling constant.

The basis in \r{eq:basis} differs from the one used in the NLO
calculations in \cite{NLOQCD2, NLOQCD3} in that some of the operators
used there are Fierz conjugates of the ones in \r{eq:basis}. This is
the case for $Q_2$, $Q_4$, $Q_6$, $Q_8$ and $Q_{10}$. As pointed out
in \cite{NLOQCD2}, the Wilson coefficients evaluated in the NDR scheme
(anticommuting $\gamma_5$ in $D\neq 4$ dimensions) depend on the form
of the operators and the standard Wilson coefficients in the NDR
scheme as given in \cite{NLOQCD2, BBL} cannot be used in conjunction
with the basis \r{eq:basis}. On the other hand the HV scheme is
Fierz-symmetric and the Wilson coefficients in this scheme calculated
in \cite{NLOQCD2, NLOQCD3} also apply to the basis \r{eq:basis}. We
will comment at the end of Section \ref{sec:general} on how the
expressions for the effective parameters given there have to be
modified when the NDR scheme with the basis of \cite{NLOQCD2,NLOQCD3}
is used.

We observe that, unless there is some huge enhancement of their matrix
elements, the contributions of the electroweak penguin operators $Q_7$
and $Q_8$ are fully negligible. The operators $Q_9$ and $Q_{10}$ play
only a role in decays in which for some dynamical reasons the matrix
elements of current-current operators and QCD-penguin operators are
strongly suppressed. Generally also the contributions of QCD-penguin
operators, especially of $Q_3$ and $Q_5$, are substantially smaller
than those of the current-current operators $Q_1$ and $Q_2$.

\section{Classification of Topologies}
\label{sec:class}

\begin{figure}   % produce figure here
    \begin{flushleft}
\input{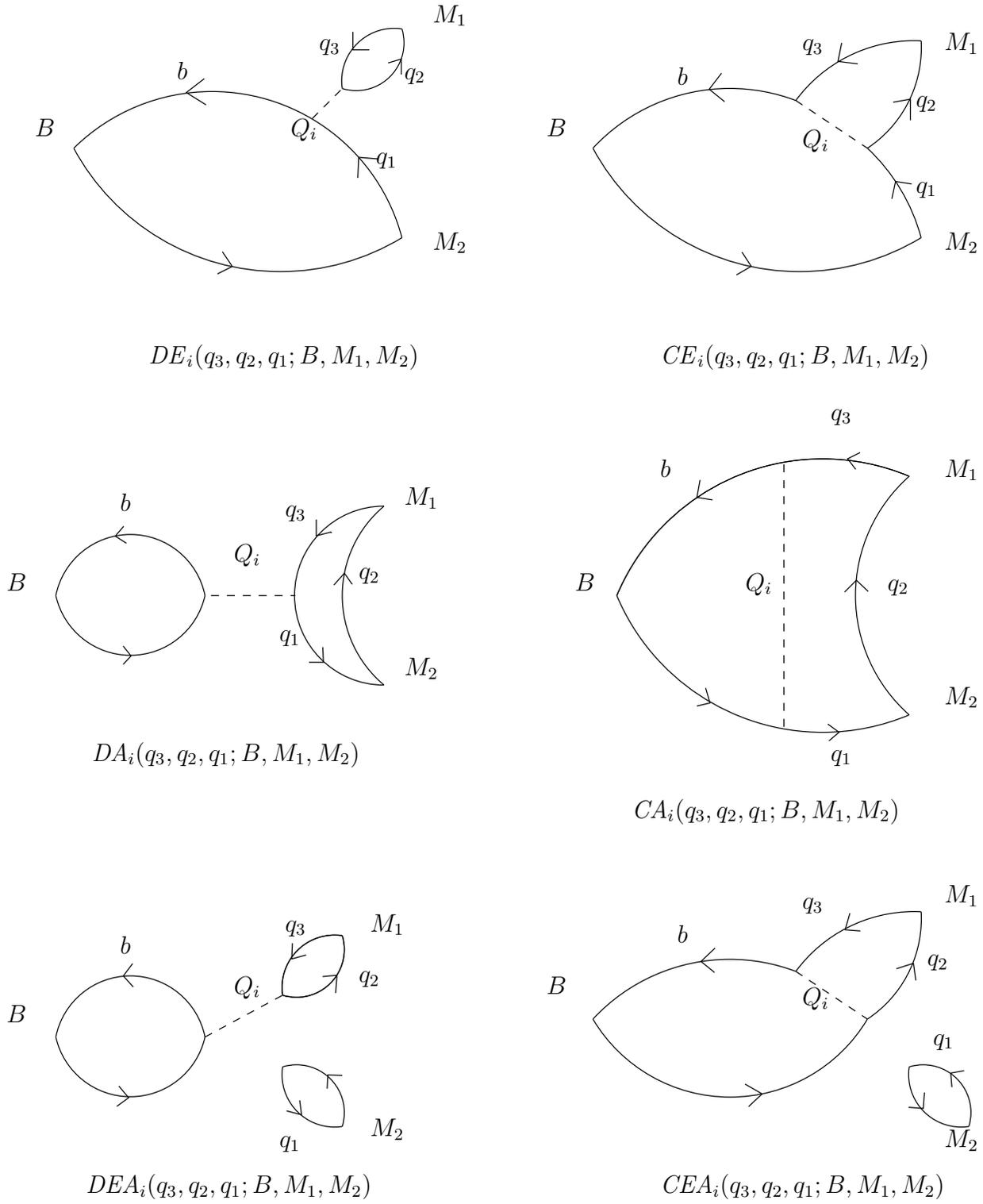}
    \end{flushleft}
    \caption[]{Emission, annihilation and emission-annihilation
      topologies of Wick contractions in the matrix elements of
      operators $Q_i$.}
    \label{fig:emiss-ann}
\end{figure}

Following and generalizing the discussion of ref.~\cite{CHARMING1}, we
classify the various contributions to the matrix elements of the
operators $Q_i$ distinguishing the different topologies of Wick
contractions as in figs.~\ref{fig:emiss-ann} and \ref{fig:penguins}.
We have emission topologies: Disconnected Emission ({\it DE}\/) and
Connected Emission ({\it CE\/}); annihilation topologies: Disconnected
Annihilation ({\it DA\/}) and Connected Annihilation ({\it CA\/});
emission-annihilation topologies: Disconnected Emission-Annihilation
({\it DEA}\/) and Connected Emission-Annihilation ({\it CEA\/});
penguin topologies: Disconnected Penguin ({\it DP\/}) and Connected
Penguin ({\it CP\/}); penguin-emission topologies: Disconnected
Penguin-Emission ({\it DPE\/}) and Connected Penguin-Emission ({\it
  CPE\/}); penguin-annihilation topologies: Disconnected
Penguin-Annihilation ({\it DPA\/}) and Connected Penguin-Annihilation
({\it CPA\/}); double-penguin-annihilation topologies: Disconnected
Double-Penguin-Annihilation ($\overline{\it DPA}$) and Connected
Double-Penguin-Annihilation ($\overline{\it CPA}$). The unlabeled line
corresponds to a $d$, $u$ or $s$ quark for $B_d$, $B^+$ and $B_s$
decays respectively. The dashed lines represent the operators. The
apparently disjoint pieces in the topologies {\it DEA}, {\it CEA},
{\it DPE}, {\it CPE}, {\it DPA}, {\it CPA}, $\overline{\it DPA}$ and
$\overline{\it CPA}$ are connected to each other by gluons or photons,
which are not explicitly shown. A comment about these topologies is
necessary at this point. 

These special topologies in which only gluons connect the disjoint
pieces are Zweig suppressed and are therefore naively expected to play
a minor role in $B$ decays. For this reason, they have been neglected
in previous studies \cite{CHARMING1}. However, as we shall demonstrate
in the next Section, they have to be included in order to define
scheme and scale independent combinations of Wilson coefficients and
matrix elements. We therefore take all of them into account in the
following discussion of the effective scheme and scale independent
parameters suitable to describe $B$-decay amplitudes. We will later
discuss some approximations that will allow us to reduce the number of
parameters necessary to parameterize non-leptonic $B$ decays. On the
other hand, if the disjoint pieces are connected by photons, these
contributions are automatically at least of $O(\alpha)$.

\begin{figure}   % produce figure here
    \begin{flushleft}
\input{penguins.pstex_t}
    \end{flushleft}
    \caption[]{Penguin, penguin-emission, penguin-annihilation and
      double-penguin-annihilation topologies of Wick contractions in
      the matrix elements of operators $Q_i$.}
    \label{fig:penguins}
\end{figure}

Our aim is to find a parameterization of the full contribution
to the amplitude, including non-factorizable contributions and
rescattering effects. Therefore, we will keep all the contributions
defined above and consider them to be complex, in order to take into
account final state interactions. 

The large number of different possible contributions, with different
chiralities and flavour structures, requires the introduction of
fourteen flavour-dependent, scale and renormalization scheme
independent complex parameters in order to be able to describe all
decay amplitudes. The fact that the number of effective parameters
equals the number of different topologies is accidental. This
relatively large number of parameters is necessary if we do not want
to make any specific assumption about non-factorizable effects and
FSI. Yet, as we will see, this approach offers a transparent
classification of various possible contributions and constitutes a
good starting point for approximations which would reduce the number
of parameters.

In general, the topologies defined above will depend on the flavour
and chiral structure, and on the initial and final states. In fact, if
factorization were to hold, all this dependence would be taken into
account by the factorized matrix element; however, non-factorizable
contributions and FSI will in general have a different flavour
dependence and cause for example a violation of the universality of
the parameters $a_1^{\rm eff}$ and $a_2^{\rm eff}$ introduced in
ref.~\cite{NS97} in the framework of generalized factorization. Since
here we take into account non-factorizable effects, we keep track of
the flavour dependence.

\section{Effective Parameters -- Generalities}
\label{sec:general}

The matrix elements of the operators $Q_i(\mu)$ depend in general on
the renormalization scale $\mu$ and on the renormalization scheme for
the operators. These unphysical dependences are cancelled by those
present in the Wilson coefficients $C_i(\mu)$. Due to the mixing under
renormalization this cancellation involves generally several
operators. From the phenomenological point of view, it is desirable to
identify those linear combinations of operator matrix elements times
the corresponding Wilson coefficients which are both scheme and scale
independent. Such combinations will define the effective parameters to
be used in phenomenological applications.

Before entering the details, let us make a few general comments on how
the scale and scheme independent effective parameters can be found. It
turns out that the flavour structure of the operators $Q_1$ and $Q_2$,
inserted in the various topologies listed above, and the known
behaviour of the operators $Q_i$ under renormalization group
transformations, allow us to identify the independent effective parameters. 

The first scale and scheme independent combinations of Wilson
coefficients and matrix elements that one can identify correspond to
the emission matrix elements of the current-current operators $Q_1$
and $Q_2$. Denoting by $\langle Q_i \rangle_{DE}$ and $\langle Q_i
\rangle_{CE}$ the insertions of $Q_i$ into ${\it DE}$ and ${\it CE}$
topologies respectively, one finds two effective parameters
\begin{eqnarray}
  E_1&=& C_1\, \langle Q_1 \rangle_{{\it DE}} + C_2 \,\langle Q_2 
  \rangle_{{\it CE}}
  \,, \nonumber \\
  E_2&=& C_1\, \langle Q_1 \rangle_{{\it CE}} + C_2 \,\langle Q_2 
  \rangle_{{\it DE}}\,.
  \label{eq:E1E20}
\end{eqnarray}
We have suppressed the flavour variables for the moment.  They will be
given explicitly in Section \ref{sec:details}. $E_1$ and $E_2$ are
generalizations of $a_1^{\rm eff}\langle Q_1 \rangle_F$ and $a_2^{\rm
  eff}\langle Q_2 \rangle_F$ in the formulation of ref.~\cite{NS97}.

The reason why two effective parameters are needed can be found by
switching off QCD effects, in which case $C_1=1$ and $C_2=0$.
Dependently then on the decay channel considered, the operator $Q_1$
may contribute either through topology ${\it DE}$ or through topology
${\it CE}$, which gives $E_1$ and $E_2$ respectively. There are of
course channels whose flavour structure allows both $\langle Q_1
\rangle_{{\it DE}}$ and $\langle Q_1 \rangle_{{\it CE}}$. These
channels can be described by a third effective parameter
which is, however, a linear combination of $E_1$ and $E_2$.
Consequently from the point of view of scale and scheme dependences
it cannot be considered as a new effective parameter.

The reason why $E_1$ and $E_2$ are scale and scheme independent can be
understood in the following manner. Consider operators $Q_1$ and $Q_2$
in the case in which all four quark flavours are different. In this
case the penguin topologies do not contribute and also the penguin
operators cannot be generated by QCD corrections. Choosing in addition
channels in which also annihilation contributions are absent, we
observe that $E_1$ and $E_2$ represent, up to CKM factors, physical
amplitudes for the particular channels in question and as such must be
scale and scheme independent.

Next, annihilation topologies must be considered. Here we can consider
channels in which both emission and penguin topologies as well as
penguin operators do not contribute. Analogous arguments as given for
$E_1$ and $E_2$ allow us to identify two new effective parameters: 
\begin{eqnarray}
  A_1&=& C_1\, \langle Q_1 \rangle_{\it DA} + C_2 \,\langle Q_2 
  \rangle_{\it CA}
  \,, \nonumber \\
  A_2&=& C_1\, \langle Q_1 \rangle_{\it CA} + C_2 \,\langle Q_2 
  \rangle_{\it DA}\,,
  \label{eq:A1A20}
\end{eqnarray}
where $\langle Q_i \rangle_{\it DA}$ and $\langle Q_i \rangle_{\it
  CA}$ denote the $Q_i$-insertions into ${\it DA}$ and ${\it CA}$
topologies respectively. Due to the flavour structure of operators
$Q_1$ and $Q_2$, $A_1$ can only contribute to $B^+$ decays while $A_2$
can only contribute to $B_{d,s}$ decays.

It should be stressed that the $A_i$ are independent of the
$E_i$. Indeed, the arguments given above demonstrate that the
cancellations of scheme and scale dependences take place separately
within emission and annihilation topologies when only current-current
operators $Q_1$ and $Q_2$ are considered.

The last class of non-penguin contractions that we consider
corresponds to the insertion of $Q_1$ and $Q_2$ into
emission-annihilation topologies, denoted by ${\it DEA}$ and ${\it
  CEA}$ in fig.~\ref{fig:emiss-ann}. Proceeding as above, we can
identify two new effective parameters:
\begin{eqnarray}
  {\it EA}_1&=& C_1\, \langle Q_1 \rangle_{\it DEA} + C_2 \,\langle Q_2 
  \rangle_{\it CEA}, \nonumber \\
  {\it EA}_2&=& C_1\, \langle Q_1 \rangle_{\it CEA} + C_2 \,\langle Q_2 
  \rangle_{\it DEA}.
  \label{eq:EA1EA20}
\end{eqnarray}
As in the case of $A_1$ and $A_2$, due to the flavour structure of
$Q_1$ and $Q_2$, ${\it EA}_1$ can only contribute to $B^+$ decays while
${\it EA}_2$ can only contribute to $B_{d,s}$ decays.

We next turn to penguin contractions of current-current operators and
matrix elements of penguin operators. Several arguments can be given
to conclude that these two types of contributions should be combined
in order to obtain scale and scheme independent effective parameters.

First of all let us recall that the Wilson coefficients $C_1(\mu)$ and
$C_2(\mu)$ are independent of the presence of penguin operators. The
same applies for the insertions of $Q_1$ and $Q_2$ into emission and
annihilation topologies. Consequently the scale and scheme independent
effective parameters $E_1$, $E_2$, $A_1$, $A_2$, $EA_1$ and $EA_2$ are
unaffected by the presence of penguin operators. This means that the
sum of the remaining contributions to a physical amplitude, that is of
penguin contractions of $Q_1$ and $Q_2$ and matrix elements of penguin
operators, should be separately scheme and scale independent.

In order to see this more clearly let us focus on $b\to s \bar d d$
decays. Consider first ${\cal H}_{\rm eff}$ to be evaluated at a scale
$\mu_1 > m_c$. At this scale, we have penguin insertions of
current-current operators containing two charm quarks ($Q_{1,2}^{s c
  c}$) in penguin, penguin-emission, penguin-annihilation and
double-penguin-annihilation topologies. In addition, we have emission,
annihilation and emission-annihilation matrix elements of penguin
operators ($Q^{s}_{3-10}$). Now take instead ${\cal H}_{\rm eff}$
computed at a scale $\mu_2 < m_c$: the operators $Q_{1,2}^{s c c}$
have disappeared, and the coefficients of operators $Q^{s}_{3-10}$
have changed in such a way as to compensate for the absence of
$Q_{1,2}^{s c c}$.

Now similarly to the sets $(E_1,E_2)$, $(A_1,A_2)$ and $({\it
  EA}_1,{\it EA}_2)$ one can find four effective
``penguin''-parameters $P_1$, $P_2$, $P_3$ and $P_4$. The explicit
expressions for them will be derived below.  Here we just want to
relate these parameters to the penguin topologies introduced in
Section \ref{sec:class}:
\begin{enumerate}
\item $P_1$ involves the insertions of $Q_1$ and $Q_2$ into ${\it CP}$
  and ${\it DP}$ topologies respectively and a particular set of
  matrix elements of QCD-penguin and electroweak penguin operators
  necessary for the cancellation of scale and scheme dependences;
  
\item $P_2$ involves the insertions of $Q_1$ and $Q_2$ into ${\it
    CPE}$ and ${\it DPE}$ topologies respectively and a suitable set
  of matrix elements of QCD-penguin and electroweak penguin operators
  necessary for the cancellation of scale and scheme dependences;
  
\item $P_3$ involves the insertions of $Q_1$ and $Q_2$ into ${\it
    CPA}$ and ${\it DPA}$ topologies respectively and the
  corresponding set of matrix elements of QCD-penguin and electroweak
  penguin operators necessary for the cancellation of scale and scheme
  dependences;
  
\item $P_4$ involves the insertions of $Q_1$ and $Q_2$ into
  $\overline{\it CPA}$ and $\overline{\it DPA}$ topologies
  respectively and the remaining matrix elements of QCD-penguin and
  electroweak penguin operators which have not been included in $P_1$,
  $P_2$ and $P_3$.
\end{enumerate}

In spite of the fact that in each case both disconnected and connected
topologies are present only four effective parameters exist and not
eight. This is related to the fact that the insertions of $Q_1$ into
disconnected penguin topologies and the insertions of $Q_2$ into
connected penguin topologies vanish because of the flavour structure
of these operators. This should be contrasted with the insertions of
these operators into emission and annihilation topologies, where in
each case both connected and disconnected topologies can contribute.

The explicit expressions for the $P_1$, $P_2$, $P_3$ and $P_4$
parameters are as follows:
\begin{eqnarray}
  P_1&=&C_1 \langle Q_1\rangle^c_{\it CP} + 
  C_2 \langle Q_2\rangle^c_{DP} 
  +\sum_{i=2}^5 \Bigl(C_{2i-1} \langle Q_{2i-1} \rangle_{{\it CE}}
  + C_{2i} \langle Q_{2i} \rangle_{{\it DE}}\Bigr) \nn \\
  &+&\sum_{i=3}^{10} \Bigl(C_{i} \langle Q_{i} \rangle_{\it CP}
  + C_{i} \langle Q_{i} \rangle_{DP}\Bigr)  
  +\sum_{i=2}^5 \Bigl(C_{2i-1} \langle Q_{2i-1} \rangle_{\it CA}
  + C_{2i} \langle Q_{2i} \rangle_{\it DA}\Bigr)\, , \label{eq:p1gen} \\
  P_2&=&C_1 \langle Q_1 \rangle^c_{\it CPE} + 
  C_2 \langle Q_2 \rangle^c_{\it DPE} 
  + \sum_{i=2}^5 \Bigl(C_{2i-1} \langle Q_{2i-1} \rangle_{{\it DE}}
  + C_{2i} \langle Q_{2i} \rangle_{{\it CE}}\Bigr) \nn \\
  &+&\sum_{i=2}^5 \Bigl(C_{2i-1} \langle Q_{2i-1} \rangle_{{\it CEA}}
  + C_{2i} \langle Q_{2i} \rangle_{{\it DEA}}\Bigr)
  + \sum_{i=3}^{10} \Bigl(C_{i} \langle Q_{i} \rangle_{\it CPE}
  + C_{i} \langle Q_{i} \rangle_{\it DPE}\Bigr)\, , \label{eq:p2gen} \\
  P_3&=&C_1 \langle Q_1 \rangle^c_{\it CPA} + 
  C_2 \langle Q_2 \rangle^c_{\it DPA} 
  + \sum_{i=2}^5 \Bigl(C_{2i-1} \langle Q_{2i-1} \rangle_{\it DA}
  + C_{2i} \langle Q_{2i} \rangle_{\it CA}\Bigr) \nn \\
  &+&\sum_{i=3}^{10} \Bigl(C_{i} \langle Q_{i} \rangle_{\it CPA}
  + C_{i} \langle Q_{i} \rangle_{\it DPA}\Bigr)\, , \label{eq:p3gen} \\
  P_4&=&C_1 \langle Q_1 \rangle^c_{\overline{\it CPA}} + 
  C_2 \langle Q_2 \rangle^c_{\overline{\it DPA}} 
  + \sum_{i=2}^5 \Bigl(C_{2i-1} \langle Q_{2i-1} \rangle_{\it DEA}
  + C_{2i} \langle Q_{2i} \rangle_{\it CEA}\Bigr) \nn \\
  &+&\sum_{i=3}^{10} \Bigl(C_{i} \langle Q_{i} \rangle_{\overline{\it CPA}}
  + C_{i} \langle Q_{i} \rangle_{\overline{\it DPA}}\Bigr)\, ,
  \label{eq:p4gen}
\end{eqnarray}
where we have denoted by $\langle Q_i\rangle^c_{\it CP}$ the insertion
of operator $Q_i$ in a {\it CP\/} topology with a $c$-quark running in
the loop (this corresponds to the charming penguin of
ref.~\cite{CHARMING1}), and analogously for ${\it DP}$, ${\it CPE}$,
${\it DPE}$, ${\it CPA}$, ${\it DPA}$, $\overline{\it CPA}$ and
$\overline{\it DPA}$ topologies.  We notice that, due to the flavour
structure of the penguin-annihilation contributions, $P_3$ and $P_4$
cannot contribute to $B^+$ decays. Moreover $P_4$ contributes only to
final states with two flavour neutral mesons $\bar q_1 q_1$ and $\bar
q_2 q_2$. Similarly $P_2$ contributes only to states with at least one
flavour neutral meson $\bar q_2 q_2$. 

We also stress that, in order to cancel the scheme and scale
dependence of the emission contractions of penguin operators in the
first line of the expression \r{eq:p2gen} for $P_2$, one is
forced to introduce the $\it{CPE}$, ${\it DPE}$, ${\it CEA}$ and ${\it
  DEA}$ topologies that also contribute to $P_2$. Analogously, to
cancel the scheme and scale dependence of the annihilation
contractions of penguin operators in the first line of the expression
\r{eq:p3gen} for $P_3$, one has to consider the ${\it CPA}$ and
${\it DPA}$ topologies that also contribute to $P_3$. Finally, when
considering the insertion of penguin operators in the ${\it CEA}$ and
${\it DEA}$ topologies (first line of the expression \r{eq:p4gen} for
$P_4$), one has to introduce $\overline{\it CPA}$ and $\overline{\it
  DPA}$ Wick contractions to ensure the scheme and scale dependence of
$P_4$.  Therefore, the Zweig-suppressed $\it{CPE}$, ${\it DPE}$, ${\it
  CEA}$, ${\it DEA}$, ${\it CPA}$, ${\it DPA}$, $\overline{\it CPA}$
and $\overline{\it DPA}$ topologies are all needed to obtain a
complete scheme and scale independent result.

The $P_1$, $P_2$, $P_3$ and $P_4$ parameters are always accompanied by
the CKM factor $V_{tb} V_{td_i}^*$, where $d_i={d,s}$. Penguin-type
matrix elements are also present in the part of ${\cal H}_{\rm eff}$
proportional to $V_{ub} V_{ud_i}^*$. They correspond to penguin
contractions of operators $Q_{1,2}$, or, more precisely, of the
differences $\Bigl(Q_1^{d_iuu}-Q_1^{d_icc}\Bigr)$ and
$\Bigl(Q_2^{d_iuu}-Q_2^{d_icc}\Bigr)$. When these combinations are
inserted into penguin topologies, they give rise to a generalization
of the GIM penguins of ref.~\cite{CHARMING1}. These scale and scheme
independent contributions are the following:
\begin{eqnarray}
  P_1^{\sss {\rm GIM}}&=& C_1 \Bigl(\langle Q_1 \rangle^c_{\it CP} -
  \langle Q_1 
  \rangle^u_{\it CP} \Bigr) 
  + C_2 \Bigl(\langle Q_2 \rangle^c_{DP} - \langle Q_2
  \rangle^u_{DP} \Bigr)\,,\nn \\
  P_2^{\sss {\rm GIM}}&=& C_1 \Bigl(\langle Q_1 \rangle^c_{\it CPE} -
  \langle Q_1 
  \rangle^u_{\it CPE} \Bigr) 
  + C_2 \Bigl(\langle Q_2 \rangle^c_{\it DPE} - \langle Q_2
  \rangle^u_{\it DPE} \Bigr)\,,\nn \\
  P_3^{\sss {\rm GIM}}&=& C_1 \Bigl(\langle Q_1 \rangle^c_{\it CPA} -
  \langle Q_1 
  \rangle^u_{\it CPA} \Bigr) 
  + C_2 \Bigl(\langle Q_2 \rangle^c_{\it DPA} - \langle Q_2
  \rangle^u_{\it DPA} \Bigr)\,,\nn \\
  P_4^{\sss {\rm GIM}}&=& C_1 \Bigl(\langle Q_1
  \rangle^c_{\overline{\it CPA}} - 
  \langle Q_1 
  \rangle^u_{\overline{\it CPA}} \Bigr) 
  + C_2 \Bigl(\langle Q_2 \rangle^c_{\overline{\it DPA}} - \langle Q_2
  \rangle^u_{\overline{\it DPA}} \Bigr)\,,
  \label{eq:gengim}
\end{eqnarray}
and they would vanish in the limit of degenerate $u$ and $c$.  The
unitarity of the CKM matrix assures that in a given decay $P_i$ is
always accompanied by $P_i^{\sss {\rm GIM}}$ with the same index
``{\em i\/}''. However, $P_i$ and $P_i^{\sss {\rm GIM}}$ are always
multiplied by different CKM factors and in order to keep the latter
factors explicitly one has to consider separately $P_i$ and $P_i^{\sss
  {\rm GIM}}$.

It should be stressed that the eight penguin parameters introduced by us
differ considerably from the eight penguin parameters $a_i^{\rm eff}$
($i$ = $3$--$10$) introduced in refs.~\cite{GNF}--\cite{Cheng}. There a
given parameter corresponds to a particular operator $Q_i$ and the
scale and renormalization scheme dependence is removed by calculating
the corresponding matrix elements perturbatively between external
quark states. Such an approach suffers from gauge and infrared
dependences as stressed by us in ref.~\cite{BSI}.

These problems are absent in our approach as each of the penguin
parameters $P_i$ receives contributions from all penguin operators
$Q_{3-10}$ which mix under renormalization. Consequently the scale and
scheme independence of the $P_i$ parameters is automatically assured
after the inclusion of penguin contractions of $Q_{1,2}$, without the
necessity for any dubious perturbative calculations of matrix
elements. Moreover we include all possible penguin topologies, while
penguin contractions in penguin-emission and
double-penguin-annihilation topologies were not considered in the
factorization approach in refs.~\cite{GNF}--\cite{Cheng}. The same
applies to emission-annihilation topologies.

If the operator basis is chosen so that the operators are Fierz
conjugates of the ones in \r{eq:basis} then the expressions for the
effective parameters given above have to be modified
appropriately. Denoting generally by $\langle Q_i \rangle_D$ and
$\langle Q_i \rangle_C$ the insertions in the disconnected and
connected topologies respectively, the modification in question is
straightforward. For each Fierz-transformed operator one simply makes
the replacements $\langle Q_i \rangle_D \leftrightarrow \langle Q_i
\rangle_C$. In particular, in the case of the NDR scheme the
modification amounts to 
\begin{equation}
  \langle Q_{2i} \rangle_D \leftrightarrow
  \langle Q_{2i} \rangle_C, \qquad i=1-5.
  \label{eq:NDRex}
\end{equation}
Since the effective parameters are renormalization scheme independent,
this transformation does not change their numerical values. On the
other hand, in the NDR scheme the operators $Q_{2i}$ appear in the
colour non-singlet form and consequently the large $N$ counting is
less transparent than in the basis \r{eq:basis}.

\section{Effective Parameters -- Flavour Dependence}
\label{sec:details}

In this Section, we discuss in detail the flavour dependence of the
effective parameters defined in the previous Section. We give explicit
expressions for the effective parameters in terms of the topologies
defined in figures \ref{fig:emiss-ann} and \ref{fig:penguins}.
Furthermore, we provide a diagrammatic derivation of the flavour
structure of the effective parameters.  In this Section, $B$ stands
generically for the $B_d$, $B^+$ and $B_s$ mesons.

\subsection{The Emission Parameters}
\label{sec:emiss}

As we said above, the first scale- and scheme-independent combinations
of Wilson coefficients and matrix elements that we can identify
correspond to the emission matrix elements of current-current
operators ($Q_1$ and $Q_2$):
\begin{eqnarray}
  E_1(q_i,q_j,q_k;B,M_1,M_2)&=& C_1 {\it DE}_{1}(q_i,q_j,q_k;B,M_1,M_2)
  + C_2 {\it CE}_{2}(q_i,q_j,q_k;B,M_1,M_2)\,, \nonumber \\
  E_2(q_i,q_j,q_k;B,M_1,M_2)&=& C_1 {\it CE}_{1}(q_i,q_j,q_k;B,M_1,M_2)
  + C_2 {\it DE}_{2}(q_i,q_j,q_k;B,M_1,M_2)\,.
  \label{eq:E1E2}
\end{eqnarray}
In order to exhibit the flavour dependence we use here and in the
following the notation of figs.~\ref{fig:emiss-ann} and
\ref{fig:penguins} instead of the short-hand notation $\langle
Q_i\rangle_T$ ($T$ = topology) used in Section \ref{sec:general}. If
factorization held, we would obtain
\begin{eqnarray}
  \label{eq:E1fact}
  {\it DE}_{1,2}(q_i,q_j,q_k;B,M_1,M_2)&=&\langle M_1 \vert (\bar{q}_j
  q_i)_{(V-A)} \vert 0 \rangle \langle M_2 \vert (\bar{b} q_k
  )_{(V-A)} \vert B \rangle\,,  \\
  {\it CE}_{1,2}(q_i,q_j,q_k;B,M_1,M_2)&=&\frac{1}{N} 
  {\it DE}_{2,1}(q_i,q_j,q_k;B,M_1,M_2)\,,
\end{eqnarray}
and consequently 
\begin{eqnarray}
  \label{eq:E1fact2}
  E_1(q_i,q_j,q_k;B,M_1,M_2)&=&
  \left(C_1 + \frac{1}{N} C_2\right)\langle M_1 \vert (\bar{q}_j
  q_i)_{(V-A)} \vert 0 \rangle \langle M_2 \vert (\bar{b} q_k
  )_{(V-A)} \vert B \rangle\ \,; \nonumber \\
  E_2(q_i,q_j,q_k;B,M_1,M_2)&=&\left(C_2 + \frac{1}{N} C_1\right)
  \langle M_1 \vert (\bar{q}_j
  q_i)_{(V-A)} \vert 0 \rangle \langle M_2 \vert (\bar{b} q_k
  )_{(V-A)} \vert B \rangle\,.
\end{eqnarray}

\subsection{The Annihilation Parameters}
\label{sec:annihi}

The next scale- and scheme-independent quantities corresponding to the
annihilation matrix elements of current-current operators are given by:
\begin{eqnarray}
  A_1(q_i,q_j,q_k;B,M_1,M_2)&=& C_1 {\it DA}_{1}(q_i,q_j,q_k;B,M_1,M_2)
  + C_2 {\it CA}_{2}(q_i,q_j,q_k;B,M_1,M_2)\,, \nonumber \\
  A_2(q_i,q_j,q_k;B,M_1,M_2)&=& C_1 {\it CA}_{1}(q_i,q_j,q_k;B,M_1,M_2)
  + C_2 {\it DA}_{2}(q_i,q_j,q_k;B,M_1,M_2)\,.
  \label{eq:A1A2}
\end{eqnarray}

If factorization held, we would obtain
\begin{eqnarray}
  \label{eq:A1fact}
  A_1(q_i,q_j,q_k;B,M_1,M_2)&=&
  \left(C_1 + \frac{1}{N} C_2\right)\langle M_1 M_2 \vert (\bar{q}_i
  q_k)_{(V-A)} \vert 0 \rangle \langle 0 \vert (\bar{b} u
  )_{(V-A)} \vert B \rangle\ \,; \nonumber \\
  A_2(q_i,q_j,q_k;B,M_1,M_2)&=&\left(C_2 + \frac{1}{N} C_1\right)
  \langle M_1 M_2 \vert (\bar{q}_i
  q_k)_{(V-A)} \vert 0 \rangle \langle 0 \vert (\bar{b} d_l
  )_{(V-A)} \vert B \rangle\,, \nn
\end{eqnarray}
where $d_l=d,s$. Due to the flavour structure, $A_1$ only contributes
to $B^+$ decays and $A_2$ to $B_{d,s}$ decays.

\subsection{The Emission-Annihilation Parameters}
\label{sec:emiss-annihi}

The last two independent combinations of non-penguin topologies are
given by the emission-annihilation matrix elements of current-current
operators:
\begin{eqnarray}
  {\it EA}_1(q_i,q_j,q_k;B,M_1,M_2)&=& C_1 {\it DEA}_{1}(q_i,q_j,q_k;B,M_1,M_2)
  + C_2 {\it CEA}_{2}(q_i,q_j,q_k;B,M_1,M_2)\,, \nonumber \\
  {\it EA}_2(q_i,q_j,q_k;B,M_1,M_2)&=& C_1 {\it CEA}_{1}(q_i,q_j,q_k;B,M_1,M_2)
  + C_2 {\it DEA}_{2}(q_i,q_j,q_k;B,M_1,M_2)\,.
  \label{eq:EA1EA2}
\end{eqnarray}

\subsection{The Penguin Parameters}
\label{sec:peng}

Each of the parameters $P_1$, $P_2$, $P_3$ and $P_4$ discussed in
Section \ref{sec:general} is
obtained by summing the following contributions, with $d_i=\{d,s\}$:
\begin{itemize}
\item[a)] two terms containing a given penguin contraction of
  $Q_{1}^{d_icc}$ and $Q_{2}^{d_icc}$;

\item[b)] all the possible terms obtained by replacing the $c$-quark loop
  with the insertion of the penguin operators $Q^{d_i}_{3-10}$ as
  explained below;
  
\item[c)] the terms obtained by replacing $Q_{1,2}^{d_icc}$ by
  $Q^{d_i}_{3-10}$ in the penguin contractions as explained below.
\end{itemize}

\begin{figure}   % produce figure here
    \begin{flushleft}
\input{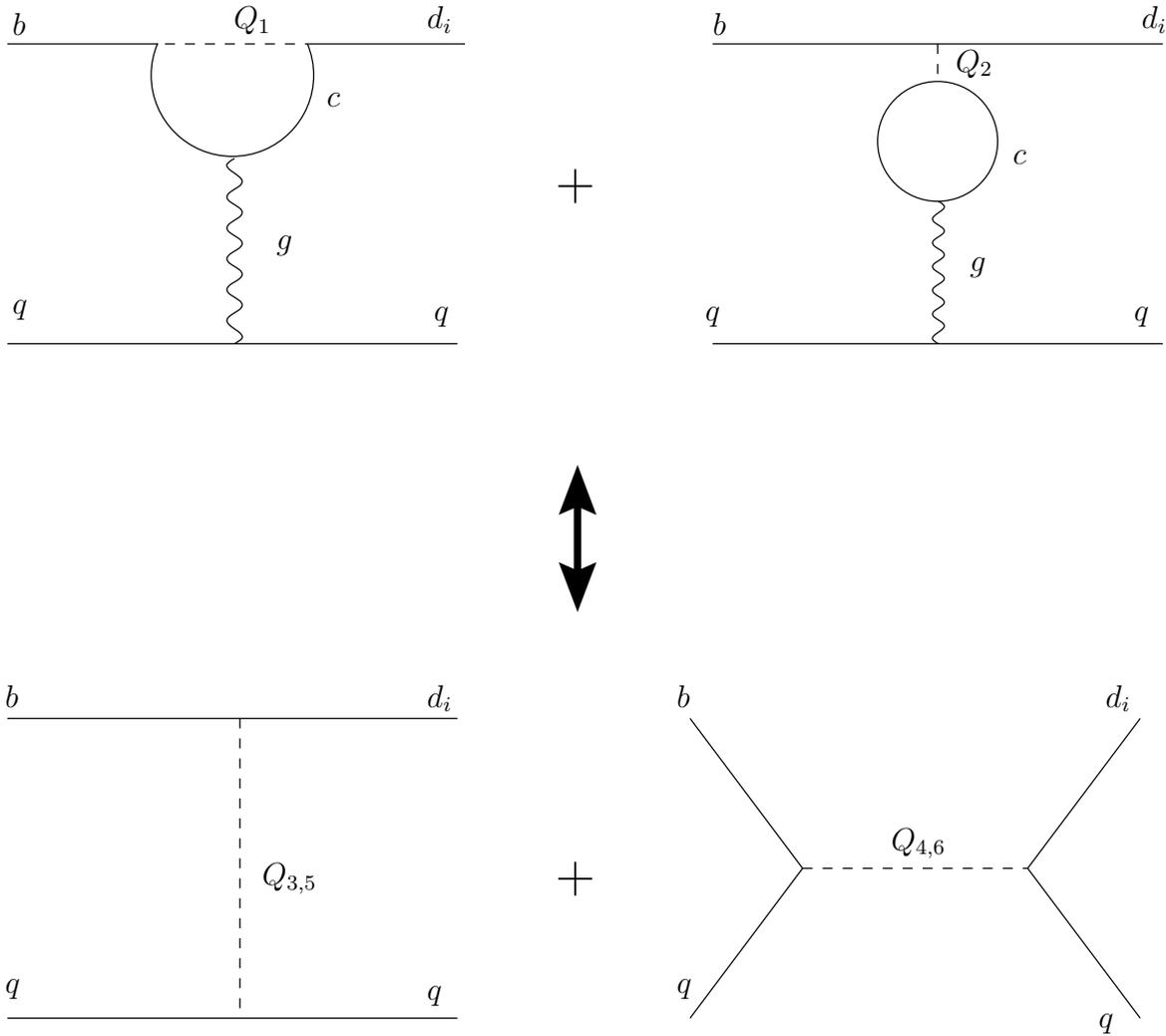}
    \end{flushleft}
    \caption[]{The basic step to relate penguin insertions of
      current-current operators to emission, annihilation and
      emission-annihilation insertions of penguin operators. Wherever
      a gluon can be exchanged between the $c$-quark loop in the penguin
      insertions and a quark line, this corresponds to the insertion
      of a penguin operator.}
    \label{fig:equiv1}
\end{figure}

\begin{figure}   % produce figure here
    \begin{flushleft}
\input{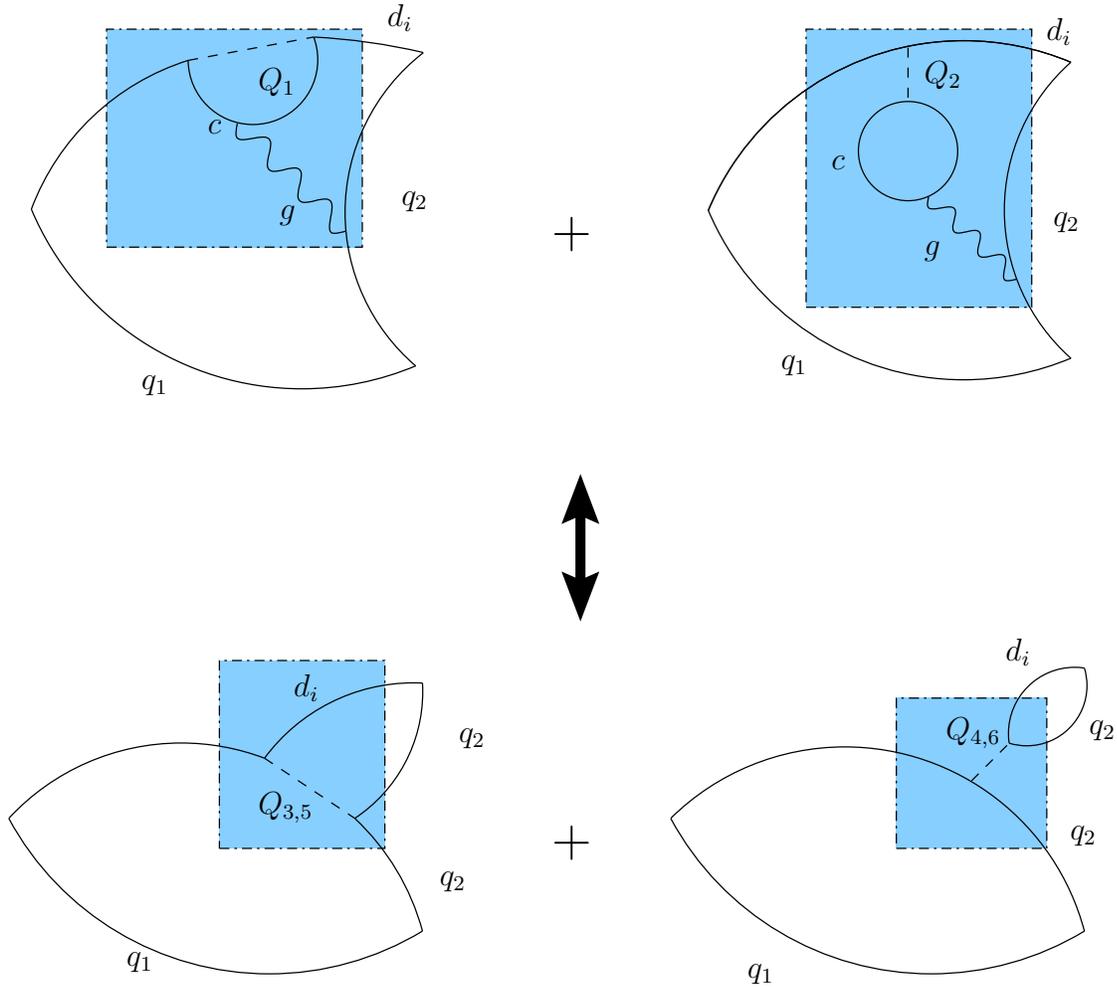}
    \end{flushleft}
    \caption[]{An example of how penguin contractions of
      current-current operators are related to insertions of penguin
      operators in emission topologies. First, one notices that a
      gluon might be exchanged between the $c$-quark loop and the
      $q_2$ line in the penguin topologies (shaded region). Next,
      using the prescription in fig.~\ref{fig:equiv1}, one replaces
      the gluon exchange in the shaded region by the corresponding
      insertion of a penguin operator.}
    \label{fig:equiv2}
\end{figure}

We will now explain how the contributions b) and c) can be obtained,
restricting our attention to QCD-penguin operators $Q_{3-6}^{d_i}$.
The generalization to electroweak penguin operators is straightforward
and will be done subsequently.  A diagrammatic representation of how
to obtain contribution b) is given in
figs.~\ref{fig:equiv1}--\ref{fig:equiv11}. Consider the insertion of a
current-current operator in a penguin-like diagram in
fig.~\ref{fig:equiv1}. This insertion is clearly scheme and scale
dependent. As we discussed above these dependences can be canceled by
adding the corresponding contributions of penguin operators. To this
end it is useful to develop a diagrammatic correspondence between the
penguin insertions of $Q_{1,2}$ and the insertions of penguin
operators.  In fig.~\ref{fig:equiv1} we show the basic procedure which
is applied in figs.~\ref{fig:equiv2}--\ref{fig:equiv11}.  Wherever a gluon
can be exchanged between the loop and a quark line, this corresponds
to the diagram obtained by replacing the $u$- or $c$-quark loop with
the insertion of penguin operators. As an example, in
fig.~\ref{fig:equiv2} we show the correspondence between penguin-type
contractions of $Q_{1,2}^{d_i c c}$ and emission-type contractions of
$Q^{d_i}_{3-6}$. The same is shown in fig.~\ref{fig:equiv3} for
annihilation-type contractions of $Q^{d_i}_{3-6}$. In
fig.~\ref{fig:equiv5} we show the correspondence between
penguin-emission contractions of $Q_{1,2}^{d_i c c}$ and emission-type
contractions of $Q^{d_i}_{3-6}$. The same is shown in
fig.~\ref{fig:equiv10} for emission-annihilation contractions of
$Q^{d_i}_{3-6}$. In fig.~\ref{fig:equiv6} we show the relation between
penguin-annihilation contractions of $Q_{1,2}^{d_i c c}$ and
annihilation-type contractions of $Q^{d_i}_{3-6}$. In
fig.~\ref{fig:equiv11} we show the relation between
double-penguin-annihilation contractions of $Q_{1,2}^{d_i c c}$ and
annihilation-emission contractions of $Q^{d_i}_{3-6}$. The single
gluon exchange in figs.~\ref{fig:equiv1}--\ref{fig:equiv11} is just a
symbolic representation for the exchange of an arbitrary number of
gluons between quark lines.

\begin{figure}   % produce figure here
    \begin{flushleft}
\input{equiv3.pstex_t}
    \end{flushleft}
    \caption[]{The same as fig.~\ref{fig:equiv2}, for another possible
      gluon exchange: in this case, the gluon connects the $c$-quark
      loop with the $q_1$ line in the penguin topology.}
    \label{fig:equiv3}
\end{figure}
\begin{figure}   % produce figure here
    \begin{center}
\input{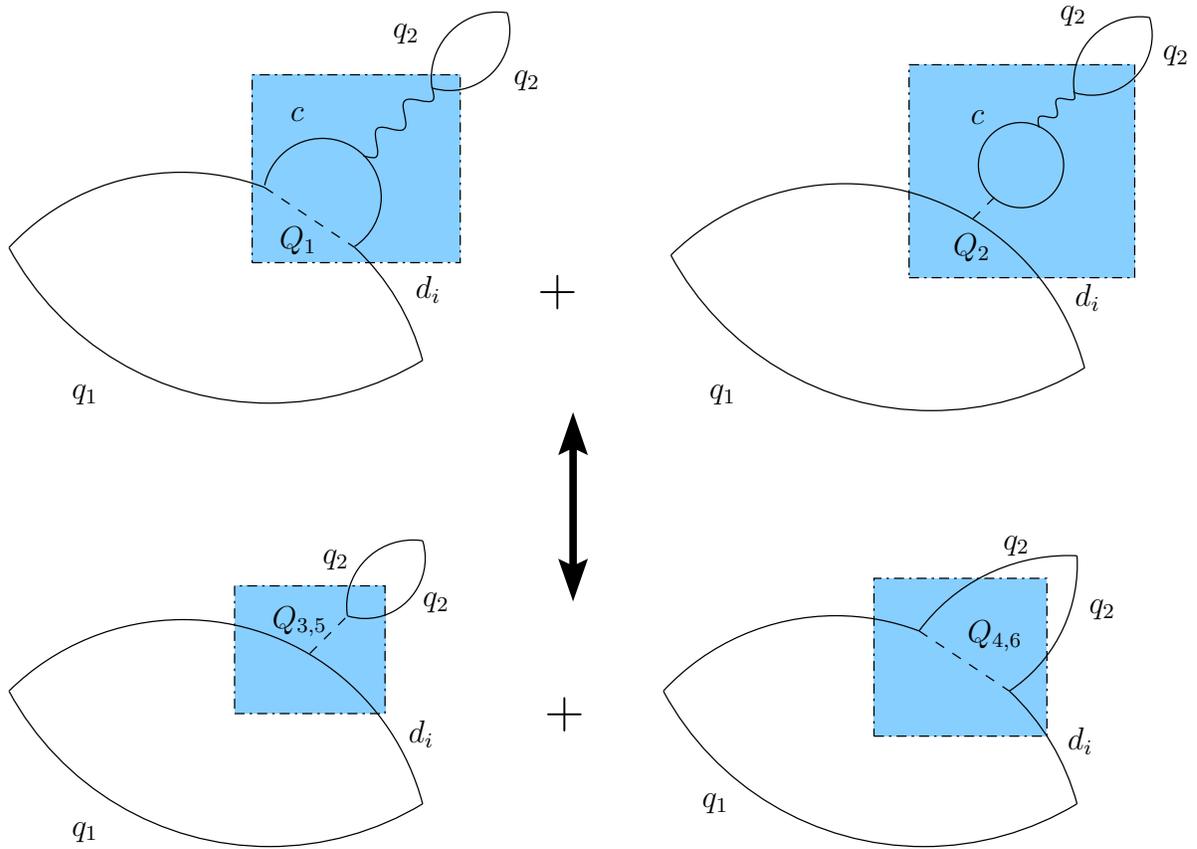}
    \end{center}
    \caption[]{The same as figs.~\ref{fig:equiv2}--\ref{fig:equiv3},
      for penguin-emission topologies. The shaded region corresponds
      to a gluon exchange in the penguin topologies, and to the
      insertion of a penguin operator in the emission topologies.}
    \label{fig:equiv5}
\end{figure}
\begin{figure}   % produce figure here
    \begin{center}
\input{equiv10.pstex_t}
    \end{center}
    \caption[]{The same as fig.~\ref{fig:equiv5}, for another possible
      gluon exchange: in this case, the gluon connects the $c$-quark
      loop with the $q_1$ line in the penguin-emission topology.}
    \label{fig:equiv10}
\end{figure}
\begin{figure}   % produce figure here
    \begin{center}
\input{equiv6.pstex_t}
    \end{center}
    \caption[]{The same as fig.~\ref{fig:equiv5}, but for
      penguin-annihilation topologies.}
    \label{fig:equiv6}
\end{figure}
\begin{figure}   % produce figure here
    \begin{center}
\input{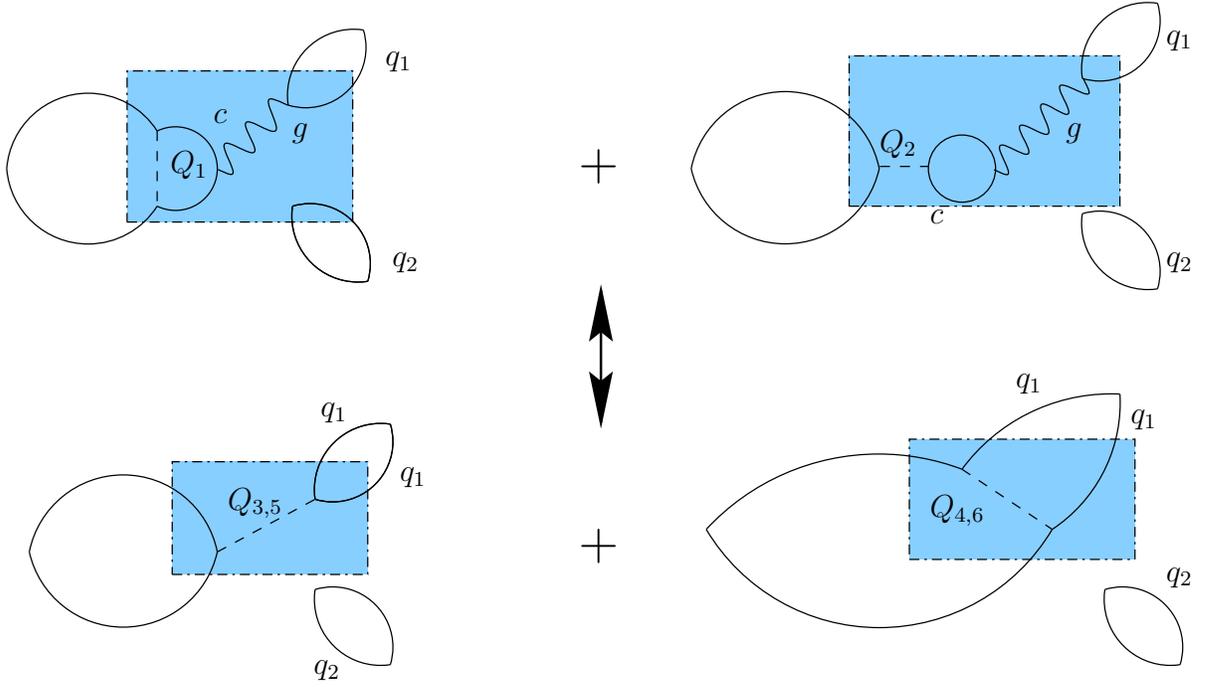}
    \end{center}
    \caption[]{The same as fig.~\ref{fig:equiv6}, but for
      double-penguin-annihilation topologies.}
    \label{fig:equiv11}
\end{figure}

\begin{figure}   % produce figure here
    \begin{center}
\input{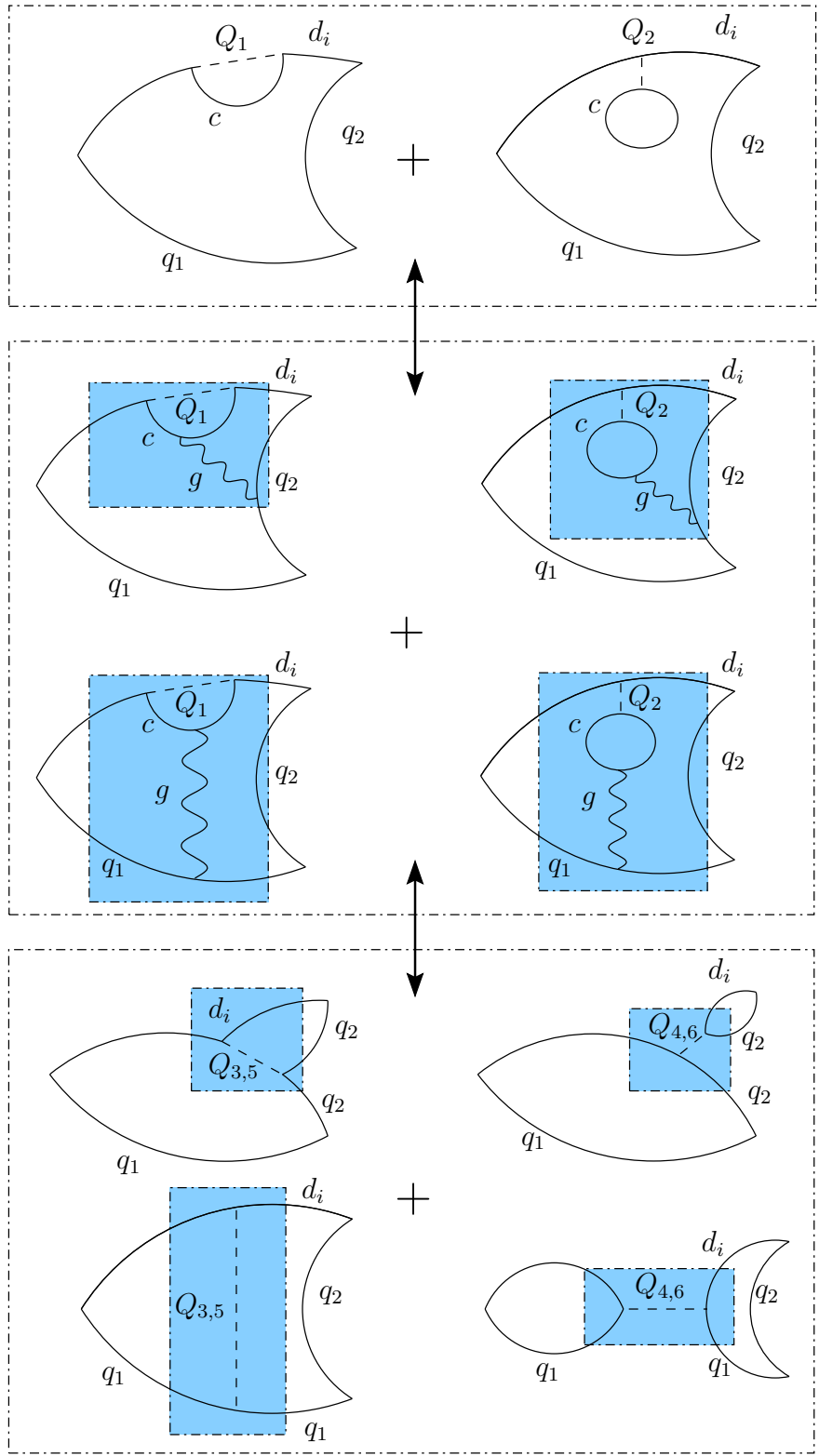}
    \end{center}
    \caption[]{A diagrammatic representation of the correspondence
      between penguin-like contractions of current-current operators
      and insertions of penguin operators. See the text for details.}
    \label{fig:equiv4}
\end{figure}

Using the correspondence rules represented in
figs.~\ref{fig:equiv1}--\ref{fig:equiv11}, we can obtain contribution
b) to the scale- and scheme-independent combinations of penguin-like
topologies. As an example, in fig.~\ref{fig:equiv4} we show
diagrammatically the construction of the b) part of the $P_1$
contribution defined in Section \ref{sec:general}. One starts by
considering penguin-like contractions of current-current operators, as
depicted in the first box in the figure, which contains the
combination $C_1 {\it CP}_{1}(c,d_i,q_2;B,M_1,M_2)+C_2 {\it
  DP}_{2}(c,d_i,q_2;B,M_1,M_2)$.  Then one considers the gluon
exchanges drawn in the second box, and substitutes the shaded regions
with the corresponding insertions of penguin operators. The resulting
third box contains the combination
\begin{eqnarray}
  && \Bigl( C_3 {\it CE}_{3}(d_i,q_2,q_2;B,M_1,M_2) +
  C_5 {\it CE}_{5}(d_i,q_2,q_2;B,M_1,M_2) \nn \\
  && + C_4 {\it DE}_{4}(d_i,q_2,q_2;B,M_1,M_2) +
  C_6 {\it DE}_{6}(d_i,q_2,q_2;B,M_1,M_2)\Bigr) \nonumber \\
  &+& \Bigl( C_3 {\it CA}_{3}(d_i,q_2,q_1;B,M_1,M_2) +
  C_5 {\it CA}_{5}(d_i,q_2,q_1;B,M_1,M_2) \nn \\ 
  && + C_4 {\it DA}_{4}(d_i,q_2,q_1;B,M_1,M_2) +
  C_6 {\it DA}_{6}(d_i,q_2,q_1;B,M_1,M_2)\Bigr)\,,  
\end{eqnarray}
see eq.~(\ref{eq:P1}), where $q_1=d$, $u$ or $s$ for $B_d$, $B^+$ and
$B_s$ decays respectively. Analogously, one can build up part b) of
$P_2$ as shown in fig.~\ref{fig:equiv9}. The starting point is the
insertion of $Q_1$ and $Q_2$ in penguin-emission diagrams, as shown in
the first box in the figure. This corresponds to the combination $C_1 {\it
  CPE}_{1}(c,d_i,q_2;B,M_1,M_2)+C_2 {\it
  DPE}_{2}(c,d_i,q_2;B,M_1,M_2)$. 
Then, a gluon can
connect the $c$-quark loop either with the $q_1$ line or with the
$q_2$ one, as shown in the second box of fig.~\ref{fig:equiv9}. In the
third box, one replaces the gluon exchange with penguin operators,
following figs.~\ref{fig:equiv5} and \ref{fig:equiv10}, and one
obtains the contribution
\begin{eqnarray}
  && \Bigl(C_3 {\it DE}_{3}(q_2,q_2,d_i;B,M_1,M_2) +
  C_5 {\it DE}_{5}(q_2,q_2,d_i;B,M_1,M_2) \nonumber \\
  && + C_4 {\it CE}_{4}(q_2,q_2,d_i;B,M_1,M_2) +
  C_6 {\it CE}_{6}(q_2,q_2,d_i;B,M_1,M_2)\Bigr) \nonumber \\
  &+& \Bigl( C_3 {\it CEA}_{3}(d_i,q_1,q_2;B,M_2,M_1) +
  C_5 {\it CEA}_{5}(d_i,q_1,q_2;B,M_2,M_1) \nonumber \\
  && + C_4 {\it DEA}_{4}(d_i,q_1,q_2;B,M_2,M_1) +
  C_6 {\it DEA}_{6}(d_i,q_1,q_2;B,M_2,M_1)\Bigr)\,,  
\end{eqnarray}
see eq.~(\ref{eq:P2}), where $q_1=d$, $u$ or $s$ for $B_d$, $B^+$ and
$B_s$ decays respectively. The same is shown in fig.~\ref{fig:equiv8}
for the $P_3$ contribution defined in Section \ref{sec:general}.  Here
one starts by the insertion of current-current operators in
penguin-annihilation diagrams (first box in the figure), which is
given by $C_1 {\it CPA}_{1}(c,q_2,q_1;B,M_1,M_2)+C_2 {\it
  DPA}_{2}(c,q_2,q_1;B,M_1,M_2)$. As in the previous cases, there are
two possible types of gluon exchanges, depending on whether the gluon
connects the blob to the $q_1$ or to the $q_2$ line (second box in
fig.~\ref{fig:equiv8}). Substituting the gluon exchange with
insertions of penguin operators, following fig.~\ref{fig:equiv6}, we
get the contribution in the third box, which corresponds to
\begin{eqnarray}
  && \Bigl( C_3 {\it DA}_{3}(q_1,q_2,q_1;B,M_1,M_2) +
  C_5 {\it DA}_{5}(q_1,q_2,q_1;B,M_1,M_2) \nonumber \\
  && + C_4 {\it CA}_{4}(q_1,q_2,q_1;B,M_1,M_2) +
  C_6 {\it CA}_{6}(q_1,q_2,q_1;B,M_1,M_2)\Bigr) \nonumber \\
  &+& \Bigl( C_3 {\it DA}_{3}(q_2,q_1,q_2;B,M_2,M_1) +
  C_5 {\it DA}_{5}(q_2,q_1,q_2;B,M_2,M_1) \nonumber \\
  && + C_4 {\it CA}_{4}(q_2,q_1,q_2;B,M_2,M_1) +
  C_6 {\it CA}_{6}(q_2,q_1,q_2;B,M_2,M_1)\Bigr)\,, 
\end{eqnarray}
see eq.~(\ref{eq:P3}).
Finally, the b) part of $P_4$ is represented in
fig.~\ref{fig:equiv12}. The starting point here is the insertion of
$Q_1$ and $Q_2$ in double-penguin-annihilation diagrams, as shown in
the first box in the figure. This corresponds to the combination $C_1
\overline{\it CPA}_{1}(c,q_1,q_2;B,M_1,M_2)+C_2 \overline{\it
  DPA}_{2}(c,q_1,q_2;B,M_1,M_2)$. Then, a gluon can connect the
$c$-quark loop either with the $q_1$ line or with the $q_2$ one, as
shown in the second box of fig.~\ref{fig:equiv12}. In the third box,
one replaces the gluon exchange with penguin operators, following
fig.~\ref{fig:equiv11}, and one obtains the contribution
\begin{eqnarray}
  && \Bigl(C_3 {\it DEA}_{3}(q_2,q_2,q_1;B,M_1,M_2) +
  C_5 {\it DEA}_{5}(q_2,q_2,q_1;B,M_1,M_2) \nonumber \\
  && + C_4 {\it CEA}_{4}(q_2,q_2,q_1;B,M_1,M_2) +
  C_6 {\it CEA}_{6}(q_2,q_2,q_1;B,M_1,M_2)\Bigr) \nonumber \\
  &+& \Bigl( C_3 {\it DEA}_{3}(q_1,q_1,q_2;B,M_2,M_1) +
  C_5 {\it DEA}_{5}(q_1,q_1,q_2;B,M_2,M_1) \nonumber \\
  && + C_4 {\it CEA}_{4}(q_1,q_1,q_2;B,M_2,M_1) +
  C_6 {\it CEA}_{6}(q_1,q_1,q_2;B,M_2,M_1)\Bigr)\,,  
\end{eqnarray}
see eq.~(\ref{eq:P4}).

\begin{figure}   % produce figure here
    \begin{center}
\input{equiv9.pstex_t}
    \end{center}
    \caption[]{A diagrammatic representation of the correspondence
      between penguin-emission contractions of current-current
      operators and insertions of penguin operators. See the text for
      details.}
    \label{fig:equiv9}
\end{figure}
\begin{figure}   % produce figure here
    \begin{center}
\input{equiv8.pstex_t}
    \end{center}
    \caption[]{A diagrammatic representation of the correspondence
      between penguin-annihilation contractions of current-current
      operators and insertions of penguin operators. See the text for
      details.}
    \label{fig:equiv8}
\end{figure}
\begin{figure}   % produce figure here
    \begin{center}
\input{equiv12.pstex_t}
    \end{center}
    \caption[]{A diagrammatic representation of the correspondence
      between double-penguin-annihilation contractions of
      current-current operators and insertions of penguin operators.
      See the text for details.}
    \label{fig:equiv12}
\end{figure}

\begin{figure}   % produce figure here
    \begin{center}
\input{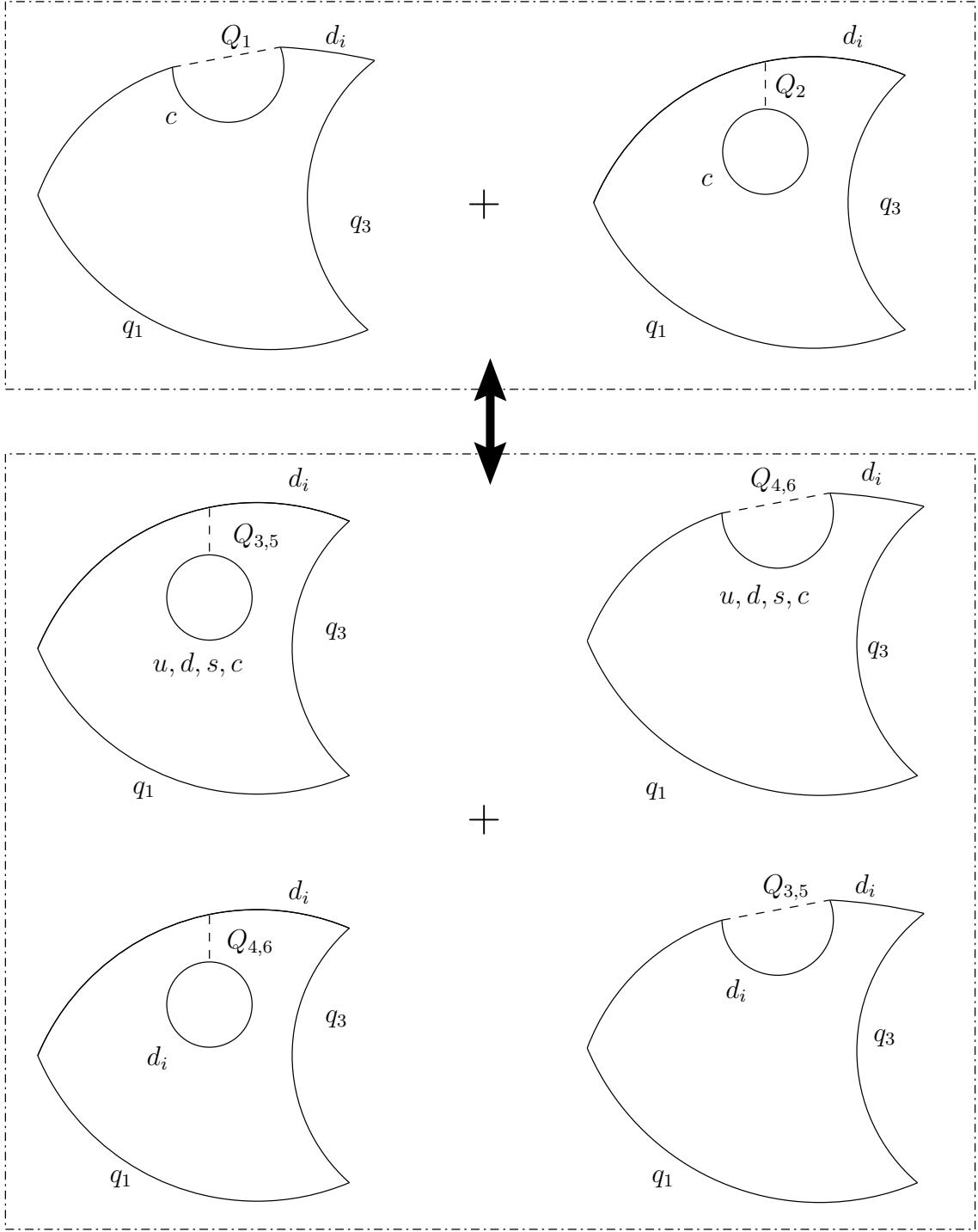}
    \end{center}
    \caption[]{A diagrammatic representation of the rule to obtain
      contribution c). See the text for details.}
    \label{fig:equiv7}
\end{figure}

Concerning the c) contribution, it can be obtained by replacing the
insertion of current-current operators in penguin topologies with the
insertion of penguin operators in the same penguin topologies. An
example is shown in fig.~\ref{fig:equiv7} for the insertion of
$Q_{1,2}^{d_icc}$ in penguin topologies, where $d_i$ can be either a
$d$ or an $s$ quark.  The c) contribution is obtained by replacing
$Q_{1,2}^{d_icc}$ with $Q_{3-6}^{d_i}$, as shown in the figure. We
remark that the $Q_{3-6}^{d_i}$ operators contain a term with the
flavour structure $(\bar b d_i)(\bar d_i d_i)$. This term contributes
with two different Wick contractions, depending on which $d_i$ field
is contracted with the $\bar d_i$. This extra Wick contraction gives
rise to the two diagrams in the last line of fig.~\ref{fig:equiv7}.

The contributions of electroweak penguins to b) and c) can be obtained
by replacing $Q^{d_i}_{3-6}$ with $Q^{d_i}_{7-10}$ using the following
correspondence rules:
\begin{eqnarray}
  Q_3 \leftrightarrow Q_9\,, &\qquad& Q_4 \leftrightarrow Q_{10}\,,
  \nonumber \\
  Q_5 \leftrightarrow Q_7\,, &\qquad& Q_6 \leftrightarrow Q_{8}\,.
  \label{eq:ewcorr}
\end{eqnarray}

Applying the above considerations, we are able to identify the scale-
and scheme-independent contributions involving penguin operators and
penguin contractions:
\begin{eqnarray} 
  \label{eq:P1}
   P_1(d_i,q_j;B,M_1,M_2)&=&  
  C_1 {\it CP}_1(c,d_i,q_j;B,M_1,M_2)+C_2 {\it DP}_2(c,d_i,q_j;B,M_1,M_2)
  \\
  &+&\sum_{l=2}^{5} \Bigl(C_{2l-1} {\it CE}_{2l-1}(d_i,q_j,q_j;B,M_1,M_2) +
  C_{2l} {\it DE}_{2l}(d_i,q_j,q_j;B,M_1,M_2)\Bigr) \nonumber \\
  &+& \sum_{l=2}^5 \Biggl[\sum_{q}\Bigl(C_{2l-1}
  {\it DP}_{2l-1}(q,d_i,q_j;B,M_1,M_2) 
  + C_{2l} {\it CP}_{2l}(q,d_i,q_j;B,M_1,M_2)\Bigr) \nonumber \\
  && \qquad+ C_{2l-1} {\it CP}_{2l-1}(d_i,d_i,q_j;B,M_1,M_2)+
  C_{2l} {\it DP}_{2l}(d_i,d_i,q_j;B,M_1,M_2)\Biggr] \nonumber \\
  &+& \sum_{l=2}^5 \Bigl( C_{2l-1} {\it CA}_{2l-1}(d_i,q_j,q_k;B,M_1,M_2) +
  C_{2l} {\it DA}_{2l}(d_i,q_j,q_k;B,M_1,M_2)\Bigr)\,,\nn
\end{eqnarray}
where $q_k=d$, $u$ or $s$ for $B_d$, $B^+$ and $B_s$ decays
respectively and $q=u$, $d$, $s$ and $c$. The second and the last line
on the r.h.s. of eq.~(\ref{eq:P1}) can be readily found by following
fig.~\ref{fig:equiv4}. The relevant quark variables $d_i$ and $q_j$ in
$P_1$ can be identified from the penguin insertion of $Q_1$ as seen on
the top of fig.~\ref{fig:equiv4}. The variable $d_i$ denotes the quark
line flowing into the penguin. The variable $q_j$ denotes the quark
line attached to $d_i$ on its right hand side.
\begin{eqnarray}
  \label{eq:P2}  
  P_2(d_i,q_j;B,M_1,M_2)&=& 
  C_1 {\it CPE}_1(c,q_j,d_i;B,M_1,M_2)+C_2 {\it DPE}_2(c,q_j,d_i;B,M_1,M_2)
  \\
  &+&\sum_{l=2}^5 \Bigl(C_{2l-1} {\it DE}_{2l-1}(q_j,q_j,d_i;B,M_1,M_2) +
  C_{2l} {\it CE}_{2l}(q_j,q_j,d_i;B,M_1,M_2)\Bigr) \nonumber \\
  &+&\sum_{l=2}^5 \Bigl(C_{2l-1} {\it CEA}_{2l-1}(d_i,q_k,q_j;B,M_2,M_1) +
  C_{2l} {\it DEA}_{2l}(d_i,q_k,q_j;B,M_2,M_1)\Bigr) \nonumber \\
  &+& \sum_{l=2}^5 \Biggl[\sum_{q}\Bigl(C_{2l-1}
  {\it DPE}_{2l-1}(q,q_j,d_i;B,M_1,M_2) 
  + C_{2l} {\it CPE}_{2l}(q,q_j,d_i;B,M_1,M_2)\Bigr) \nonumber \\
  && \qquad+ C_{2l-1} {\it CPE}_{2l-1}(d_i,q_j,d_i;B,M_1,M_2)+
  C_{2l} {\it DPE}_{2l}(d_i,q_j,d_i;B,M_1,M_2)\Biggr]\,, \nonumber
\end{eqnarray}
where $q_k=d$, $u$ or $s$ for $B_d$, $B^+$ and $B_s$ decays
respectively. The second and the third line on the r.h.s. of
eq.~(\ref{eq:P2}) can be found by following fig.~\ref{fig:equiv9}. The
relevant quark variables $d_i$ and $q_j$ in $P_2$ can be identified
from the relevant penguin insertion of $Q_1$ as seen on the top of
fig.~\ref{fig:equiv9}. The variable $d_i$ denotes the quark line
flowing into the penguin. The variable $q_j$ denotes the flavour in
the neutral ``blob'' representing the meson $\bar q_j q_j$. 
\begin{eqnarray}
  \label{eq:P3}  
  P_3(q_j,q_k;B,M_1,M_2)&=&
  C_1 {\it CPA}_1(c,q_j,q_k;B,M_1,M_2)+C_2 {\it DPA}_2(c,q_j,q_k;B,M_1,M_2) \\
  &+& \sum_{l=2}^5 \Bigl( C_{2l-1} {\it DA}_{2l-1}(q_k,q_j,q_k;B,M_1,M_2) +
  C_{2l} {\it CA}_{2l}(q_k,q_j,q_k;B,M_1,M_2)\Bigr)  
  \nonumber \\
  &+& \sum_{l=2}^5 \Bigl( C_{2l-1} {\it DA}_{2l-1}(q_j,q_k,q_j;B,M_2,M_1) +
  C_{2l} {\it CA}_{2l}(q_j,q_k,q_j;B,M_2,M_1)\Bigr)  
  \nonumber \\
  &+& \sum_{l=2}^5 \Biggl[\sum_{q}\Bigl(C_{2l-1}
  {\it DPA}_{2l-1}(q,q_j,q_k;B,M_1,M_2) 
  + C_{2l} {\it CPA}_{2l}(q,q_j,q_k;B,M_1,M_2)\Bigr) \nonumber \\
  && \qquad + C_{2l-1} {\it CPA}_{2l-1}(d_i,q_j,q_k;B,M_1,M_2)+
  C_{2l} {\it DPA}_{2l}(d_i,q_j,q_k;B,M_1,M_2)\Biggr]\,, \nonumber
\end{eqnarray}
where $d_i=d$ or $s$ for $B_d$ and $B_s$ decays respectively (due to
the flavour structure, $P_3$ cannot contribute to charged B
decays). The second and third lines on the r.h.s. of eq.~(\ref{eq:P3})
can be found using fig.~\ref{fig:equiv8}. The relevant quark variables
$q_j$ and $q_k$ in $P_3$ can be identified from the relevant penguin
insertion of $Q_1$ as seen on the top of fig.~\ref{fig:equiv8}. The
variable $q_j$ denotes the external quark line in the right hand side
of the diagram. The variable $q_k$ denotes the internal quark line in
this part of the diagram. 
\begin{eqnarray}
  \label{eq:P4}  
  P_4(q_j,q_k;B,M_1,M_2) &=&
  C_1 \overline{\it CPA}_1(c,q_j,q_k;B,M_1,M_2)+C_2 \overline{\it
  DPA}_2(c,q_j,q_k;B,M_1,M_2) 
  \\
  &+& \sum_{l=2}^5 \Bigl( C_{2l-1} {\it DEA}_{2l-1}(q_j,q_j,q_k;B,M_1,M_2) +
  C_{2l} {\it CEA}_{2l}(q_j,q_j,q_k;B,M_1,M_2)\Bigr)  
  \nonumber \\
  &+& \sum_{l=2}^5 \Bigl( C_{2l-1} {\it DEA}_{2l-1}(q_k,q_k,q_j;B,M_2,M_1) +
  C_{2l} {\it CEA}_{2l}(q_k,q_k,q_j;B,M_2,M_1)\Bigr)  
  \nonumber \\
  &+& \sum_{l=2}^5 \Biggl[\sum_{q}\Bigl(C_{2l-1}
  \overline{\it DPA}_{2l-1}(q,q_j,q_k;B,M_1,M_2) 
  + C_{2l} \overline{\it CPA}_{2l}(q,q_j,q_k;B,M_1,M_2)\Bigr) \nonumber \\
  && \qquad + C_{2l-1} \overline{\it CPA}_{2l-1}(d_i,q_j,q_k;B,M_1,M_2)+
  C_{2l} \overline{\it DPA}_{2l}(d_i,q_j,q_k;B,M_1,M_2)\Biggr]\,, \nonumber
\end{eqnarray}
where $d_i=d$ or $s$ for $B_d$ and $B_s$ decays respectively
(analogously to $P_3$, due to the flavour structure, $P_4$ cannot
contribute to charged B decays). The second and third lines on the
r.h.s. of eq.~(\ref{eq:P4}) can be found using fig.~\ref{fig:equiv12}.
The relevant quark variables $q_j$ and $q_k$ in $P_4$ can be
identified from the relevant penguin insertion of $Q_1$ as seen on the
top of fig.~\ref{fig:equiv12}. The variable $q_j$ denotes the flavour
in the upper neutral ``blob'' representing the meson $\bar q_j
q_j$. The variable $q_k$ denotes the corresponding flavour in the
lower neutral meson $\bar q_k q_k$. 

If factorization held, the penguin-type matrix elements would vanish
and one would be left with the factorized emission and annihilation
matrix elements of penguin operators.

\subsection{The GIM-Penguin Parameters}
\label{sec:gimpen}

The last four effective parameters are the GIM-penguin ones,
$P_1^{\sss {\rm GIM}}$, $P_2^{\sss {\rm GIM}}$, $P_3^{\sss {\rm GIM}}$
and $P_4^{\sss {\rm GIM}}$, already introduced in
eq.~(\ref{eq:gengim}). Their explicit flavour structure is given as
follows:
\begin{eqnarray}  
  \label{eq:flavgim}
  P_1^{\sss {\rm GIM}}(d_i,q_j;B,M_1,M_2)&=&C_1
  \Bigl({\it CP}_1(c,d_i,q_j;B,M_1,M_2)- 
  {\it CP}_1(u,d_i,q_j;B,M_1,M_2)\Bigr) \\
  &+& C_2 \Bigl({\it DP}_2(c,d_i,q_j;B,M_1,M_2)-
  {\it DP}_2(u,d_i,q_j;B,M_1,M_2)\Bigr), \nn \\
  P_2^{\sss {\rm GIM}}(d_i,q_j;B,M_1,M_2)&=&C_1
  \Bigl({\it CPE}_1(c,q_j,d_i;B,M_1,M_2)- 
  {\it CPE}_1(u,q_j,d_i;B,M_1,M_2)\Bigr)\nn \\
  &+& C_2 \Bigl({\it DPE}_2(c,q_j,d_i;B,M_1,M_2)-
  {\it DPE}_2(u,q_j,d_i;B,M_1,M_2)\Bigr), \nn \\  
  P_3^{\sss {\rm GIM}}(q_j,q_k;B,M_1,M_2)&=&C_1
  \Bigl({\it CPA}_1(c,q_j,q_k;B,M_1,M_2)- 
  {\it CPA}_1(u,q_j,q_k;B,M_1,M_2)\Bigr)\nn \\
  &+& C_2 \Bigl({\it DPA}_2(c,q_j,q_k;B,M_1,M_2)-
  {\it DPA}_2(u,q_j,q_k;B,M_1,M_2)\Bigr), \nn \\  
  P_4^{\sss {\rm GIM}}(q_j,q_k;B,M_1,M_2)&=&C_1
  \Bigl(\overline{\it CPA}_1(c,q_j,q_k;B,M_1,M_2)- 
  \overline{\it CPA}_1(u,q_j,q_k;B,M_1,M_2)\Bigr)\nn \\
  &+& C_2 \Bigl(\overline{\it DPA}_2(c,q_j,q_k;B,M_1,M_2)-
  \overline{\it DPA}_2(u,q_j,q_k;B,M_1,M_2)\Bigr). \nn
\end{eqnarray}

\section{Hierarchies}
\label{sec:approx}

In the previous Sections we have introduced the effective scheme and
scale independent parameters suitable for the description of two-body
$B$ decays, and we have discussed in detail their flavour structure.
In this Section, we show, using the large $N$ approximation, that a
hierarchy is expected between the various parameters which we
previously introduced.

Using the large $N$ classification one has in units of $\sqrt{N}$ the
following hierarchy for various topologies:
\begin{eqnarray}
  \label{eq:hier1}
  {\it DE},\,{\it DA}&:& \,\,\ord (1), \\
  {\it CE},\,{\it CA},\,{\it DEA},\,{\it CP},\,{\it DPA},\,{\it
  DPE}&:&\,\, \ord(1/N), \nn \\
  {\it CEA},\,{\it DP},\,{\it CPA},\,{\it CPE},\,{\it
  \overline{DPA}}&:&\,\, \ord(1/N^2), \nn \\ 
  {\it \overline{CPA}}&:&\,\, \ord (1/N^3). \nn
\end{eqnarray}
On the other hand the Wilson coefficients $C_{1-6}$ have in the strict
$1/N$ classification the following dependence:
\begin{eqnarray}
  \label{eq:hierc}
  C_1\,&:&\,\, \ord (1), \\
  C_2\,,\,\,C_4\,,\,\,C_6\,&:&\,\, \ord(1/N),\nn \\
  C_3\,,\,\,C_5\,&:&\,\, \ord(1/N^2). \nn
\end{eqnarray}
The coefficients $C_{7-10}$ are all $\ord(\alpha)$ and consequently
the large $N$ classification of $C_{7-10}$ with respect to
$C_{1-6}$ is not useful here. In discussing the $1/N$ hierarchy of
the effective parameters we will therefore not include the electroweak
penguin operators.

Two additional comments on \r{eq:hier1} and \r{eq:hierc} should be
made. If there is an s-channel resonance, as in the case of
annihilation contributions, there is an additional enhancement factor
$N$ as in the large $N$ limit the width of a resonance behaves as
$\Gamma \sim 1/N$. On the other hand annihilation is suppressed by
other dynamical factors and it is justified to ignore this additional
factor of $N$ in the following discussion.

Concerning \r{eq:hierc}, the Wilson coefficients can be enhanced by
large logarithms due to the renormalization group evolution from
$M_W$ down to $m_b$. Since $\ln M_W^2/m_b^2 \sim 6$, these logarithms
may over-compensate the $1/N$ suppression. On the other hand as seen in
\r{eq:coeffi} $C_1 > C_2 > C_{4,6} > C_{3,5}$. Thus except for $C_2$
being substantially higher than $C_{4,6}$ the hierarchy in
\r{eq:hierc} is roughly respected and we will consider it to be valid
in the following discussion.

Using \r{eq:hier1} and \r{eq:hierc} we find the following large $N$
classification of the effective parameters in units of $\sqrt{N}$:
\begin{eqnarray}
  \label{eq:classhier}
  E_1,\,\, A_1&:& \,\, \ord(1),\\
  E_2,\,\,A_2,\,\,{\it EA}_1,\,\,P_1,\,\,P_1^{\sss {\rm
  GIM}}&:&\,\,\ord(1/N), \nn\\
  {\it EA}_2,\,\,P_2,\,\,P_3,\,\,P_2^{\sss {\rm
  GIM}},\,\,P_3^{\sss {\rm
  GIM}}&:&\,\,\ord(1/N^2), \nn\\
  P_4,\,\,P_4^{\sss {\rm
  GIM}}&:&\,\, \ord(1/N^3). \nn
\end{eqnarray}
As one can see, within the large $N$ classification the annihilation
amplitude $A_1$ is expected to be larger than $E_2$, which contradicts
the usual expectations. Future measurements of channels dominated by
$A_1$ or $E_2$ will teach us whether this hierarchy is consistent with
the data.

Next, the decay amplitudes are linear combinations of effective
parameters multiplied by CKM factors. The hierarchy of the latter can
be roughly described in terms of the Wolfenstein parameter
$\lambda=\vert V_{us}\vert=0.22$. In particular we have:
\begin{equation}
  \label{eq:hierlam}
  \begin{array}{cccc}
    V_{us}V^*_{ub}=\ord \left(\lambda^4\right),&V_{ts} V_{tb}^*=\ord
    \left(\lambda^2\right), &   V_{ud}V_{ub}^* =\ord
    \left(\lambda^3\right),& V_{td}V_{tb}^* =\ord \left(\lambda^3\right).
\end{array}
\end{equation}

When calculating branching ratios one could for instance neglect
$V_{us} V_{ub}^*$ with respect to $V_{ts} V_{tb}^*$ provided this is
also supported by the hierarchy in \r{eq:classhier}. On the other hand
when studying CP-asymmetries it is important to keep all CKM factors
and we will do so in the following. 

In tables \ref{tab:classbd}--\ref{tab:classbs} we collect a large number
of two-body $B_d$, $B^+$ and $B_s$ decays respectively. We list there
the effective parameters contributing to each decay, the size of each
parameter according to the large $N$ classification and the order in
$\lambda$ of the CKM parameters multiplying it. These tables should be
useful in identifying the most suitable decays for the determination
of the effective parameters and for approximations which would reduce
the number of parameters.

In the following Sections we will present explicit expressions for the
amplitudes of the two-body $B$ decays listed in tables
\ref{tab:classbd}--\ref{tab:classbs}. In order to simplify the
presentation we will omit in Sections \ref{sec:classific} and
\ref{sec:classificbs}:
\begin{enumerate}
\item the contributions proportional to the parameters $P_4$ and
  $P_4^{\sss {\rm GIM}}$;
\item the contributions which vanish in the $SU(2)$ limit, unless their
  neglect would make a CKM factor disappear from the amplitude.
\end{enumerate}

The contributions omitted in Sections \ref{sec:classific} and
\ref{sec:classificbs} are listed for completeness in Appendix
\ref{sec:appew}.

\begin{table}[htbp]
  \begin{center}
    \begin{tabular}{||c|c|c|c|c|c|c|c|c|c|c|c|c|c||}
      \hline \hline
      Channel & Cl. & $E_1$ & $E_2$ & ${\it EA}_2$ & $A_2$ & $P_1$ &
      $P_2$ & $P_3$ & $P_1^{\sss {\rm GIM}}$ & $P_2^{\sss {\rm GIM}}$
      & $P_3^{\sss {\rm GIM}}$ & $P_4$ & $P_4^{\sss {\rm GIM}}$\\
      & & $ 1$ & $ \frac{1}{N}$ & $
      \frac{1}{N^2}$ & $ \frac{1}{N}$ &
      $ \frac{1}{N}$ & $
      \frac{1}{N^2}$ & $ \frac{1}{N^2}$
      & $ \frac{1}{N}$ & $
      \frac{1}{N^2}$ & $
      \frac{1}{N^2}$ & $ \frac{1}{N^3}$
      & $ \frac{1}{N^3}$ \\ \hline    
      $B_d \to D^- \pi^+$ & {\bf A} & $\lambda^2$ & -- & -- & 
      $\lambda^2$ & -- & -- & -- & -- & -- & -- & -- & -- \\ \hline
      $B_d \to \bar D^0 \pi^0$ & {\bf A} & -- & $\lambda^2$ 
      & $\left[\lambda^2 \right]$ & $\lambda^2$ & -- & -- & -- & -- &
      -- & -- & -- & -- \\ \hline 
      $B_d \to D^- K^+$ & {\bf B} & $\lambda^3$ & -- &
      -- & -- & -- & -- & -- & -- & -- & -- & -- & -- \\ \hline
      $B_d \to \bar D^0 K^0$ & {\bf B} & -- & $\lambda^3$ &
      -- & -- & -- & -- & -- & -- & -- & -- & -- & -- \\ \hline
      $B_d \to D^0 K^0$ & {\bf B} & -- & $\lambda^3$ &
      -- & -- & -- & -- & -- & -- & -- & -- & -- & -- \\ \hline
      $B_d \to D_s^+ \pi^-$ & {\bf B} & $\lambda^3$ & -- &
      -- & -- & -- & -- & -- & -- & -- & -- & -- & -- \\ \hline
      $B_d \to J/\psi K^0$ & {\bf C} & -- & $\lambda^2$ &
      -- & -- & -- & $\lambda^2$ & -- & -- & $\lambda^4$ & -- & -- &
      -- \\ \hline 
      $B_d \to D_s^+ D^-$ & {\bf C} & $\lambda^2$ & -- &
      -- & -- & $\lambda^2$ & -- & -- & $\lambda^4$ & -- & -- & -- &
      -- \\ \hline
      $B_d \to \pi^0 \pi^0$ & {\bf D} & -- & $\lambda^3$ &
      $\left[\lambda^3 \right]$ & $\lambda^3$ & $\lambda^3$ &
      $\left[\lambda^3 \right]$ & 
      $\lambda^3$ & $\lambda^3$ 
      & $\left[\lambda^3 \right]$ & $\lambda^3$ & $\left[\lambda^3
      \right]$ & $\left[\lambda^3 \right]$ \\  \hline 
      $B_d \to \pi^+ \pi^-$ & {\bf D} & $\lambda^3$ & -- &
      -- & $\lambda^3$ & $\lambda^3$ & -- & $\lambda^3$ & $\lambda^3$
      & -- & $\lambda^3$ & -- & -- \\  \hline
      $B_d \to D^+ D^-$ & {\bf D} & $\lambda^3$ & -- &
      -- & $\lambda^3$ & $\lambda^3$ & -- & $\lambda^3$ & $\lambda^3$
      & -- & $\lambda^3$ & -- & -- \\  \hline
      $B_d \to \pi^0 J/\psi$ & {\bf D} & -- & $\lambda^3$ &
      $\lambda^3$ & -- & -- & $\lambda^3$ & -- & -- &
      $\lambda^3$ 
      & -- & $\left[\lambda^3\right]$ & $\left[\lambda^3\right]$ \\
      \hline 
      $B_d \to K^+ \pi^-$ & {\bf E} & $\lambda^4$ & -- &
      -- & -- & $\lambda^2$ & -- & -- & $\lambda^4$ & -- & -- & -- &
      -- \\ \hline 
      $B_d \to K^0 \pi^0$ & {\bf E} & -- & $\lambda^4$ &
      -- & -- & $\lambda^2$ & $\left[\lambda^2 \right]$ & -- &
      $\lambda^4$ & $\left[\lambda^4 \right]$ & -- & -- & -- \\ \hline 
      $B_d \to K^0 \phi$ & {\bf E} & -- & -- &
      -- & -- & $\lambda^2$ & $\lambda^2$ & -- & $\lambda^4$ &
      $\lambda^4$ & -- & -- & -- \\ \hline 
      $B_d \to K^0 \bar K^0$ & {\bf F} & -- & -- &
      -- & -- & $\lambda^3$ & -- & $\lambda^3$ & $\lambda^3$ &
      -- & $\lambda^3$ & -- & -- \\ \hline 
      $B_d \to \phi \pi^0$ & {\bf F} & -- & -- &
      $\lambda^3$ & -- & -- & $\lambda^3$ & -- & -- &
      $\lambda^3$ & -- & $\left[\lambda^3 \right]$ & $\left[\lambda^3
      \right]$ \\ \hline  
      $B_d \to \phi \phi$ & {\bf F} & -- & -- &
      -- & -- & -- & -- & $\lambda^3$ & -- &
      -- & $\lambda^3$ & $\lambda^3$ & $\lambda^3$\\ \hline 
      $B_d \to D_s^- K^+$ & {\bf G} & -- & --  
      & -- & $\lambda^2$ & -- & -- & -- & -- & -- & -- & -- & -- \\ \hline
      $B_d \to \bar D^0 J/\psi$ & {\bf G} & -- & --  
      & $\lambda^2$ & $\lambda^2$ & -- & -- & -- & -- & -- & -- & -- &
      -- \\ \hline 
      $B_d \to \bar D^0 \phi$ & {\bf G} & -- & --  
      & $\lambda^2$ & -- & -- & -- & -- & -- & -- & -- & -- & -- \\ \hline 
      $B_d \to K^+ K^-$ & {\bf G} & -- & --  
      & -- & $\lambda^3$ & -- & -- & $\lambda^3$ & -- & -- & $\lambda^3$ &
      -- & -- \\ \hline 
      $B_d \to D_s^+ D_s^-$ & {\bf G} & -- & --  
      & -- & $\lambda^3$ & -- & -- & $\lambda^3$ & -- & -- & $\lambda^3$ &
      -- & -- \\ \hline 
      $B_d \to D^0 \bar D^0$ & {\bf G} & -- & --  
      & -- & $\lambda^3$ & -- & -- & $\lambda^3$ & -- & -- & $\lambda^3$ &
      -- & -- \\ 
      \hline \hline
    \end{tabular}
    \caption{Summary of the classification of two-body $B_d$ decays
      discussed in Section \ref{sec:classific}. For each channel, the
      order in the Wolfenstein parameter $\lambda$ of the relevant
      contributions is given. The contributions in squared brackets
      are of order $\alpha$ and they vanish in the limit of $SU(2)$
      symmetry (see Appendix \ref{sec:appew}); these contributions,
      together with the ones in the last two columns, have been
      neglected in the analysis of Section \ref{sec:classific}
      wherever possible.}
    \label{tab:classbd}
  \end{center}
\end{table}

\begin{table}[htbp]
  \begin{center}
    \begin{tabular}{||c|c|c|c|c|c|c|c|c|c||}
      \hline \hline
      Channel & Class & $E_1$ & $E_2$ & $A_1$ & ${\it EA}_1$ & $P_1$ &
      $P_2$ & $P_1^{\sss {\rm GIM}}$ & $P_2^{\sss {\rm GIM}}$ \\
      & & $ 1$ & $ \frac{1}{N}$ & $1$ & 
      $ \frac{1}{N}$ & $
      \frac{1}{N}$  & $ \frac{1}{N^2}$ &
      $ \frac{1}{N}$ & $
      \frac{1}{N^2}$ \\ \hline 
      $B^+ \to \bar D^0 \pi^+$ & {\bf A} & $\lambda^2$ & $\lambda^2$ & 
      -- & -- & -- & -- & -- & -- \\ \hline
      $B^+ \to \bar D^0 K^+$ & {\bf B} & $\lambda^3$ & $\lambda^3$ &
      -- & -- & -- & -- & -- & -- \\ \hline
      $B^+ \to D^0 K^+$ & {\bf B} & -- & $\lambda^3$ &
      $\lambda^3$ & -- & -- & -- & -- & -- \\ \hline
      $B^+ \to D_s^+ \pi^0$ & {\bf B} & $\lambda^3$ & -- &
      -- & $\left[\lambda^3 \right]$ & -- & -- & -- & -- \\ \hline
      $B^+ \to J/\psi K^+$ & {\bf C} & -- & $\lambda^2$ &
      -- & $\lambda^4$ & -- & $\lambda^2$ & -- & $\lambda^4$ \\ \hline
      $B^+ \to D_s^+ \bar D^0$ & {\bf C} & $\lambda^2$ & -- &
      $\lambda^4$  & -- & $\lambda^2$ & -- & $\lambda^4$ & -- \\  \hline
      $B^+ \to \pi^+ \pi^0$ & {\bf D} & $\lambda^3$ & $\lambda^3$ &
      $\left[\lambda^3 \right]$ & $\left[\lambda^3 \right]$ &
      $\left[\lambda^3 \right]$ & 
      $\left[\lambda^3 \right]$ & $\left[ \lambda^3 \right]$
      & $\left[ \lambda^3 \right]$ \\  \hline
      $B^+ \to K^+ \bar K^0$ & {\bf D} & -- & -- &
      $\lambda^3$ & -- & $\lambda^3$ &
      -- & $\lambda^3$
      & -- \\  \hline
      $B^+ \to \pi^+ J/\psi$ & {\bf D} & -- & $\lambda^3$ &
      -- & $\lambda^3$ & -- &
      $\lambda^3$ & --
      & $\lambda^3$ \\  \hline
      $B^+ \to D^+ \bar D^0$ & {\bf D} & $\lambda^3$ & -- &
      $\lambda^3$ & -- & $\lambda^3$ &
      -- & $\lambda^3$
      & -- \\  \hline
      $B^+ \to K^0 \pi^+$ & {\bf E} & -- & -- & $\lambda^4$ & -- & 
      $\lambda^2$ & -- & $\lambda^4$ & -- \\ \hline
      $B^+ \to K^+ \pi^0$ & {\bf E} &  $\lambda^4$ & $\lambda^4$ &
      $\lambda^4$ & $\left[\lambda^4 \right]$ & $\lambda^2$ &
      $\left[\lambda^2 \right]$ 
      & $\lambda^4$ & $\left[\lambda^4 \right]$ \\ \hline 
      $B^+ \to K^+ \phi$ & {\bf E} & -- & -- &
      $\lambda^4$ & $\lambda^4$ & $\lambda^2$ & $\lambda^2$ & $\lambda^4$ &
      $\lambda^4$ \\ \hline 
      $B^+ \to \phi \pi^+$ & {\bf F} & -- & -- &
      -- & $\lambda^3$ & -- & $\lambda^3$ & -- &
      $\lambda^3$ \\ \hline 
      $B^+ \to D^+ K^0$ & {\bf G} & -- & -- &
      $\lambda^3$ & -- & -- & -- & -- & -- \\ \hline
      $B^+ \to D_s^+ \phi$ & {\bf G} & -- & -- &
      $\lambda^3$ & $\lambda^3$ & -- & -- & -- & -- \\ \hline
      $B^+ \to D_s^+ J/\psi$ & {\bf G} & -- & -- &
      $\lambda^3$ & $\lambda^3$ & -- & -- & -- & -- \\
      \hline \hline
    \end{tabular}
    \caption{Summary of the classification of two-body $B^+$ decays
      discussed in Section \ref{sec:classific}. For each channel, the
      order in the Wolfenstein parameter $\lambda$ of the relevant
      contributions is given. The contributions in squared brackets
      are of order $\alpha$ and they vanish in the limit of $SU(2)$
      symmetry (see Appendix \ref{sec:appew}); they have been
      neglected in the analysis of Section \ref{sec:classific}
      wherever possible.}
    \label{tab:classbp}
  \end{center}
\end{table}

\begin{table}[htbp]
  \begin{center}
    \begin{tabular}{||c|c|c|c|c|c|c|c|c|c|c|c|c|c||}
      \hline \hline
      Channel & Cl. & $E_1$ & $E_2$ & ${\it EA}_2$ & $A_2$ & $P_1$ &
      $P_2$ & $P_3$ & $P_1^{\sss {\rm GIM}}$ & $P_2^{\sss {\rm GIM}}$
      & $P_3^{\sss {\rm GIM}}$ & $P_4$ & $P_4^{\sss {\rm GIM}}$ \\
      & & $1$ & $ \frac{1}{N}$ & $
      \frac{1}{N^2}$ & $ \frac{1}{N}$ &
      $ \frac{1}{N}$ & $
      \frac{1}{N^2}$ & $ \frac{1}{N^2}$
      & $ \frac{1}{N}$ & $
      \frac{1}{N^2}$ & $ \frac{1}{N^2}$
      & $ \frac{1}{N^3}$ & $
      \frac{1}{N^3}$  \\ \hline
      $B_s \to D_s^- \pi^+$ & {\bf A} & $\lambda^2$ & -- & -- & -- 
      & -- & -- & -- & -- & -- & -- & -- & --\\ \hline
      $B_s \to \bar D^0 \bar K^0$ & {\bf A} & -- & $\lambda^2$ 
      & -- & -- & -- & -- & -- & -- & -- & -- & -- & --\\ \hline
      $B_s \to D_s^- K^+$ & {\bf B} & $\lambda^3$ & -- &
      -- & $\lambda^3$ & -- & -- & -- & -- & -- & -- & -- & --\\ \hline
      $B_s \to D_s^+ K^-$ & {\bf B} & $\lambda^3$ & -- &
      -- & $\lambda^3$ & -- & -- & -- & -- & -- & -- & -- & --\\ \hline
      $B_s \to \bar D^0 \phi$ & {\bf B} & -- & $\lambda^3$ &
      $\lambda^3$ & -- & -- & -- & -- & -- & -- & -- & -- & --\\ \hline
      $B_s \to D^0 \phi$ & {\bf B} & -- & $\lambda^3$ &
      $\lambda^3$ & -- & -- & -- & -- & -- & -- & -- & -- & --\\ \hline
      $B_s \to D^+ K^-$ & {\bf B} & $\lambda^4$ & -- &
      -- & -- & -- & -- & -- & -- & -- & -- & -- & --\\ \hline
      $B_s \to D^0 \bar K^0$ & {\bf B} & -- & $\lambda^4$ &
      -- & -- & -- & -- & -- & -- & -- & -- & -- & --\\ \hline
      $B_s \to \phi \pi^0$ & {\bf B} & -- & $\lambda^4$ &
      $\lambda^4$ & -- & -- & $\left[\lambda^2 \right]$ & -- & -- &
      $\left[\lambda^4 \right]$ & -- & $\left[\lambda^2 \right]$
      & $\left[\lambda^4 \right]$ \\ \hline  
      $B_s \to J/\psi \phi$ & {\bf C} & -- & $\lambda^2$ &
      $\lambda^2$ & -- & -- & $\lambda^2$ & -- & -- & $\lambda^4$ & --
      & $\lambda^2$ & $\lambda^4$ \\ \hline
      $B_s \to D_s^+ D_s^-$ & {\bf C} & $\lambda^2$ & -- &
      -- & $\lambda^2$ & $\lambda^2$ & -- & $\lambda^2$ & $\lambda^4$
      & -- & $\lambda^4$ & -- & --\\ \hline 
      $B_s \to \pi^0 \bar K^0$ & {\bf D} & -- & $\lambda^3$ &
      -- & -- & $\lambda^3$ & $\left[\lambda^3 \right]$ &
      -- & $\lambda^3$ & $\left[\lambda^3 \right]$ & -- & -- & --\\  \hline
      $B_s \to \pi^+ K^-$ & {\bf D} & $\lambda^3$ & -- &
      -- & -- & $\lambda^3$ & -- & -- & $\lambda^3$ & -- & -- & -- &
      --\\  \hline 
      $B_s \to J/\psi \bar K^0$ & {\bf D} & -- & $\lambda^3$ &
      -- & -- & -- & $\lambda^3$ & -- & -- & $\lambda^3$ & -- & -- &
      --\\  \hline 
      $B_s \to D^+ D_s^-$ & {\bf D} & $\lambda^3$ & -- &
      -- & -- & $\lambda^3$ & -- & -- & $\lambda^3$ & -- & -- & -- &
      --\\  \hline 
      $B_s \to K^0 \bar K^0$ & {\bf E} & -- & -- &
      -- & -- & $\lambda^2$ & -- & $\lambda^2$ & $\lambda^4$ & -- &
      $\lambda^4$ & -- & 
      --\\ \hline 
      $B_s \to \phi \phi$ & {\bf E} & -- & -- &
      -- & -- & $\lambda^2$ & $\lambda^2$ & $\lambda^2$ &
      $\lambda^4$ & $\lambda^4$ & $\lambda^4$ & $\lambda^2$ &
      $\lambda^4$ \\ \hline  
      $B_s \to \pi^+ \pi^-$ & {\bf E} & -- & -- &
      -- & $\lambda^4$ & -- & -- & $\lambda^2$ & -- &
      -- & $\lambda^4$ & -- & -- \\ \hline 
      $B_s \to \pi^0 \pi^0$ & {\bf E} & -- & -- &
      $\left[\lambda^4 \right]$ & $\lambda^4$ & -- & -- & $\lambda^2$ & -- &
      -- & $\lambda^4$ & $\left[\lambda^2 \right]$ & $\left[\lambda^4
      \right]$\\ \hline  
      $B_s \to K^+ K^-$ & {\bf E} & $\lambda^4$ & -- &
      -- & $\lambda^4$ & $\lambda^2$ & -- & $\lambda^2$ &
      $\lambda^4$ & -- & $\lambda^4$ & -- & -- \\ \hline 
      $B_s \to \bar K^0 \phi$ & {\bf F} & -- & -- &
      -- & -- & $\lambda^3$ & $\lambda^3$ & -- & $\lambda^3$ &
      $\lambda^3$ & -- & -- & -- \\ \hline 
      $B_s \to D^{\stackrel{-}{0}}
      \stackrel{(-)}{D} \,^{\stackrel{+}{0}}$ & {\bf G} & -- & --   
      & -- & $\lambda^2$ & -- & -- & $\lambda^2$ & -- & -- &
      $\lambda^4$ & -- & -- \\ \hline 
      $B_s \to \pi^0 J/\psi$ & {\bf G} & -- & --  
      & $\left[\lambda^2 \right]$-$\lambda^4$  & -- & -- &
      -- & -- & -- & -- & 
      -- & $\left[\lambda^2 \right]$  & $\left[\lambda^4 \right]$  \\ \hline 
      $B_s \to \pi^{\pm} D^{\mp}$ & {\bf G} & -- & --  
      & -- & $\lambda^3$ & -- & -- & -- & -- & -- & -- & -- & -- \\ \hline
      $B_s \to \pi^0 \stackrel{(-)}{D} \,^0$ & {\bf G} & -- & --  
      & $\left[\lambda^3 \right]$ & $\lambda^3$ & -- & -- & -- & -- &
      -- & -- & -- & -- \\  \hline
      $B_s \to J/\psi D^0$ & {\bf G} & -- & --  
      & $\lambda^3$ & $\lambda^3$ & -- & -- & -- & -- &
      -- & -- & -- & -- \\  \hline
      $B_s \to J/\psi \bar D^0$ & {\bf G} & -- & --  
      & $\lambda^3$ & $\lambda^3$ & -- & -- & -- & -- &
      -- & -- & -- & -- \\  
      \hline \hline
    \end{tabular}
    \caption{Summary of the classification of two-body $B_s$ decays
      discussed in Section \ref{sec:classificbs}. For each channel,
      the order in the Wolfenstein parameter $\lambda$ of the relevant
      contributions is given. The contributions in squared brackets
      are of order $\alpha$ and they vanish in the limit of $SU(2)$
      symmetry (see Appendix \ref{sec:appew}); they have been
      neglected in the analysis of Section \ref{sec:classificbs},
      together with the contributions in the last two columns,
      wherever possible.}
    \label{tab:classbs}
  \end{center}
\end{table}

\section{Classification of Two-Body $B$ Decays}
\label{sec:classific}

Using the effective parameters defined above, we can classify the
two-body $B$ decay channels according to the effective parameters
entering in the decay amplitude. This classification enables us to
identify subsets of channels that, when measured, would allow us to
directly extract the effective parameters previously defined, making
no assumption about non-factorizable contributions and rescattering. 
We postpone the discussion of channels with $\eta$ and $\eta^\prime$
in the final state to a future publication.

As we discussed in the previous Section, we neglect here, wherever it
is possible, contributions proportional to the parameters $P_4$ and
$P_4^{\sss {\rm GIM}}$, and contributions that vanish in the $SU(2)$
symmetric limit. The expressions for the neglected terms, denoted in
this Section by $\Delta {\cal A}$, can be found in Appendix
\ref{sec:appew}.

We use the following conventions for the flavour content of mesons:
\begin{equation}
  \label{eq:states}
  \begin{array}{cccccc}
    B^+=\bar b u, & B_d = \bar b d, &
    B_s=\bar b s, & \pi^+ = \bar d u, &
    \pi^0 = \frac{1}{\sqrt{2}} \left(\bar d d -\bar u u\right), & \pi^-
    = - \bar u d,\\
    K^+ = \bar s u, & K^0 = \bar s d, &
    \bar K^0 = \bar d s, & K^- = -\bar u s,&
    D_s^+ = \bar s c, & D_s^-=\bar c s,\\
    D^+ = \bar d c, &  D^-=\bar c d,&
    D^0 = -\bar u c, & \bar D^0 = \bar c u,&
    \phi =  \bar s s, & J/\psi = \bar c c,
  \end{array}
\end{equation}
which agree with refs.~\cite{diagr2}.

\subsection*{Class A decays}

Class {\bf A} decays are CKM-allowed penguin-free decay channels.
These are particularly interesting since they would allow us to
extract $E_1$ and $E_2$. A typical example is given by $B \to D
\pi$ decays. We have (here and in the following we give results in
units of $G_F/\sqrt{2}$):
\begin{eqnarray}
  {\cal A}(B_d \to D^- \pi^+) &=& V_{ud} V_{cb}^*
  \Bigl(E_1(d,u,c;B_d,\pi^+,D^-) + A_2(c,d,u;B_d,D^-,\pi^+)\Bigr)\,;
  \nn \\
  {\cal A}(B_d \to \bar{D}^0 \pi^0) &=& \frac{V_{ud} V_{cb}^*}{\sqrt{2}}
  \Bigl(E_2(c,u,d;B_d,\bar{D}^0,\pi^0) - 
  A_2(c,u,u;B_d,\bar{D}^0,\pi^0)\Bigr) + \Delta {\cal A}(B_d \to \bar D^0
  \pi^0)\,; 
  \nn \\
  {\cal A}(B^+ \to \bar{D}^0 \pi^+) &=& V_{ud} V_{cb}^*
  \Bigl(E_1(d,u,c;B^+,\pi^+,\bar{D}^0) + E_2(c,u,d;B^+,\bar{D}^0,\pi^+)  
  \Bigr)\,.
  \label{eq:bdpi}
\end{eqnarray}

Other channels in class {\bf A} are obtained by replacing $D$ by $D^*$
and/or $\pi$ by $\rho$.

\subsection*{Class B decays}

These are penguin-free CKM-suppressed decay channels. 
A typical example of class {\bf B} channels is given by $B \to D K$ decays.
We have:
\begin{eqnarray}
  {\cal A}(B_d \to D^- K^+) &=& V_{us} V_{cb}^*
  \, E_1(s,u,c;B_d,K^+,D^-)\,; \nn
  \\
  {\cal A}(B_d \to \bar{D}^0 K^0) &=& V_{us} V_{cb}^*
  \, E_2(c,u,s;B_d,\bar{D}^0,K^0)\,;
  \nn \\
  {\cal A}(B^+ \to \bar{D}^0 K^+) &=& V_{us} V_{cb}^*
  \Bigl(E_1(s,u,c;B^+,K^+,\bar{D}^0) + 
  E_2(c,u,s;B^+,\bar{D}^0,K^+)\Bigr)\,.  
  \label{eq:bdk1}
\end{eqnarray}

The above channels are very interesting since they are also
annihilation-free and therefore would provide us with a cleaner
measurement of emission matrix elements with respect to class {\bf A}
decays. The same holds for the analogous channels with vector mesons.

Other interesting class {\bf B} decay channels are the following:
\begin{eqnarray}
  {\cal A}(B_d \to D^0 K^0) &=& - V_{cs} V_{ub}^*
  \,E_2(u,c,s;B_d,D^0,K^0)\,,
  \nn \\
  {\cal A}(B^+ \to D^0 K^+) &=& -V_{cs} V_{ub}^*
  \Bigl(E_2(u,c,s;B^+,D^0,K^+) + A_1(s,u,c;B^+,K^+,D^0)\Bigr)\,,
  \nn \\
  {\cal A}(B_d \to D_s^+ \pi^-) &=& - V_{cs} V_{ub}^*
  \,E_1(s,c,u;B_d,D_s^+,\pi^-)\,,
  \nn \\
  {\cal A}(B^+ \to D_s^+ \pi^0) &=& - \frac{V_{cs} V_{ub}^*}{\sqrt{2}}
  E_1(s,c,u;B^+,D_s^+,\pi^0) + \Delta{\cal A}(B^+ \to D_s^+ \pi^0)\,,  
  \label{eq:bdk2}
\end{eqnarray}
plus the corresponding ones with vector mesons. 

\subsection*{Class C decays}

Class {\bf C} decays are CKM-allowed channels in which penguin
contributions are present (but not dominant). Here are some typical
examples:
\begin{eqnarray}
  {\cal A}(B_d \to J/\psi K^0) &=& V_{cs} V_{cb}^*
  \,E_2(c,c,s;B_d,J/\psi,K^0) - V_{ts} V_{tb}^*
  \,P_2(s,c;B_d,J/\psi,K^0)-\nn \\ 
  &&V_{us} V_{ub}^* P_2^{\sss {\rm GIM}}(s,c;B_d,J/\psi,K^0)\,;
  \nn \\
  {\cal A}(B^+ \to J/\psi K^+) &=& V_{cs} V_{cb}^*\,
  E_2(c,c,s;B^+,J/\psi,K^+) - V_{ts} V_{tb}^*
  \,P_2(s,c;B^+,J/\psi,K^+)-\nn \\ 
  &&V_{us} V_{ub}^* \Bigl(P_2^{\sss {\rm GIM}}(s,c;B^+,J/\psi,K^+) -
  {\it EA}_1(s,u,c;B^+,K^+,J/\psi)\Bigr)\,;
  \nn \\
  {\cal A}(B_d \to D_s^+ D^-) &=& V_{cs} V_{cb}^*\,
  E_1(s,c,c;B_d,D_s^+,D^-) - V_{ts} V_{tb}^* \,P_1(s,c;B_d,D_s^+,D^-)- \nn
  \\
  &&V_{us} V_{ub}^* \,P_1^{\sss {\rm GIM}}(s,c;B_d,D_s^+,D^-)\,; 
  \nn \\
  {\cal A}(B^+ \to D_s^+ \bar{D}^0) &=& V_{cs} V_{cb}^*\,
  E_1(s,c,c;B^+,D_s^+,\bar{D}^0) -
  V_{ts} V_{tb}^* \,
  P_1(s,c;B^+,D_s^+,\bar{D}^0)+\nn \\
  &&V_{us} V_{ub}^* \Bigl(
  A_1(s,c,u;B^+,D_s^+,\bar{D}^0) - P_1^{\sss {\rm
  GIM}}(s,c;B^+,D_s^+,\bar{D}^0)\Bigr)\,.
  \label{eq:psik}
\end{eqnarray}

\subsection*{Class D decays}

Class {\bf D} decays are CKM-suppressed decays in which penguin
contributions are present. Well-known examples of this kind are $B
\to \pi \pi$ decays:
\begin{eqnarray}
  \sqrt{2}{\cal A}(B_d \to \pi^0 \pi^0) = &-& V_{ud} V_{ub}^*
  \Bigl(E_2(u,u,d;B_d,\pi^0,\pi^0) -
  A_2(u,u,u;B_d,\pi^0,\pi^0) + P_1^{\sss {\rm
  GIM}}(d,d;B_d,\pi^0,\pi^0) \nn \\
  &&\qquad \qquad 
  +\frac{1}{2} P_3^{\sss {\rm GIM}}(d,d;B_d,\pi^0,\pi^0) +
  \frac{1}{2} P_3^{\sss {\rm GIM}}(u,u;B_d,\pi^0,\pi^0) \Bigr) \nn \\
  &-& V_{td} V_{tb}^* \Bigl(P_1(d,d;B_d,\pi^0,\pi^0)+ 
  \frac{1}{2} P_3(d,d;B_d,\pi^0,\pi^0) +
  \frac{1}{2} P_3(u,u;B_d,\pi^0,\pi^0) \Bigr) \nn \\ 
  &+& \Delta{\cal A}(B_d \to \pi^0 \pi^0)\,;  \nn \\
  {\cal A}(B_d \to \pi^+ \pi^-) = &-& V_{ud} V_{ub}^*
  \Bigl( E_1(d,u,u;B_d,\pi^+,\pi^-) +
  A_2(u,d,u;B_d,\pi^-,\pi^+) -\nn \\
  && \qquad \qquad P_1^{\sss {\rm GIM}}(d,u;B_d,\pi^+,\pi^-) -
  P_3^{\sss {\rm GIM}}(u,d;B_d,\pi^+,\pi^-) \Bigr)\nn \\
  &+& V_{td} V_{tb}^* \Bigl(P_1(d,u;B_d,\pi^+,\pi^-) +
  P_3(u,d;B_d,\pi^+,\pi^-)\Bigr)\nn \\
  {\cal A}(B^+ \to \pi^+ \pi^0) = &-& \frac{V_{ud} V_{ub}^*}{\sqrt{2}}
  \Bigl( E_1(d,u,u;B^+,\pi^+,\pi^0) + E_2(u,u,d;B^+,\pi^0,\pi^+)
  \Bigr) \nn \\
  &-& \frac{V_{td} V_{tb}^*}{\sqrt{2}} \Biggl(\Bigl[P_1(d,d;B^+,\pi^0,\pi^+) -
  P_1(d,u;B^+,\pi^+,\pi^0) \Bigr] +\nn \\
  && \qquad \qquad \Bigl[P_2(d,d;B^+,\pi^0,\pi^+) -
  P_2(d,u;B^+,\pi^0,\pi^+) \Bigr]\Biggr)\nn \\
  &+& \Delta{\cal A}(B^+ \to \pi^+ \pi^0)\,,
  \label{eq:bpipi}
\end{eqnarray}
where the quantities in square brackets in ${\cal A}(B^+
\to \pi^+ \pi^0)$ vanish in the limit of exact $SU(2)$ symmetry. 

The presence of many different contributions with different weak
phases and potentially different strong phases implies that it will be
very difficult to extract $\sin 2 \alpha$ from the measurement of the
asymmetry in $B_d \to \pi^+ \pi^-$ \cite{CHARMING1,sin2al}.

Other channels in this class are the following:
\begin{eqnarray}
  \label{eq:classd2}
  {\cal A}(B_d \to \pi^0 J/\psi) = &+& \frac{V_{cd} V_{cb}^*}{\sqrt{2}}
  E_2(c,c,d;B_d,J/\psi,\pi^0) - \frac{V_{td} V_{tb}^* }{\sqrt{2}}
  P_2(d,c;B_d,J/\psi,\pi^0)\nn \\
  &-&\frac{V_{ud} V_{ub}^*}{\sqrt{2}} \Bigl(P_2^{\sss {\rm
  GIM}}(d,c;B_d,J/\psi,\pi^0) -
  {\it EA}_2(u,u,c;B_d,\pi^0,J/\psi)\Bigr) \nn \\
  &+& \Delta{\cal A}(B_d \to \pi^0 J/\psi) \,; 
  \nn \\
  {\cal A}(B_d \to D^+ D^-) = &+& V_{cd} V_{cb}^*
  \Bigl( E_1(d,c,c;B_d,D^+,D^-) +
  A_2(c,d,c;B_d,D^-,D^+)\Bigr) \nn \\
  &-&V_{ud} V_{ub}^* \Bigl(P_1^{\sss {\rm GIM}}(d,c;B_d,D^+,D^-) +
  P_3^{\sss {\rm GIM}}(d,c;B_d,D^-,D^+) \Bigr)\nn \\
  &-& V_{td} V_{tb}^* \Bigl(P_1(d,c;B_d,D^+,D^-) +
  P_3(d,c;B_d,D^-,D^+)\Bigr) \,; \nn \\
  {\cal A}(B^+ \to K^+ \bar K^0) = &+& V_{ud} V_{ub}^* \Bigl(
  A_1(d,s,u;B^+,\bar K^0,K^+) -  P_1^{\sss {\rm GIM}}(d,s;B^+,\bar
  K^0,K^+)\Bigr) \nn \\
  &-& V_{td}V_{tb}^* \,P_1(d,s;B^+,\bar K^0,K^+)\,;\nn \\
  {\cal A}(B^+ \to \pi^+ J/\psi) = &+& V_{cd} V_{cb}^*\,
  E_2(c,c,d;B^+,J/\psi,\pi^+) - V_{td} V_{tb}^*\,
  P_2(d,c;B^+,J/\psi,\pi^+) \nn \\
  &-&V_{ud} V_{ub}^* \Bigl(P_2^{\sss {\rm
  GIM}}(d,c;B^+,J/\psi,\pi^+) -
  {\it EA}_1(d,u,c;B^+,\pi^+,J/\psi)\Bigr)\,; 
  \nn \\
  {\cal A}(B^+ \to D^+ \bar D^0) = &+& V_{cd} V_{cb}^*\, 
  E_1(d,c,c;B^+,D^+, \bar D^0) - V_{td}V_{tb}^* \,P_1(d,c;B^+,D^+,\bar
  D^0) \nn \\
  &+& V_{ud} V_{ub}^* \Bigl(
  A_1(d,c,u;B^+,D^+,\bar D^0) -  P_1^{\sss {\rm GIM}}(d,c;B^+,D^+,\bar
  D^0) \Bigr)\, ,
\end{eqnarray}
and the corresponding channels with vector mesons
replacing the pseudoscalars. 

\subsection*{Class E decays}

This very interesting class consists of (charming) penguin dominated
channels. These decays would be doubly CKM suppressed if there
were no penguins; since the penguin contributions are instead CKM
allowed, they are expected to dominate the decay amplitude.

These modes have recently received a lot of attention after the
observation at CLEO of $B \to K \pi$ decays \cite{CLEO}. As an
example, we write down here the expression for these measured
channels:
\begin{eqnarray}
  {\cal A}(B_d \to K^+ \pi^-) = &-& V_{us} V_{ub}^*
  \Bigl(E_1(s,u,u;B_d,K^+,\pi^-) - P_1^{\sss {\rm
  GIM}}(s,u;B_d,K^+,\pi^-)\Bigr)\nn \\ 
  &+& V_{ts} V_{tb}^* \,P_1(s,u;B_d,K^+,\pi^-)\,; 
  \nn \\
  {\cal A}(B^+ \to K^+ \pi^0) = &-& \frac{V_{us} V_{ub}^*}{\sqrt{2}}
  \Bigl( E_1(s,u,u;B^+,K^+,\pi^0) + E_2(u,u,s;B^+,\pi^0,K^+) - \nn \\
  && \qquad \qquad P_1^{\sss {\rm
  GIM}}(s,u;B^+,K^+,\pi^0) + A_1(s,u,u;B^+,K^+,\pi^0) \Bigr)\nn \\ 
  &+&\frac{ V_{ts} V_{tb}^*}{\sqrt{2}} P_1(s,u;B^+,K^+,\pi^0)
  + \Delta{\cal A}(B^+ \to K^+ \pi^0) \,; \nn \\
  {\cal A}(B^+ \to K^0 \pi^+) = &+& V_{us} V_{ub}^*
  \Bigl(A_1(s,d,u;B^+,K^0,\pi^+) -  P_1^{\sss {\rm
  GIM}}(s,d;B^+,K^0,\pi^+)\Bigr)\nn \\ 
  &-& V_{ts} V_{tb}^* \,P_1(s,d;B^+,K^0,\pi^+)\,.
  \label{eq:bKpi}
\end{eqnarray}

Other modes in this class are the following ones:
\begin{eqnarray}
  {\cal A}(B_d \to K^0 \pi^0) = &-& \frac{V_{us} V_{ub}^*}{\sqrt{2}}
  \Bigl( E_2(u,u,s;B_d,\pi^0,K^0) + P_1^{\sss {\rm
  GIM}}(s,d;B_d,K^0,\pi^0)\Bigr)\nn \\ 
  &-&\frac{ V_{ts} V_{tb}^*}{\sqrt{2}} P_1(s,d;B_d,K^0,\pi^0)
  + \Delta{\cal A}(B_d \to K^0 \pi^0)\,;  
  \nn \\
  {\cal A}(B_d \to K^0 \phi) =  
  &-& V_{us} V_{ub}^* \Bigl(P_1^{\sss {\rm GIM}}(s,s;B_d,\phi,K^0) +
  P_2^{\sss {\rm GIM}}(s,s;B_d,\phi,K^0)\Bigr)  \nn \\
  &-&
  V_{ts} V_{tb}^* \Bigl(P_1(s,s;B_d,\phi,K^0) +
  P_2(s,s;B_d,\phi,K^0)\Bigr)\,;  \nn \\
  {\cal A}(B^+ \to K^+ \phi) = &-& V_{us} V_{ub}^*
  \Bigl(P_1^{\sss {\rm GIM}}(s,s;B^+,\phi,K^+)+ P_2^{\sss {\rm
  GIM}}(s,s;B^+,\phi,K^+)-\nn \\ 
  &&\qquad \qquad A_1(s,s,u;B^+,\phi,K^+) - {\it
  EA}_1(s,u,s;B^+,K^+,\phi)\Bigr)\nn \\ 
  &-& V_{ts} V_{tb}^* \Bigl(P_1(s,s;B^+,\phi,K^+)+ P_2(s,s;B^+,\phi,K^+)
  \Bigr)\,.
  \label{eq:bKpi2}
\end{eqnarray}

\subsection*{Class F decays}

These are CKM-suppressed pure penguin decays. Observation of these
channels would also give us a measurement of penguin contributions.
As an example, we write down the amplitudes for $B \to K^0 \bar K^0$,
 $B \to \phi \pi$ and $B \to \phi \phi$ decays:
\begin{eqnarray}
  {\cal A}(B_d \to K^0 \bar{K}^0) = &-&V_{ud} V_{ub}^* \Bigl(
  P_1^{\sss {\rm GIM}}(d,s;B_d,\bar{K}^0,K^0) + 
  P_3^{\sss {\rm GIM}}(d,s;B_d,K^0,\bar{K}^0) \Bigr) \nn \\
  &-& V_{td} V_{tb}^*
  \Bigl( P_1(d,s;B_d,\bar{K}^0,K^0) +
  P_3(d,s;B_d,K^0,\bar{K}^0)\Bigr)\,; 
  \nn \\
  {\cal A}(B_d \to \phi \pi^0) =  &-& \frac{V_{ud} V_{ub}^*}{\sqrt{2}}
  \Bigl(P_2^{\sss {\rm GIM}}(d,s;B_d,\phi,\pi^0) +
  {\it EA}_2(u,u,s;B_d,\pi^0,\phi)\Bigr) \nn \\ 
  &-&\frac{V_{td} V_{tb}^*}{\sqrt{2}}
  P_2(d,s;B_d,\phi,\pi^0) 
  + \Delta{\cal A}(B_d \to \phi \pi^0)\,;
  \nn \\
  {\cal A}(B^+ \to \phi \pi^+) = &-& V_{ud} V_{ub}^* \Bigl(
  P_2^{\sss {\rm GIM}}(d,s;B^+,\phi,\pi^+) - {\it
  EA}_1(d,u,s;B^+,\pi^+,\phi)\Bigr)\nn \\
   &-& V_{td} V_{tb}^*\,
  P_2(d,s;B^+,\phi,\pi^+)\,; \nn \\ 
  \sqrt{2} {\cal A}(B_d \to \phi \phi) = &-& V_{ud} V_{ub}^*
  P_3^{\sss {\rm GIM}}(s,s;B_d,\phi,\phi)\Bigr) 
  -  V_{td} V_{tb}^*
  P_3(s,s;B_d,\phi,\phi)
  + \Delta{\cal A}(B_d \to \phi \phi)\,.
  \label{eq:bKK}
\end{eqnarray}

\subsection*{Class G decays}

These decays proceed only via annihilations: a measurement of these
channels would provide us with a direct determination of annihilation
amplitudes. 

The best examples of class {\bf G} decays are these 
CKM-allowed channels:
\begin{eqnarray}
  {\cal A}(B_d \to D_s^- K^+) &=& V_{ud} V_{cb}^*
  \,A_2(c,s,u;B_d,D_s^-,K^+)\,; 
  \nn \\
  {\cal A}(B_d \to \bar{D}^0 J/\psi) &=& V_{ud} V_{cb}^*
  \Bigl(A_2(c,c,u;B_d,J/\psi,\bar{D}^0) + {\it
  EA}_2(c,u,c;B_d,\bar{D}^0,J/\psi)\,; \nn \\  
  {\cal A}(B_d \to \bar{D}^0 \phi) &=&  V_{ud} V_{cb}^*
  \,{\it
  EA}_2(c,u,s;B_d,\bar{D}^0,\phi)\,,
\label{eq:bann1}
\end{eqnarray}
which would allow the extraction of $A_2$ and ${\it EA}_2$. To be able
to extract $A_1$ one has to measure CKM suppressed decays such as
the following:
\begin{eqnarray}
  {\cal A}(B^+ \to D^+ K^0) &=& V_{cs} V_{ub}^*
  A_1(s,d,c;B^+,K^0,D^+)\,; 
  \nn \\
  {\cal A}(B^+ \to D_s^+ \phi) &=&  V_{cs} V_{ub}^*
  \Bigl(A_1(s,s,c;B^+,\phi,D_s^+) + {\it EA}_1(s,c,s;B^+,D_s^+,\phi)
  \Bigr)\,,\nn \\
  {\cal A}(B^+ \to D_s^+ J/\psi) &=& V_{cs} V_{ub}^*
  \Bigl(A_1(s,c,c;B^+,D_s^+,J/\psi) + {\it EA}_1(s,c,c;B^+,D_s^+,J/\psi)
  \Bigr)\,.   
  \label{eq:bann3}
\end{eqnarray}

An example of decays proceeding through annihilations and
penguin-annihilations is given by the following channels:
\begin{eqnarray}
  {\cal A}(B_d \to D^0 \bar D^0) &=&
  -V_{cd}V_{cb}^*  A_2(c,u,c;B_d,\bar D^0,D^0)+ V_{td}V_{tb}^*
  P_3(u,c;B_d,\bar D^0,D^0) \nn \\
  &&-V_{ud}V_{ub}^* \Bigl(A_2(u,c,u;B_d,D^0,\bar D^0)- P_3^{\sss {\rm
      GIM}}(u,c;B_d,\bar D^0,D^0) \Bigr)\,;\nn \\
  {\cal A}(B_d \to D_s^+ D_s^-) &=&
  V_{cd}V_{cb}^*  \,A_2(c,s,c;B_d,D_s^-,D_s^+) - V_{ud}V_{ub}^* \,
  P_3^{\sss {\rm
      GIM}}(c,s;B_d,D_s^+,D_s^-) \nn \\
  &&- V_{td}V_{tb}^* \,P_3(c,s;B_d,D_s^+,D_s^-)\,;\nn \\
  {\cal A}(B_d \to K^+ K^-) &=&
  V_{ud}V_{ub}^* \Bigl( - A_2(u,s,u;B_d,K^-,K^+) + P_3^{\sss {\rm
      GIM}}(u,s;B_d,K^+,K^-) \Bigr) \nn \\
  &&+ V_{td}V_{tb}^* \,P_3(u,s;B_d,K^+,K^-)\,.
  \label{eq:kpkm}
\end{eqnarray}

The classification of $B_d$ and $B^+$ decays is summarized in Tables
\ref{tab:classbd} and \ref{tab:classbp} respectively.

\section{Classification of Two-Body $B_s$ Decays}
\label{sec:classificbs}

The same formalism introduced in Sections \ref{sec:details} and
\ref{sec:classific} can be applied to $B_s$ decays, and a
classification of the various decay channels can be made according to
the CKM structure and to the effective parameters entering the
amplitude.

As in the previous Section, we present here the results obtained by
neglecting terms that vanish in the limit of exact $SU(2)$ symmetry
and discarding the effective
parameters  $P_4$ and
$P_4^{\sss {\rm GIM}}$, wherever it is justified to do so. The
neglected terms, denoted in this Section by $\Delta {\cal A}$, can be
found in Appendix \ref{sec:appew}.

\subsection*{Class A decays}

Class {\bf A} decays are CKM-allowed penguin-free decay channels.
We have:
\begin{eqnarray}
  {\cal A}(B_s \to D_s^- \pi^+) &=& V_{ud} V_{cb}^*
  \,E_1(d,u,c;B_s,\pi^+,D_s^-)\,;
  \nn \\
  {\cal A}(B_s \to \bar{D}^0 \bar{K}^0) &=& V_{ud} V_{cb}^*
  \,E_2(c,u,d;B_s,\bar{D}^0,\bar{K}^0)\,.
  \label{eq:bsA}
\end{eqnarray}

Other channels in class {\bf A} are obtained by replacing pseudoscalar
mesons with vector ones.

\subsection*{Class B decays}

These are penguin-free CKM-suppressed decay channels. 
We have:
\begin{eqnarray}
  {\cal A}(B_s \to D_s^- K^+) &=& V_{us} V_{cb}^*
  \Bigl(E_1(s,u,c;B_s,K^+,D_s^-)+ A_2(c,s,u;B_s,D_s^-,K^+)\Bigr)\,,
  \nn \\
  {\cal A}(B_s \to D_s^+ K^-) &=& -\,V_{cs} V_{ub}^*
  \Bigl(E_1(s,c,u;B_s,D_s^+,K^-) + 
  A_2(u,s,c;B_s,K^-,D_s^+)\Bigr)\,,\nn \\
  {\cal A}(B_s \to \bar{D}^0 \phi) &=&  V_{us} V_{cb}^*
  \Bigl(E_2(c,u,s;B_s,\bar D^0,\phi) + {\it EA}_2(c,u,s;B_s,\bar
  D^0,\phi) \Bigr)\,,
  \nn \\
  {\cal A}(B_s \to D^0 \phi) &=& - V_{cs} V_{ub}^*
  \Bigl(E_2(u,c,s;B_s,D^0,\phi) + {\it EA}_2(u,c,s;B_s,D^0,\phi)
  \Bigr)\,.
  \label{eq:bsB1}
\end{eqnarray}
Other class {\bf B} decays are the following
doubly-CKM-suppressed transitions:
\begin{eqnarray}
  {\cal A}(B_s \to D^+ K^-) &=& -\,V_{cd} V_{ub}^*
  E_1(d,c,u;B_s,D^+,K^-)\,;
  \nn \\
  {\cal A}(B_s \to D^0 \bar K^0) &=& -\,V_{cd} V_{ub}^*
  \,E_2(u,c,d;B_s,D^0,\bar K^0)\,;
  \nn \\
  {\cal A}(B_s \to \pi^0 \phi) = &-& \frac{V_{us} V_{ub}^*}{\sqrt{2}}
  \Bigl(E_2(u,u,s;B_s,\pi^0,\phi) + {\it EA}_2(u,u,s;B_s,\pi^0,\phi)
  \Bigr)  \nn \\ 
  &-& \frac{V_{ts} V_{tb}^*}{\sqrt{2}}
  \Bigl[P_2(s,d;B_s,\pi^0,\phi) -
  P_2(s,u;B_s,\pi^0,\phi) \Bigr] + \Delta{\cal A}(B_s \to \pi^0 \phi)\,,  
  \label{eq:bsB2}
\end{eqnarray}
plus the corresponding ones with vector mesons. The term in square
brackets in ${\cal A}(B_s \to \pi^0 \phi)$ vanishes in the limit of
exact isospin symmetry.

\subsection*{Class C decays}

Class {\bf C} decays are CKM-allowed channels in which penguin
contributions are present (but not dominant). Here are some typical
examples:
\begin{eqnarray}
  {\cal A}(B_s   \to D_s^+ D_s^-) &=& V_{cs} V_{cb}^*
  \Bigl(E_1(s,c,c;B_s,D_s^+,D_s^-)+A_2(c,s,c;B_s,D_s^-,D_s^+)\Bigr) - \nn \\
  &&V_{ts} V_{tb}^*
  \Bigl(P_1(s,c;B_s,D_s^+,D_s^-)+P_3(c,s;B_s,D_s^+,D_s^-)\Bigr) 
  -\nn \\
  &&V_{us} V_{ub}^*\Bigl(P_1^{\sss {\rm
  GIM}}(s,c;B_s,D_s^+,D_s^-)+P_3^{\sss {\rm GIM}}
  (c,s;B_s,D_s^+,D_s^-)\Bigr) \,;  
  \nn \\
  {\cal A}(B_s \to J/\psi \phi) &=& V_{cs} V_{cb}^*
  \Bigl(E_2(c,c,s;B_s,J/\psi,\phi) + {\it
  EA}_2(c,c,s;B_s,J/\psi,\phi)\Bigr) \nn \\
  &-& V_{ts} V_{tb}^*\,
  P_2(s,c;B_s,J/\psi,\phi) - V_{us} V_{ub}^* \,P_2^{\sss {\rm
  GIM}}(s,c;B_s,J/\psi,\phi) \nn \\
  &+& \Delta{\cal A}(B_s \to J/\psi \phi) \,. 
  \label{eq:bsC}
\end{eqnarray}

\subsection*{Class D decays}

Class {\bf D} decays are CKM-suppressed decays in which penguin
contributions are present:
\begin{eqnarray}
  {\cal A}(B_s \to \pi^0 \bar K^0) = &-& \frac{V_{ud} V_{ub}^*}{\sqrt{2}}
  \Bigl( E_2(u,u,d;B_s,\pi^0,\bar K^0) + P_1^{\sss {\rm
  GIM}}(d,d;B_s,\pi^0,\bar K^0)\Bigr) \nn \\ 
  &-& \frac{V_{td} V_{tb}^*}{\sqrt{2}} 
  P_1(d,d;B_s,\pi^0,\bar K^0) + \Delta{\cal A}(B_s
  \to \pi^0 \bar K^0)\,; 
  \nn \\
  {\cal A}(B_s \to \pi^+ K^-) = &-& V_{ud} V_{ub}^*
  \Bigl( E_1(d,u,u;B_s,\pi^+,K^-) -
  P_1^{\sss {\rm GIM}}(d,u;B_s,\pi^+,K^-) \Bigr)\nn \\
  &+& V_{td} V_{tb}^* \,P_1(d,u;B_s,\pi^+,K^-)\,; \nn \\
  {\cal A}(B_s   \to D^+ D_s^-) &=& V_{cd} V_{cb}^*
  \,E_1(d,c,c;B_s,D^+,D_s^-)- V_{td} V_{tb}^*\, P_1(d,c;B_s,D^+,D_s^-)-\nn \\
  &&V_{ud} V_{ub}^* \,P_1^{\sss {\rm GIM}}(d,c;B_s,D^+,D_s^-)\,; 
  \nn \\
  {\cal A}(B_s \to \bar{K}^0 J/\psi) &=& V_{cd} V_{cb}^*
  \,E_2(c,c,d;B_s,J/\psi,\bar K^0)- V_{td} V_{tb}^*\,
  P_2(d,c;B_s,J/\psi,\bar K^0)-\nn \\ 
  &&V_{ud} V_{ub}^*\, P_2^{\sss {\rm GIM}}(d,c;B_s,J/\psi,\bar K^0)\,. 
  \label{eq:bsD1}
\end{eqnarray}

\subsection*{Class E decays}

This class contains channels in which penguin contributions are
CKM-allowed, while non-penguin ones are either absent:
\begin{eqnarray}
  {\cal A}(B_s \to K^0 \bar K^0) = &-& V_{us} V_{ub}^*
  \Bigl(P_1^{\sss {\rm GIM}}(s,d;B_s,K^0,\bar K^0) + P_3^{\sss {\rm
  GIM}}(d,s;B_s,K^0,\bar 
  K^0)\Bigr)\nn \\ 
  &-& V_{ts} V_{tb}^* \Bigl(P_1(s,d;B_s,K^0,\bar K^0) + P_3(d,s;B_s,K^0,\bar
  K^0)\Bigr)\,;  
  \nn \\
  \sqrt{2}{\cal A}(B_s \to \phi \phi) = &-& V_{us} V_{ub}^*
  \Bigl(2 P_1^{\sss {\rm GIM}}(s,s;B_s,\phi,\phi) + 2 P_2^{\rm
    GIM}(s,s;B_s,\phi,\phi) +   
  P_3^{\sss {\rm GIM}}(s,s;B_s,\phi,\phi)\Bigr) \nn \\ 
  &-& V_{ts} V_{tb}^* \Bigl(2
  P_1(s,s;B_s,\phi,\phi) + 2 P_2(s,s;B_s,\phi,\phi) +
  P_3(s,s;B_s,\phi,\phi)\Bigr)\nn \\ 
  &+& \Delta{\cal A}(B_s \to \phi \phi) \,; 
  \label{eq:bsE1}
\end{eqnarray}
or doubly CKM-suppressed:
\begin{eqnarray}
  {\cal A}(B_s \to \pi^+ \pi^-) = &+& V_{us} V_{ub}^*
  \Bigl(P_3^{\sss {\rm GIM}}(u,d;B_s,\pi^+,\pi^-) -
  A_2(u,d,u;B_s,\pi^-,\pi^+)\Bigr)\nn \\ 
  &+& V_{ts} V_{tb}^*\, P_3(u,d;B_s,\pi^+,\pi^-) \,; 
  \nn \\
  \sqrt{2}{\cal A}(B_s \to \pi^0 \pi^0) = &-& V_{us} V_{ub}^*
  \Bigl(\frac{1}{2} P_3^{\sss {\rm GIM}}(u,u;B_s,\pi^0,\pi^0) +
  \frac{1}{2} P_3^{\sss {\rm GIM}}(d,d;B_s,\pi^0,\pi^0)\nn \\
  && \qquad \qquad -
  A_2(u,u,u;B_s,\pi^0,\pi^0)\Bigr)\nn \\ 
  &-& V_{ts} V_{tb}^* \Bigl(\frac{1}{2} P_3(u,u;B_s,\pi^0,\pi^0) +
  \frac{1}{2} P_3(d,d;B_s,\pi^0,\pi^0) \Bigr)
  + \Delta{\cal A}(B_s \to \pi^0 \pi^0) \,; 
  \nn \\
  {\cal A}(B_s \to K^+ K^-) = &-& V_{us} V_{ub}^*
  \Bigl(E_1(s,u,u;B_s,K^+,K^-) + A_2(u,s,u;B_s,K^-, K^+) - \nn \\
  && \qquad \qquad 
  P_1^{\sss {\rm GIM}}(s,u;B_s,K^+, K^-) - P_3^{\sss {\rm GIM}}
  (u,s;B_s,K^+,K^-) \Bigr) \nn \\  
  &+& V_{ts} V_{tb}^* \Bigl(P_1(s,u;B_s,K^+,K^-) + P_3(u,s;B_s,K^+,K^-)
  \Bigr)\,. 
  \label{eq:bsE2}
\end{eqnarray}

\subsection*{Class F decays}

The following channel is an example of a CKM-suppressed, 
pure penguin decay:
\begin{eqnarray}
  {\cal A}(B_s \to \bar{K}^0 \phi) = &-& V_{td} V_{tb}^* \Bigl(
  P_1(d,s;B_s,\bar K^0,\phi) + P_2(d,s;B_s,\phi,\bar K^0)\Bigr)\nn \\
  &-&V_{ud} V_{ub}^*\Bigl(
  P_1^{\sss {\rm GIM}}(d,s;B_s,\bar K^0,\phi) + P_2^{\sss {\rm
  GIM}}(d,s;B_s,\phi,\bar 
  K^0)\Bigr)  \,. 
  \label{eq:bsD2}  
\end{eqnarray}

\subsection*{Class G decays}

The following channels are CKM-allowed decays that
proceed only through annihilation and penguin-annihilation:
\begin{eqnarray}
  {\cal A}(B_s \to D^- D^+) &=& V_{cs} V_{cb}^*
  \,A_2(c,d,c;B_s,D^-,D^+) - V_{us} V_{ub}^*
  \,P_3^{\sss {\rm GIM}}(d,c;B_s,D^-,D^+)\nn \\
  &&-V_{ts} V_{tb}^* \,P_3(d,c;B_s,D^-,D^+)\,; \nn \\ 
  {\cal A}(B_s \to D^0 \bar D^0) &=& - V_{cs} V_{cb}^*
  A_2(c,u,c;B_s,\bar D^0,D^0) +V_{ts} V_{tb}^* P_3(u,c;B_s,\bar
  D^0,D^0)\nn \\ 
  &&+ V_{us} V_{ub}^*
  \Bigl(P_3^{\sss {\rm GIM}}(u,c;B_s,\bar D^0,D^0) - A_2(u,c,u;B_s,D^0,\bar
  D^0)\Bigr)\,.
  \label{eq:bsG0}
\end{eqnarray}

$B_s \to \pi^0 J/\psi$ is an annihilation channel dominated by
electroweak (isospin-breaking) effects:
\begin{eqnarray}
  {\cal A}(B_s \to \pi^0 J/\psi) = &+& \frac{V_{cs} V_{cb}^*}{\sqrt{2}}
  \Bigl[{\it EA}_2(c,c,d;B_s,J/\psi,\pi^0) - {\it
  EA}_2(c,c,u;B_s,J/\psi,\pi^0) \Bigr]\nn \\
  &-& \frac{V_{ts} V_{tb}^*}{\sqrt{2}} \Bigl[P_4(d,c;B_s,\pi^0,J/\psi) -
  P_4(u,c;B_s,\pi^0,J/\psi)\Bigr] \nn \\
  &-& \frac{V_{us} V_{ub}^*}{\sqrt{2}}
  {\it EA}_2(u,u,c;B_s,\pi^0,J/\psi)
  + \Delta{\cal A}(B_s \to \pi^0 J/\psi)\,,
  \label{eq:bsGew}
\end{eqnarray}
where the quantities in square brackets vanish in the $SU(2)$ limit.

Here are some examples of CKM-suppressed pure annihilation decays:
\begin{eqnarray}
  {\cal A}(B_s \to \pi^- D^+) &=& - V_{cs} V_{ub}^*
  \,A_2(u,d,c;B_s,\pi^-,D^+)\,; 
  \nn \\
  {\cal A}(B_s \to \pi^+ D^-) &=& V_{us} V_{cb}^*
  \,A_2(c,d,u;B_s,D^-,\pi^+)\,; 
  \nn \\
  {\cal A}(B_s \to J/\psi D^0) &=& - V_{cs} V_{ub}^*
  \Bigl(A_2(u,c,c;B_s,D^0,J/\psi) + {\it
  EA}_2(u,c,c;B_s,D^0,J/\psi)\Bigr)\,;  
  \nn \\
  {\cal A}(B_s \to J/\psi \bar D^0) &=&  V_{us} V_{cb}^*
  \Bigl(A_2(c,c,u;B_s,J/\psi,\bar D^0) + {\it
  EA}_2(c,u,c;B_s,\bar D^0,J/\psi)\Bigr)\,;  
  \nn \\
  {\cal A}(B_s \to \pi^0 D^0) &=& \frac{V_{cs} V_{ub}^*}{\sqrt{2}}
  A_2(u,u,c;B_s,\pi^0,D^0)+ \Delta{\cal A}(B_s \to \pi^0 D^0)\,; 
  \nn \\
  {\cal A}(B_s \to \pi^0 \bar D^0) &=& -\frac{V_{us} V_{cb}^*}{\sqrt{2}}
  A_2(c,u,u;B_s,\bar D^0,\pi^0)+  \Delta{\cal A}(B_s \to \pi^0 \bar D^0)\,.
  \label{eq:bsG}
\end{eqnarray}

The classification of $B_s$ decays is summarized in Table~\ref{tab:classbs}.

\section{Strategies for the experimental determination of the
  effective parameters}
\label{sec:pheno}

In this Section we would like to make general comments on the
determination of the effective parameters from the data. A detailed
numerical analysis will be presented in a subsequent work. The
inspection of tables \ref{tab:classbd}--\ref{tab:classbs} allows to
identify the most suitable decays for this determination. We have:
\begin{itemize}
\item[i)] $|E_1|$ can be best determined from $B_d \to D^- K^+$,
  $D_s^+ \pi^-$, $B^+ \to D_s^+ \pi^0$, $B_s \to D_s^- \pi^+$;

\item[ii)] $|E_2|$ can be extracted from $B_d \to D^0 K^0$, $\bar D^0
  K^0$ and $B_s \to \bar D^0 \bar K^0$;

\item[iii)] $|A_1|$ can be determined from $B^+ \to D^+ K^0$;

\item[iv)] $|A_2|$ can be taken from $B_s \to \pi^- D^+$, $\pi^+ D^-$,
  $\pi^0 D^0$, $\pi^0 \bar D^0$;

\item[v)] $|{\it EA}_2|$ can be extracted from $B_d \to \bar D^0
  \phi$;

\item[vi)] $|P_1|$ can be determined from $B_d \to K^+ \pi^-$, $K^0
  \pi^0$ and $B^+ \to K^0 \pi^+$, $K^+ \pi^0$;

\item[vii)] $|P_3|$ can be extracted from $B_d \to \phi \phi$ and $B_s
  \to \pi^+ \pi^-$.
\end{itemize}
In all these cases the parameter in question dominate a given decay or
constitute the only contribution.

An independent determination of ${\it EA}_1$ and $P_2$ is harder:
\begin{itemize}
\item[viii)] ${\it EA}_1$ could play some role in $B^+ \to \pi^+
  J/\psi$, $\phi \pi^+$, $D_s^+ \phi$, $D_s^+ J/\psi$;

\item[ix)] $P_2$ could be non-negligible in $B^+ \to J/\psi K^+$,
  $\phi \pi^+$ and in particular in $B_d \to J/\psi K^0$, $\phi
  \pi^0$, $B_s \to J/\psi \phi$ and $\phi \phi$. On the other hand,
  this contribution being $\ord (1/N^2)$, it appears that this
  determination will only be possible by comparing various decays.  
\end{itemize}

The determination of $P_i^{\sss {\rm GIM}}$ is complicated by the fact
that these contributions are often suppressed by $\ord(\lambda^2)$
relatively to $P_i$ contributions. However, in situations in which
$P_i^{\sss {\rm GIM}}$ are multiplied by CKM factors with large
complex phases, as $V_{us} V_{ub}^*$, CP-asymmetries could be useful
in this respect. 

Finally we do not think that the parameters $P_4$ and $P_4^{\sss {\rm
    GIM}}$ will be determined in the near future, as they are expected
to be very small and one would need very precise data to investigate
their effect.

When data will be available on most of the channels discussed in
Sections \ref{sec:classific} and \ref{sec:classificbs}, it will be
possible to study the flavour dependence of the effective
parameters and to learn more on final state interactions by extracting
the phases of the effective parameters from the data.

\section{Comparison with the Diagrammatic Approach}
\label{sec:comparison}

We will now compare our approach with the diagrammatic approach of
refs.~\cite{diagr1,diagr2}, in particular with the formulation given
in \cite{diagr2}.

The main difference between these two approaches is the following one.
Whereas the diagrammatic approach of \cite{diagr2} is formulated in
terms of diagrams with full W, Z and top quark exchanges, the
approach presented here is formulated directly in terms of local
operators, the basic objects of the effective theory. The main
advantages of formulating non-leptonic decays in terms of
operators are as follows:
\begin{itemize}
\item[i)] The effective parameters introduced in Section
  \ref{sec:general} are directly expressed in terms of matrix elements
  of local operators. Therefore they can be in principle calculated
  in QCD by means of suitable non-perturbative methods. Consequently
  the comparison of phenomenologically extracted effective parameters
  from the data with the values calculated in QCD may offer some
  useful tests and teach us something about strong interactions. Such
  tests are clearly not possible in the approach of \cite{diagr2} as no
  prescription is given on how the phenomenological parameters in this
  approach could be calculated in QCD. 

\item[ii)] The inclusion of QCD perturbative corrections can be
  consistently performed by using the NLO Wilson coefficients in
  ${\cal H}_{\rm eff}$. 

\item[iii)] The issues of renormalization scheme dependences,
  non-factorizable contributions, flavour dependences and in
  particular the important issue of final state interactions can be
  addressed transparently in the operator approach, whereas the
  diagrammatic approach can be in this context sometimes even
  misleading. 
\end{itemize}

Several of these advantages cannot be fully appreciated yet in view of
the limitations of present non-perturbative methods, but this may
improve in the future. 

In spite of these basic differences between our approach and the one
in refs.~\cite{diagr1,diagr2}, it is possible to establish some
connections between them and to show explicitly where they differ from
each other.

In the approach of ref.~\cite{diagr2} the basic parameters for
strangeness-preserving decays are the ``Tree'' (colour favoured)
amplitude $T$, the ``colour suppressed'' amplitude $C$, the
``penguin'' amplitude $P$, the ``exchange'' amplitude $E$, the
``annihilation'' amplitude $A$ and the ``penguin annihilation''
amplitude ${\it PA}$. For strangeness-changing decays one has
$T^\prime$, $C^\prime$, $P^\prime$, $E^\prime$, $A^\prime$, ${\it
  PA}^\prime$. If $Z$-penguins are taken into account one introduces
in addition the ``colour-allowed'' $Z$-penguin $P_{\sss {\rm EW}}$ and
the ``colour-suppressed'' $Z$-penguin $P^C_{\sss {\rm EW}}$. Similarly
$P_{\sss {\rm EW}}^\prime$ and $P_{\sss {\rm EW}}^{\prime C}$ are
introduced for strangeness-changing decays. All these amplitudes
include the relevant CKM factors, whereas in our approach the CKM
factors are not included in the effective parameters. While this
distinction is important for phenomenological applications, we will
omit the CKM parameters in the discussion below, as far as possible.

Let us investigate how the parameters of ref.~\cite{diagr2} are
related to the parameters introduced in Section \ref{sec:general}. It
is sufficient to consider ``unprimed'' amplitudes. The discussion of
``primed'' amplitudes is completely analogous.

As stated above, the amplitudes $T$, $C$, $P$, $E$, $A$, ${\it PA}$,
$P_{\sss {\rm EW}}$ and $P^C_{\sss {\rm EW}}$ are defined in terms of
diagrams in the full theory which contain explicit $W$, $Z$ and top
propagators, whereas our approach is formulated in terms of operators
and diagrams in the effective theory. The connection between these two
approaches can be established by noting that in all diagrams of
ref.~\cite{diagr2} a $W$ propagator is present, whereas in our
approach the operator $Q_1$ contributes to all the effective
parameters. Since a $W$-exchange between two quark lines is
represented in the effective theory by $Q_1$ this is not surprising.

Now, the situation is of course more complicated as QCD corrections to
a $W$-exchange generate other operators for which no diagrams exist in
the diagrammatic approach of ref.~\cite{diagr2}. In the limit $\alpha_s
=0$, $\alpha =0$ we have, however, $C_1=1$ and $C_i=0$ $(i \ne 1)$.
Consequently in this particular limit the effective parameters in our
approach are entirely given in terms of matrix elements of $Q_1$ and
the correspondence between our approach and the one of
ref.~\cite{diagr2} is easier to establish. One can then think that when
QCD and QED corrections are included, the contributions of the
operators $Q_i$ $(i\ne 1)$ are added properly to those of $Q_1$ so
that the effective parameters are scale and renormalization scheme
independent. 

Proceeding in this manner it is easy to establish first the following
correspondence 
\begin{equation}
  \label{eq:correa}
  T \leftrightarrow E_1, \,\, C \leftrightarrow E_2, \,\, A
  \leftrightarrow A_1, \,\, E \leftrightarrow A_2.
\end{equation}
The Zweig-suppressed contributions represented in our approach by
${\it EA}_1$ and ${\it EA}_2$ have not been taken into account in
\cite{diagr2}, although it is straightforward to draw the corresponding
diagrams. In the case of charmless $B$-decays to two pseudoscalars,
considered in \cite{diagr2}, it is very plausible that these
contributions can be neglected as seen in tables
\ref{tab:classbd}--\ref{tab:classbs}. On the other hand ${\it EA}_1$
could be important for $B^+ \to \phi \pi^+$ and in particular in
decays with charm in the final state such as $B^+ \to \pi^+ J/\psi$,
$D_s^+ \phi$ and $D_s^+ J/\psi$. ${\it EA}_2$ is fully responsible for
$B_d \to \bar D^0 \phi$ and could be significant in $B_d \to \pi^0
J/\psi$, $B_s \to \bar D^0 \phi$, $D^0 \phi$,
$J/\psi \phi$, $J/\psi D^0$.  

The case of $P$, ${\it PA}$, $P_{\sss {\rm EW}}$ and $P^C_{\sss {\rm
    EW}}$ is more involved. If one sets $\alpha =0$ we have roughly
speaking
\begin{equation}
  \label{eq:diagp}
  P \leftrightarrow P_1, \qquad {\it PA} \leftrightarrow P_3,
\end{equation}
but this correspondence is a bit oversimplified and requires some
explanation. It is sufficient to discuss only the first relation. 

Let us write $P$ as 
\begin{equation}
  \label{eq:peq}
  P = V_{ud} V^*_{ub} \left( P_u - P_c \right) + V_{td} V^*_{tb}
  \left( P_t - P_c \right) ,
\end{equation}
where we have used the unitarity of the CKM matrix. In the language of
ref.~\cite{diagr2} $P_u$, $P_c$ and $P_t$ denote QCD penguin diagrams
with internal $u$, $c$ and $t$ exchanges. As discussed already in
\cite{bfm,flei}, in the operator approach $P_t$ is represented by the
contributions of the QCD penguin operators $Q_{3-6}$ whereas $P_u$ and
$P_c$ by the matrix elements $\langle Q_1 \rangle^u_{CP}$ ($\langle
Q_2 \rangle^u_{DP}$)  and $\langle Q_1 \rangle^c_{CP}$ ($\langle
Q_2 \rangle^c_{DP}$) respectively. Thus we can establish the relation 
\begin{equation}
  \label{eq:pcpt}
  P_c - P_t \leftrightarrow P_1, \qquad P_c - P_u \leftrightarrow
  P_1^{\sss {\rm GIM}}.
\end{equation}
At this point, following \cite{bfm}, it should be emphasized that
whereas $P_1$ and $P_1^{\sss {\rm GIM}}$ are $\mu$ and renormalization
scheme independent, this is not the case for $P_u$, $P_c$ and
$P_t$. Consequently while $P_1^{\sss {\rm GIM}}$ could possibly be
neglected with respect to $P_1$, the neglect of $P_c$ with respect to
$P_t$ or of $P_u$ with respect to $P_c$ would automatically introduce
unphysical scheme dependences.

In this context we would like to recall that the impact of $P_c$ and
$P_u$ on the extraction of CKM-phases has been investigated for the
first time in \cite{BFcharm}. In the operator language, the importance
of the charm contribution to $P_1$ (``charming penguins'') in
connection with the CLEO data on $B \to K \pi$ has been stressed in
\cite{CHARMING1} and analyzed in detail in \cite{CHARMING2}. Finally
the role of $P_u$ in connection with final state interactions in $B$
decays has been pointed out in \cite{bfm} and analyzed subsequently in
\cite{gewe}-\cite{groro-FSI}, \cite{neubert}.

Let us now discuss the issue of electroweak penguin contributions
which in the approach of ref.~\cite{diagr2} are represented by
$P_{\sss {\rm EW}}$ and $P^C_{\sss {\rm EW}}$. It is sufficient to
discuss $P_{\sss {\rm EW}}$ only. As in the case of $P$ one can write
\begin{equation}
  \label{eq:pew}
  P_{\sss {\rm EW}} = V_{ud} V^*_{ub} \left( P^u_{\sss {\rm EW}} - 
    P^c_{\sss {\rm EW}}
 \right)
 + V_{td} V^*_{tb}  \left( P^t_{\sss {\rm EW}} - P^c_{\sss {\rm EW}} \right) ,
\end{equation}
where in the language of ref.~\cite{bfm,flei,BF98} $ P^u_{\sss {\rm
    EW}}$, $ P^c_{\sss {\rm EW}}$ and $ P^t_{\sss {\rm EW}}$ denote
$Z$-penguin diagrams with internal $u$, $c$ and $t$ exchanges. Strictly
speaking $ P^i_{\sss {\rm EW}}$ have to include also the box
contributions to make them gauge independent.

In our approach the electroweak penguin contributions are represented
by the electroweak penguin operators $Q_{7-10}$, which are included in
$P_1$ and in the other penguin parameters $P_i$. The point is that the
contributions of $Q_{7-10}$ are by themselves scale and
renormalization scheme dependent and have to be considered
simultaneously with other operators to obtain physical scheme
independent results. Consequently when electroweak penguin
contributions and generally ${\cal O} (\alpha)$ effects are included
the correspondences \r{eq:pcpt} generalize to 
\begin{equation}
  P_c - P_t + P^c_{\sss {\rm EW}} - P^t_{\sss {\rm EW}} 
  \leftrightarrow P_1,\qquad 
  P_c - P_u + P^c_{\sss {\rm EW}} - P^u_{\sss {\rm EW}} 
  \leftrightarrow P_1^{\sss
  {\rm GIM}}\,, 
  \label{eq:pcptew}
\end{equation}
with 
\begin{equation}
  P_c + P^c_{\sss {\rm EW}} \leftrightarrow C_1 \langle Q_1 \rangle^c_{\it
  CP} + C_2 \langle Q_2 \rangle^c_{\it DP},\qquad
  P_u + P^u_{\sss {\rm EW}} \leftrightarrow C_1 \langle Q_1 \rangle^u_{\it
  CP} + C_2 \langle Q_2 \rangle^u_{\it DP}\,.
  \label{eq:pcq}
\end{equation}
Here $\langle Q_1 \rangle^c_{\it CP}$ includes the insertion of $Q_1$
in the ${\it CP}$ penguin topology both with gluon exchanges and with
a single photon exchange. It can be considered as a sum of QCD
charming penguins and QED charming penguins. Similar comments apply to
$\langle Q_2 \rangle^c_{\it DP}$, $\langle Q_1 \rangle^u_{\it CP}$ and
$\langle Q_2 \rangle^u_{\it DP}$. The QED charming penguins as well as
the QED $u$-penguins have been identified in \cite{flei} but neglected
with respect to top penguins represented in our approach dominantly by
the contributions of the operators $Q_9$ and $Q_{10}$. Note that
$C_{1,2} = {\cal O}(1)$ whereas $\langle Q_1 \rangle^{c,u}_{\it CP}$
and $\langle Q_2 \rangle^{c,u}_{\it DP}$ with a single photon exchange
are ${\cal O}(\alpha)$. On the other hand $C_9$ and $C_{10}$ are
${\cal O}(\alpha)$ but $\langle Q_9 \rangle_{\it CE}$ and $\langle
Q_{10} \rangle_{\it DE}$ contributing to $P_1$ in
\r{eq:p1gen} are $
{\cal O}(1)$. Thus from the point of view of an expansion in $\alpha$
the QED $c$- and $u$-penguin insertions of $Q_1$ and $Q_2$ are of the
same order as the $Q_9$ and the $Q_{10}$ contributions. In fact, as we
stated above, they have to be both included in order to obtain scheme
independent results. On the other hand one could argue that the strong
enhancement of $C_{9,10}$ through the large top quark mass and the
suppression of $\langle Q_1 \rangle^{c,u}_{\it CP}$ and $\langle Q_2
\rangle^{c,u}_{\it DP}$ through $1/16\pi^2$ factors present in a
perturbative evaluation of these matrix elements makes the neglect of
QED $c$- and $u$-penguin insertions of $Q_1$ and $Q_2$ plausible.
This is supported to some extent by the very weak scheme dependence of
$C_9$ and $C_{10}$. Still one should keep in mind that perturbative
arguments for the smallness of ${\cal O}(\alpha)$ corrections to
$\langle Q_1 \rangle^{c,u}_{\it CP}$ and $\langle Q_2 \rangle^{c,u}_{\it
  DP}$ may not apply and the relevance of these contributions has been
possibly underestimated in the literature so far.

The Zweig-suppressed penguin contributions represented in our approach
by $P_2$ and $P_4$ have not been taken into account in \cite{diagr2}.
As in the case of ${\it EA}_1$ and ${\it EA}_2$ their role in
charmless $B$ decays to two pseudoscalars is expected to be very
small. As seen in tables \ref{tab:classbd}-\ref{tab:classbs}, $P_2$
could play some role in a number of decays with charm in the final
state, in $B_d \to \phi \pi^0$, $K^0 \phi$ and analogous channels in
$B^+$ and $B_s$ decays. $P_4$ being doubly Zweig-suppressed is most
probably negligible in all decays.

Finally we would like to comment on the parameters $P$, $T$,
$P^C_{\sss {\rm EW}}$, $C$ etc. used in refs.~\cite{bfm,flei,BF98}. These
parameters should not be confused with the ones discussed above. They
have been introduced in order to make the analysis of the extraction
of the angle $\gamma$ from $B \to K \pi$ decays more transparent. As
an example we show how the parameters $P$, $T$ and $P^C_{\sss {\rm EW}}$
defined through
\begin{equation}
  \label{eq:padef}
  {\cal A}(B^+ \to \pi^+ K^0) = P\,, \qquad
  {\cal A}(B_d \to \pi^- K^+) = - (P+T+P^C_{\sss {\rm EW}}) 
\end{equation}
are given in terms of the effective parameters introduced in Sections
\ref{sec:general} and \ref{sec:details}. One has 
\begin{eqnarray}
  P &=& V_{us} V_{ub}^* \left( A_1 (s,d,u;B^+,K^0,\pi^+) - P_1^{\sss
  {\rm GIM}}(s,d;B^+,K^0,\pi^+)\right) 
  - V_{ts} V_{tb}^* \,P_1(s,d;B^+,K^0,\pi^+)\,, \nn \\
  P^C_{\sss {\rm EW}} &=& - V_{cs} V_{cb}^*
  \left[ P_1(s,d;B^+,K^0,\pi^+) - P_1(s,u;B_d,K^+,\pi^-) \right]\,,\nn \\
  T &=& V_{us} V_{ub}^* \biggl\{ E_1(s,u,u;B_d, K^+,\pi^-) -
  A_1(s,d,u;B^+,K^0,\pi^+) \nn \\
  && \qquad + \left[P_1^{\sss {\rm GIM}}(s,d;B^+,K^0,\pi^+) -
  P_1^{\sss {\rm GIM}}(s,u;B_d,K^+,\pi^-)\right]\nn \\
  && \qquad - \left[P_1(s,d;B^+,K^0,\pi^+) -
  P_1(s,u;B_d,K^+,\pi^-)\right]\biggr\}\,,
  \label{eq:ptrad}  
\end{eqnarray}
where $\lambda$ and $A$ are the Wolfenstein parameters.

Analogous expressions can be found for other $B \to K \pi$ decays.

\section{Summary}
\label{sec:sum}

In the present paper we have proposed a general framework for
analyzing non-leptonic two-body $B$-decays which combines the operator
language with the diagrammatic language. Following and generalizing the
discussion of ref.~\cite{CHARMING1} we have classified the 
contributions to the matrix elements of the relevant operators in
terms of different topologies of Wick contractions. Subsequently we
have introduced a set of effective parameters which are both
renormalization scale and renormalization scheme independent. As such
they are convenient for phenomenological applications.

On the other hand, being linear combinations of Wilson coefficients
and particular Wick contractions of local operators, these effective
parameters are in principle calculable in QCD. This feature
distinguishes our approach from the diagrammatic approach of
refs.~\cite{diagr1, diagr2} in which no reference to local operators
is made and no prescription for the calculation of the corresponding
parameters is given. 

The formulation given here allows to describe in general terms the
flavour dependence of non-leptonic two-body decays including
non-factorizable contributions and final state interactions. It is
therefore particularly useful for a general model-independent study of
CP violation in $B$ decays.

In the present paper we did not use any symmetry arguments like
$SU(2)$ or $SU(3)$ flavour symmetries. On the other hand we have used
the $1/N$ expansion to indicate a possible hierarchy among the
effective parameters. In particular we have included in our discussion
a number of Zweig-suppressed topologies, which have not been
discussed in the literature.  We have shown that these topologies have
to be included in order to obtain a consistent description of
non-leptonic decays with respect to scale and renormalization scheme
dependences. While such topologies are suppressed in the large N
limit, the role of Zweig-suppressed contributions in non-leptonic
decays is an interesting and important issue, which requires further
theoretical and phenomenological investigations.

As a preparation for phenomenological applications of our formalism we
have presented a classification of two-body $B$ decay channels
according to the effective parameters entering in the decay
amplitudes. This classification enabled us to identify subsets of
channels that, when measured, would allow a direct determination of
the effective parameters making no assumption about non-factorizable
contributions and rescattering. The execution of this program is
deferred to a subsequent work.

The approach presented here allows for a phenomenological description
of non-leptonic decays, once a large number of channels have been
measured. Such a description may teach us about the role of
non-factorizable contributions and about the flavour structure of
non-leptonic $B$ decays. One should hope that in this manner some
regularities will be found. However, without a dynamical input the
formulation presented so far can be considered only as the most
suitable language to describe non-leptonic decays in a manner
consistent with QCD. In order to be predictive some additional input
involving symmetry arguments and dynamical assumptions is needed. We
will return to these issues in a subsequent publication. 

\section*{Acknowledgments}
\label{sec:ack}

We would like to thank Stefan Bosch for invaluable comments on the
manuscript and for checking many equations. We also thank A.
Khodjamirian for discussions.

\appendix

\section{Remaining Contributions}
\label{sec:appew}

We collect here the contributions that were neglected in the analyses
of Sections \ref{sec:classific} and \ref{sec:classificbs}. These
include the terms proportional to $P_4$ and $P_4^{\sss {\rm GIM}}$,
and some of the terms vanishing in the limit of $SU(2)$. The $SU(2)$
breaking effects come on the one hand from the matrix elements of
electroweak penguin operators, and on the other hand from $O(\alpha)$
effects in the matrix elements of operators $Q_{1-6}$. We stress that,
for consistency, if one takes into account the effects of electroweak
penguin operators, one should also include $O(\alpha)$ effects in the
matrix elements of operators $Q_{1-6}$. Therefore, in this case
operators $Q_{3-6}$ can also contribute to $\Delta I=1$ (or $3/2$)
transitions, due to $O(\alpha)$ effects in the matrix elements. These
contributions will be, for example, of the form
$P_1(d,u;B^+,\pi^+,\pi^0)-P_1(d,d;B^+,\pi^0,\pi^+)$, vanishing in the
$SU(2)$ symmetric limit, and will be proportional to $\alpha$. Thus,
while the individual $P_i$ penguin parameters are of order $\alpha_s$,
differences of penguin parameters vanishing in the $SU(2)$ symmetric
limit are at least of order $\alpha$.

\subsection*{Class A decays}

We have (here and in the following we give results in
units of $G_F/\sqrt{2}$):
\begin{eqnarray}
  \Delta{\cal A}(B_d \to \bar{D}^0 \pi^0) &=& \frac{V_{ud} V_{cb}^*}{\sqrt{2}}
  \Bigl[ {\it EA}_2(c,u,d;B_d,\bar{D}^0,\pi^0)-
  {\it EA}_2(c,u,u;B_d,\bar{D}^0,\pi^0)\Bigr]\,.
  \label{eq:bdpia}
\end{eqnarray}

\subsection*{Class B decays}

We have:
\begin{eqnarray}
  \Delta{\cal A}(B^+ \to D_s^+ \pi^0) &=& - \frac{V_{cs} V_{ub}^*}{\sqrt{2}}
  \Bigl[{\it EA}_1(s,c,u;B^+,D_s^+,\pi^0) - {\it
  EA}_1(s,c,d;B^+,D_s^+,\pi^0) \Bigr]\,,\nn \\
  \Delta{\cal A}(B_s \to \pi^0 \phi) = &-& \frac{V_{us} V_{ub}^*}{\sqrt{2}}
  \Biggl(\Bigl[P_2^{\sss {\rm GIM}}(s,d;B_s,\pi^0,\phi) -
  P_2^{\sss {\rm GIM}}(s,u;B_s,\pi^0,\phi) \Bigr]  \nn \\
  &+& \Bigl[P_4^{\sss {\rm GIM}}(d,s;B_s,\pi^0,\phi) -
  P_4^{\sss {\rm GIM}}(u,s;B_s,\pi^0,\phi) \Bigr] \Biggr)  \nn \\
  &-& \frac{V_{ts} V_{tb}^*}{\sqrt{2}}
  \Bigl[P_4(d,s;B_s,\pi^0,\phi) -
  P_4(u,s;B_s,\pi^0,\phi) \Bigr]\,.  
  \label{eq:bsB2a}
\end{eqnarray}

\subsection*{Class C decays}

We have:
\begin{eqnarray}
  \Delta{\cal A}(B_s \to J/\psi \phi) &=& - V_{ts} V_{tb}^* 
  P_4(c,s;B_s,J/\psi,\phi) - V_{us} V_{ub}^* 
  P_4^{\sss {\rm GIM}}(c,s;B_s,J/\psi,\phi) \,. 
  \label{eq:bsCa}
\end{eqnarray}

\subsection*{Class D decays}

We have:
\begin{eqnarray}
  \Delta{\cal A}(B_d \to \pi^0 \pi^0) = &-& V_{ud} V_{ub}^*
  \Biggl(\Bigl[{\it EA}_2(u,u,d;B_d,\pi^0,\pi^0) -
  {\it EA}_2(u,u,u;B_d,\pi^0,\pi^0)\Bigr] \nn \\
  &+&
  \Bigl[\frac{1}{2} P_4^{\sss {\rm GIM}}(u,u;B_d,\pi^0,\pi^0) - 
  P_4^{\sss {\rm GIM}}(u,d;B_d,\pi^0,\pi^0) +
  \frac{1}{2} P_4^{\sss {\rm GIM}}(d,d;B_d,\pi^0,\pi^0)\Bigr]  \nn \\
  &-& \Bigl[P_2^{\sss {\rm GIM}}(d,u;B_d,\pi^0,\pi^0) -
  P_2^{\sss {\rm GIM}}(d,d;B_d,\pi^0,\pi^0) \Bigr] \Biggr) \nn \\
  &-& V_{td} V_{tb}^* \Biggl(
  \Bigl[\frac{1}{2} P_4(u,u;B_d,\pi^0,\pi^0)  - 
   P_4(u,d;B_d,\pi^0,\pi^0) +
  \frac{1}{2} P_4(d,d;B_d,\pi^0,\pi^0)\Bigr]\nn \\
  &-& \Bigl[P_2(d,u;B_d,\pi^0,\pi^0) -
  P_2(d,d;B_d,\pi^0,\pi^0) \Bigr] \Biggr)\,; 
  \nn \\
  \Delta{\cal A}(B_d \to \pi^0 J/\psi) = &+& \frac{V_{cd} V_{cb}^*}{\sqrt{2}}
  \Bigl[{\it EA}_2(c,c,d;B_d,J/\psi,\pi^0) -
  {\it EA}_2(c,c,u;B_d,J/\psi,\pi^0)\Bigr]\nn \\
  &-&\frac{V_{ud} V_{ub}^*}{\sqrt{2}}
  \Bigl[P_4^{\sss {\rm GIM}}(d,c;B_d,\pi^0,J/\psi) - 
  P_4^{\sss {\rm GIM}}(u,c;B_d,\pi^0,J/\psi)\Bigr] \nn \\
  &+& \frac{V_{td} V_{tb}^* }{\sqrt{2}}
  \Bigl[P_4(u,c;B_d,\pi^0,J/\psi)  -
  P_4(d,c;B_d,\pi^0,J/\psi)\Bigr] \,; 
  \nn \\
  \Delta{\cal A}(B^+ \to \pi^+ \pi^0) = &+& \frac{V_{ud} V_{ub}^*}{\sqrt{2}}
  \Biggl(\Bigl[A_1(d,d,u;B^+,\pi^0,\pi^+) -
  A_1(d,u,u;B^+,\pi^+,\pi^0) \Bigr]+ \nn \\
  &&\qquad \qquad \Bigl[ {\it EA}_1(d,u,d;B^+,\pi^+,\pi^0) -
  {\it EA}_1(d,u,u;B^+,\pi^+,\pi^0) \Bigr]- \nn \\ 
  &&\qquad \qquad \Bigl[ P_1^{\sss {\rm GIM}}(d,d;B^+,\pi^0,\pi^+) -
  P_1^{\sss {\rm GIM}}(d,u;B^+,\pi^+,\pi^0) \Bigr] +\nn \\
  && \qquad \qquad \Bigl[P_2^{\sss {\rm GIM}}(d,u;B^+,\pi^0,\pi^+) -
  P_2^{\sss {\rm GIM}}(d,d;B^+,\pi^0,\pi^+) \Bigr] \Biggr)\,; \nn \\
  \Delta{\cal A}(B_s \to \pi^0 \bar K^0) = &-& \frac{V_{ud} V_{ub}^*}{\sqrt{2}}
  \Bigl[P_2^{\sss {\rm
  GIM}}(d,d;B_s,\pi^0,\bar K^0) - P_2^{\sss {\rm
  GIM}}(d,u;B_s,\pi^0,\bar K^0)\Bigr] \nn \\ 
  &-& \frac{V_{td} V_{tb}^*}{\sqrt{2}} \Bigl[P_2(d,d;B_s,\pi^0,\bar K^0) - 
  P_2(d,u;B_s,\pi^0,\bar K^0)\Bigr]\,.
  \label{eq:bsD1a}
\end{eqnarray}

\subsection*{Class E decays}

We have:
\begin{eqnarray}
  \Delta{\cal A}(B_d \to K^0 \pi^0) = &-& \frac{V_{us} V_{ub}^*}{\sqrt{2}}
  \Bigl[ P_2^{\sss {\rm
  GIM}}(s,d;B_d,\pi^0,K^0)- P_2^{\sss {\rm
  GIM}}(s,u;B_d,\pi^0,K^0) \Bigr] \nn \\ 
  &-&\frac{ V_{ts} V_{tb}^*}{\sqrt{2}} \Bigl[ P_2(s,d;B_d,\pi^0,K^0)-
  P_2(s,u;B_d,\pi^0,K^0) \Bigr]\,;  
  \nn \\
  \Delta{\cal A}(B^+ \to K^+ \pi^0) = &-& \frac{V_{us} V_{ub}^*}{\sqrt{2}}
  \Biggl(\Bigl[ {\it EA}_1(s,u,u;B^+,K^+,\pi^0) -{\it
  EA}_1(s,u,d;B^+,K^+,\pi^0) \Bigr] + \nn \\ 
  && \qquad \qquad \Bigl[ P_2^{\sss {\rm
  GIM}}(s,d;B^+,\pi^0,K^+)- P_2^{\sss {\rm
  GIM}}(s,u;B^+,\pi^0,K^+) \Bigr] \Biggr)\nn \\ 
  &-&\frac{ V_{ts} V_{tb}^*}{\sqrt{2}} \Bigl[P_2(s,d;B^+,\pi^0,K^+)-
  P_2(s,u;B^+,\pi^0,K^+) \Bigr] \Biggr)\,; \nn \\
  \Delta{\cal A}(B_s \to \phi \phi) = &-& V_{us} V_{ub}^*
  \,P_4^{\sss {\rm
  GIM}}(s,s;B_s,\phi,\phi)  
  - V_{ts} V_{tb}^*\, P_4(s,s;B_s,\phi,\phi)\,; \nn \\
  \Delta{\cal A}(B_s \to \pi^0 \pi^0) = &-& V_{us} V_{ub}^*
  \Biggl(\Bigl[\frac{1}{2} P_4^{\sss {\rm GIM}}(u,u;B_s,\pi^0,\pi^0) +
  \frac{1}{2} P_4^{\sss {\rm GIM}}(d,d;B_s,\pi^0,\pi^0) - \nn \\ 
  && P_4^{\sss {\rm GIM}}(u,d;B_s,\pi^0,\pi^0) \Bigr] + 
  \Bigl[{\it EA}_2(u,u,d;B_s,\pi^0,\pi^0)- {\it
  EA}_2(u,u,u;B_s,\pi^0,\pi^0) \Bigr]\Biggr)\nn \\ 
  &-& V_{ts} V_{tb}^* \Bigl[\frac{1}{2} P_4(u,u;B_s,\pi^0,\pi^0) +
  \frac{1}{2} P_4(d,d;B_s,\pi^0,\pi^0) - 
  P_4(u,d;B_s,\pi^0,\pi^0) \Bigr]\,. 
  \nn \\
  \label{eq:bsE2a}
\end{eqnarray}

\subsection*{Class F decays}

We have:
\begin{eqnarray}
  \Delta{\cal A}(B_d \to \phi \pi^0) = &+&\frac{V_{td} V_{tb}^*}{\sqrt{2}}
  \Bigl[P_4(u,s;B_d,\pi^0,\phi) - P_4(d,s;B_d,\pi^0,\phi)
  \Bigr] \nn \\
  &+& \frac{V_{ud} V_{ub}^*}{\sqrt{2}}
  \Bigl[P_4^{\sss {\rm GIM}}(u,s;B_d,\pi^0,\phi) - P_4^{\sss {\rm
  GIM}}(d,s;B_d,\pi^0,\phi) 
  \Bigr]\,;
  \nn \\
  \Delta{\cal A}(B_d \to \phi \phi) = &-& V_{td} V_{tb}^*
  \,P_4(s,s;B_d,\phi,\phi) -  V_{ud} V_{ub}^* \,P_4^{\sss {\rm
  GIM}}(s,s;B_d,\phi,\phi) \,. 
  \label{eq:bsD2a}  
\end{eqnarray}

\subsection*{Class G decays}

We have:
\begin{eqnarray}
  \Delta{\cal A}(B_s \to \pi^0 J/\psi) = 
  &-& \frac{V_{us} V_{ub}^*}{\sqrt{2}}
  \Bigl[P_4^{\sss {\rm GIM}}(d,c;B_s,\pi^0,J/\psi) -
  P_4^{\sss {\rm GIM}}(u,c;B_s,\pi^0,J/\psi)\Bigr] \,;\nn \\
  \Delta{\cal A}(B_s \to \pi^0 D^0) &=& \frac{V_{cs} V_{ub}^*}{\sqrt{2}}
  \Bigl[{\it EA}_2(u,c,u;B_s,D^0,\pi^0) - 
  {\it EA}_2(u,c,d;B_s,D^0,\pi^0)\Bigr]\,; 
  \nn \\
  \Delta{\cal A}(B_s \to \pi^0 \bar D^0) &=& -\frac{V_{us} V_{cb}^*}{\sqrt{2}}
  \Bigl[{\it EA}_2(c,u,u;B_s,\bar D^0,\pi^0) - 
  {\it EA}_2(c,u,d;B_s,\bar D^0,\pi^0)\Bigr]\,.
  \label{eq:bsGa}
\end{eqnarray}

\vfill\eject
 
\end{document}